\documentclass[%
 reprint,
showkeys,
 amsmath,
 amssymb,
 pra,
floatfix,
longbibliography
]{revtex4-1}

\usepackage{graphicx}
\usepackage{amsmath}
\usepackage{algorithm}
\usepackage{algorithmicx}
\usepackage{algpseudocode}
\usepackage{dcolumn}
\begin{document}

\preprint{APS/123-QED}


\title{Finding Hadamard Matrices by a Quantum Annealing Machine}

\author{Andriyan Bayu Suksmono}
\affiliation{School of Electrical Engineering and Informatics\\
Institut Teknologi Bandung, Jl. Ganesha No.10, Bandung, Indonesia
}%

\author{Yuichiro Minato}
\affiliation{MDR Inc., Hongo 2-40-14-3F, Bunkyo-ku, Tokyo, Japan
}%

\date{\today}
%
\begin{abstract}
 Finding a Hadamard matrix (H-matrix) among the set of all binary matrices of corresponding order is a hard problem, which potentially can be solved by quantum computing. We propose a method to formulate the Hamiltonian of finding H-matrix problem and address its implementation limitation on existing quantum annealing machine (QAM) that allows up to quadratic terms, whereas the problem naturally introduces higher order ones. For an $M$-order H-matrix, such a limitation increases the number of variables from $M^2$ to $(M^3+M^2-M)/2$, which makes the formulation of the Hamiltonian too exhaustive to do by hand. We use symbolic computing techniques to manage this problem. Three related cases are discussed: (1) finding $N<M$ orthogonal binary vectors, (2) finding $M$-orthogonal binary vectors, which is equivalent to finding a H-matrix, and (3) finding $N$-deleted vectors of an $M$-order H-matrix. Solutions of the problems by a $2$-body simulated annealing software and by an actual quantum annealing hardware are also discussed.
\end{abstract}

\maketitle

\section{Introduction}
Solving a hard problem is one of the most important issues in computational science. This kind of problem is characterized by its complexity; which is the required number of computing resource for doing the computation, which grows beyond polynomial against the input's size. Researchers have put a lot of effort to solve such a problem, among others by employing quantum mechanics in the machinery of the computation process.

In a microscopic level, nature works under quantum mechanical principles which is hardly possible to be simulated by classical computing machines \cite{feynman1982}. This phenomenon drives the progress of quantum computing, both on the theory at the beginning \cite{Deutsch1985,Shor1997} and then is followed by the implementation of the quantum computer itself \cite{Chuang2001,johnson2011}. At present, a few kinds of early quantum computer models have been proposed and built, which mainly can be categorized into either a quantum gate model or a quantum annealing processor. Referring to these two classes, we will call a quantum computing machine either a QGM (Quantum Gate Machine) or a QAM (Quantum Annealing Machine), respectively.

In this paper, we will discuss a problem of finding a Hadamard matrix, denoted by \emph{H-SEARCH} and its related problems, especially the formulation of their Hamiltonians for implementation on a QAM and experimenting with them using both of a simulator and a real quantum annealer. In \cite{suksmono2018} and \cite{suksmono2012} we have suggested that finding a Hadamard matrix (H-matrix) among the set of all possible binary matrices of corresponding order, i.e. the H-SEARCH, is a hard problem. First proposed by Sylvester \cite{sylvester1867} and then explored by Hadamard \cite{hadamard1893}, an $M$-order H-matrix can be defined as an orthogonal binary $\{-1,+1\}$ matrix of size $M\times M$, where $M=1,\ 2,\ 4,\ 8, \ \cdots, 4k,  \cdots$ \cite{hedayat1978},  \cite{horadam2007}. The H-matrix is an important discrete structure in scientific fields and engineering applications \cite{seberry2005}, \cite{garg2007}. Construction of a $2^n$ order H-matrix, for any positive integer $n$, can be done easily by using Sylvester's method. Several construction methods for other values that different from $M = 2^n$ have also been proposed \cite{dadegoldberg1959, williamson1944, bush1971a, bush1971b}. Nevertheless, there is no general method for constructing (nor finding) a $4k$ order H-matrix which can be applied to every positive integer $k$. Although no proof yet exists, it is conjectured that there is a H-matrix of order $4k$ for every positive integer $k$ \cite{paley1933, wallis1976}. 

Existing Hadamard matrix construction methods, including the Sylvester's and other's methods proposed in \cite{dadegoldberg1959, williamson1944, bush1971a, bush1971b}, can be considered as \emph{analytical} methods. We have formulated a tentative method that can be categorized as a \emph{probabilistic} one, which is based on the SA (simulated annealing) \cite{metropolis1953,kirkpatrick1983, cerny1985} and later on SQA (simulated quantum annealing) \cite{battaglia2005, kadowaki1988, santoro2002}. We have successfully found some low-order H-matrices that cannot trivially be constructed by the Sylvester method, either by SA \cite{suksmono2012} or the SQA \cite{suksmono2018}. However, direct implementation of the method on existing QAM is hindered by unrealizable absolute terms in the energy function (Hamiltonian). Changing the absolute terms into their equivalent square terms will generate quartic terms, whereas existing QAM only allows up to quadratic terms to be implemented. A possible solution is by transforming the energy function containing high order terms into ones with up to two-body interaction terms using Boolean reduction \cite{biamonte2008}, \cite{perdomo2008}. In our case of H-SEARCH problem, however, it involves a large number of terms where the mathematical manipulation by hand is not an easy task.

In this paper, we also extend the H-SEARCH into a problem of finding a set of $N<M$ orthogonal (ortho-set) of binary vectors. Along with H-SEARCH, which is equivalent to finding $M$ ortho-set of $M$-order binary vectors, we also address H-matrix completion problem of finding $N$-deleted vectors of a given $M$-order H-matrix. The large number of terms in the Hamiltonian of these problems requires both a systematic and automated solution. We propose a method to systematically perform Boolean reduction on a large number of terms and encourage the usage of symbolic computation to formulate the energy function which leads to the Hamiltonian of the problems. We present some examples of finding low-order H-matrices to enlight the proposed method. Additionally, we use D-Wave \emph{neal} package to find the solutions of the formulated 2-body interacting Hamiltonian of the problems by using simulated annealing and also do the implementation on an actual quantum annealer by using the D-Wave's DW2000Q quantum processor.

The rests of the paper are organized as follows. In Section II, we briefly discuss the QAM, finding H-matrix by energy minimization, problem of transforming higher-order terms into two-terms Hamiltonian with large number of terms, and describe the proposed method. Section III presents the computation case of low order H-matrices and analysis of corresponding experimental results, both by simulation and implementation on a quantum annealer. The last Section presents discussion and concludes the paper.
\section{Methods}
\subsection{Quantum Annealing Machines}
We refer a QAM or an adiabatic quantum computing machine as a configurable or a programmable quantum Ising systems $\hat{H}_{pot}$, whose transverse magnetic field $\hat{H}_{kin}$ can be controlled and the state of its spins can be read individually upon completion of an adiabatic quantum evolution. The Hamiltonians of such a system; for a given spin configuration $\left\{ \hat{\sigma}^{\alpha}_k \right\} \equiv \hat{\sigma}$; where $\alpha \in \{x, y, z\}$ , $k \in K = \{1, 2, \cdots, i,j, \cdots \}$ is the set of lattice's indices, is given by
\begin{equation}
  \hat{H}_{pot} \left( \hat{\sigma} \right) \equiv -\sum_{i\neq j} J_{ij} \hat{\sigma}_i^z \hat{\sigma}_j^z -\sum_i h_i \hat{\sigma}_i^z
  \label{EQ_Hpot}
\end{equation}
and,
\begin{equation}
  \hat{H}_{kin}\left( \hat{\sigma} \right)\equiv -\Gamma \sum_i \hat{\sigma}_i^x    
  \label{EQ_Hkin}
\end{equation}
where $J_{ij}$ is a coupling constant or interaction strength between a spin at site $i$ with a spin at site $j$, $h_j$ is magnetic strength at site $j$, and $\{\hat{\sigma}_i^z, \hat{\sigma}_i^x\}$ are Pauli's matrices at site-$i$. In the QA \cite{battaglia2005, jiang2017, hormozi2017, kadowaki1988, santoro2002, boixo2014, heim2015, isakov2016, ronnow2014, mazzola2017a, martonak2004, titiloye2011, zick2015}, quantum fluctuation is elaborated by introducing a transverse magnetic field $\Gamma$. To solve a problem by using QAM, we have to encode the variables into spins with their corresponding Ising coefficients $\{h_i, J_{ij}\}$. Then it is executed by the following quantum adiabatic evolution 
\begin{equation}
    \hat{H}_{QA}\left( \hat{\sigma}, t \right)=\left(1-\frac{t}{\tau} \right) \hat{H}_{kin} \left( \hat{\sigma} \right) + \frac{t}{\tau}\hat{H}_{pot}\left( \hat{\sigma} \right) 
  \label{EQ_HQA}
\end{equation}
where $t \in [0,\tau]$. By keeping the system in an adiabatic condition during the process, the ground-state at the end of the evolution of the system will represent a solution of the problem.

We can see from Eq.(\ref{EQ_Hpot}) that the Hamiltonian includes up to quadratic terms, so that in principle it only allows encoding of quadratic (binary) problems. When the problem contains higher order terms than quadratic, we have to find a way to convert it into expressions that only include up to quadratic. Additionally, since the number of the spins/ qubits are related to the number of binary variables, it further constraints the size of the problem that can be handled and therefore limits the machine's capability. 

Some efforts to implement the QAM have been initiated, among others is the construction of quantum annealer where the spin is manufactured  as a superconducting quantum device called RF-SQUID (Radio Frequency-Super Conducting Quantum Interference Device) \cite{johnson2011}. The scalability of the device makes it possible for the number of spins (qubits) grows very rapidly, whose last generation at the time of this writing achieves more than 2000. This device has been applied to solve various kinds of problem, such as, quantum factorization \cite{Jiang2018}, hand written digit recognition \cite{Benedetti2017}, computational biology \cite{Li2018}, and hydrologic inverse analysis \cite{OMalley2018}.
\subsection{Finding a Hadamard Matrix By Energy Minimization}
Consider an $M=2,\ 4,\ 8,\ 12, \cdots, 4k$ order binary matrix $B$ whose elements are $b_{i,j} \in \{-1,+1\}$, with $k$ a positive integer (we have omitted $M=1$ case due to its triviality). By writing the $i^{th}$ column vector of $B$ as 
%
\begin{equation}
  \vec{b}_i=\left( b_{0,i} \ b_{1,i} \ \cdots \ b_{M-1,i}\right)^T
  \label{EQ_bvect}  
\end{equation}
%
where $(\cdot)^T$ denotes matrix transpose operation, we can express the matrix $B = \left( \vec{b}_0 \ \vec{b}_1 \cdots \ \vec{b}_{M-1} \right)$ as

\begin{equation}
    B = 
    \begin{pmatrix} 
     b_{0,0} & b_{0,1} & \cdots & b_{0,M-1} \\
     b_{1,0} & b_{1,1} & \cdots & b_{1,M-1} \\
     \cdots  & \cdots  & \cdots & \cdots    \\
     b_{M-1,0} & b_{M-1,1} & \cdots & b_{M-1,M-1} \\
    \end{pmatrix}
    \nonumber
\end{equation}
%
To indicate the orthogonality relationship among the column vectors of $B$, we define a matrix $D \equiv B^{T}B$, which explicitly can be written as
\begin{equation}
    \nonumber
    D=
    \begin{pmatrix} 
    \langle\vec{b}_0, \vec{b}_0 \rangle & \langle\vec{b}_0, \vec{b}_1 \rangle &\cdots & \langle\vec{b}_0, \vec{b}_{M-1} \rangle \\
    \langle\vec{b}_1, \vec{b}_0 \rangle & \langle\vec{b}_1, \vec{b}_1 \rangle &\cdots & \langle\vec{b}_1, \vec{b}_{M-1} \rangle \\
    \cdots & \cdots & \cdots \\
    \langle\vec{b}_{M-1}, \vec{b}_0 \rangle & \langle\vec{b}_{M-1}, \vec{b}_1 \rangle &\cdots & \langle\vec{b}_{M-1}, \vec{b}_{M-1} \rangle \\
        \end{pmatrix}
\end{equation}
where $\langle\vec{b}_i, \vec{b}_j \rangle = \vec{b}_i^T\cdot \vec{b}_j $ is the inner product between (column) vector $\vec{b}_i$ and $\vec{b}_j$. By denoting $d_{ij} \equiv \langle \vec{b}_i, \vec{b}_j \rangle$ and knowing that $\langle \vec{b}_i, \vec{b}_i \rangle =M$, the indicator matrix $D$ can be rewritten as
\begin{equation}
    D=
    \begin{pmatrix} 
    M & d_{0,1} &\cdots & d_{0,M-1} \\
    d_{1,0} & M &\cdots & d_{1,M-1} \\
    \cdots & \cdots & \cdots \\
    d_{M-1,0} & d_{M-1,1} &\cdots & M
    \end{pmatrix}
    \label{EQ_Dmatx}
\end{equation}
When all of $d_{i,j}=0$ in Eq.(\ref{EQ_Dmatx}), then, by definition, $B$ is an orthogonal matrix; which due to its elements of being $\{-1,1 \}$, is also a H-matrix. Consequently, we can define the energy function as the sum of absolute values of the off-diagonal elements of $D$, which implies that a zero energy value corresponds to all of the column vectors being orthogonal to each other, whereas a non-zero value indicates that there is at least a pair of non-orthogonal vectors among them. Since $D$ is a symmetric matrix, it is sufficient to consider only an upper- (or lower-) diagonal part of $D$, i.e., we can define the energy function for a given set of column vectors of $\left\{\vec{b}_i \right\}$ as

\begin{equation}
    E_a\left(\left\{\vec{b}_i \right\} \right) \equiv \sum_{i<j} \left| \langle \vec{b}_i, \vec{b}_j \rangle \right|=\sum_{i<j} \left| d_{i,j}\right|
    \label{E_abs}
\end{equation}

Furthermore, since we need to express the energy function as products of binary variables $b_{i,j}$'s, we have to change the absolute function into a square function. Then, Eq.(\ref{E_abs}) becomes
\begin{equation}
    E_s\left(\left\{\vec{b}_i \right\} \right)= \sum_{i<j} \left( \langle \vec{b}_i, \vec{b}_j \rangle \right)^2
    \label{E_sqr}
    \nonumber
\end{equation}
Considering Eq.(\ref{EQ_bvect}), we can show that the square of the inner product between two binary vectors $d_{i,j}^2 = \langle \vec{b}_i,\vec{b}_j \rangle ^2$ are given by

\begin{equation}
    d_{i,j}^2 = \left( b_{0,i}b_{0,j} + b_{1,i}b_{1,j} \cdots + b_{M-1,i}b_{M-1,j} \right)^2
    \label{EQ_dij2}
    \nonumber
\end{equation}
Expansion of the square terms yields the following expression
\begin{widetext}
 \begin{align*}
  d_{i,j}^2 = 
   b_{0,i}^2b_{0,j}^2+b_{1,i}^2b_{1,j}^2 + \cdots +b_{M-1,i}^2b_{M-1,j}^2 \\
    + 2b_{0,i}b_{0,j}b_{1,i}b_{1,j} + 2b_{0,i}b_{0,j}b_{2,i}b_{2,j} + \cdots + 2b_{0,i}b_{0,j}b_{M-1,i}b_{M-1,j} \\
    + 2b_{1,i}b_{1,j}b_{2,i}b_{2 ,j} + 2b_{1,i}b_{1,j}b_{3,i}b_{3,j} + \cdots + b_{1,i}b_{1,j}b_{M-1,i}b_{M-1,j} \\ 
    + \cdots \cdots \\
    + 2b_{M-2,i}b_{M-2,j}b_{M-1,i}b_{M-1,j}
 \end{align*}
\end{widetext}
Since $b_{i,j} \in \{-1,+1\}$, then $b_{i,j}^2 = 1$. Therefore, we can simplify $d_{i,j}^2$ into
\begin{equation}
  d_{i,j}^2 = M + 2\sum_{m<n<M} b_{m,i}b_{m,j}b_{n,i}b_{n,j}
\end{equation}      
Finally, the energy function related to orthogonality condition of all pairs of the column vectors in $B$ can be expressed as

\begin{widetext}
  \begin{equation}
    E_s\left( \{\vec{b}_i\} \right)=\sum_{i<j} d_{i,j}^2 = \frac{M^2(M-1)}{2}+ 2\sum_{m<n<M,i<j<M}  b_{m,i}b_{m,j}b_{n,i}b_{n,j} 
  \label{E_quartic}
  \end{equation}  
\end{widetext}

In our previous papers \cite{suksmono2012, suksmono2018}, we have employed energy function that is similar to Eq.(\ref{E_abs}). For implementation in a QAM, we need a modified form of Eq.(\ref{E_quartic}). First, we introduce a spin variable $s_i=\{-1,+1\}$ and a (Boolean) binary variable $q_i=\{0,1\}$. They are related by the following transforms
\begin{equation}
    s_i=\frac{1}{2}\left( 1-q_i \right)
    \label{EQ_s2q}
\end{equation}

\begin{equation}
    q_i=\left( 1-2s_i \right)
    \label{EQ_q2s}
\end{equation}

Considering that the elements of a H-matrix are $\{-1,1\}$, it is natural to formulate the energy function of H-SEARCH in the $s$-domain. Therefore, first we will  express the energy function in this domain. We also reassign the index of the variables from the row-column format to a single contiguous indices ranging from $0$ to $(M-1)^2$, i.e., we prefer to use a single indexed variable $s_i$ rather than the previously double indexed $b_{i,j}$. The notation of its related energy function is changed by $E_s\left( \{\vec{b}_i\} \right) \rightarrow E_k\left(\{s_i\} \right) \equiv E_k(s_i)$. Accordingly, Eq.(\ref{E_quartic}) is changed into 

\begin{equation}
  E_k\left(s_i\right)=\frac{M^2(M-1)}{2}+ 2\sum_{i<j<m<n<(M-1)^2} s_i s_j s_m s_n
  \label{E_ks}
\end{equation}
Considering the implementation in a QAM, we further need to transform the $k$-body energy function of Eq.(\ref{E_ks}) to a 2-body energy function, which normally is formulated in the $q$-domain. Following the formulation described in \cite{biamonte2008, perdomo2008}, a $k$-body interaction can be converted into a 2-body interaction by substitution and an additional compensation term as follows
\begin{equation}
    q_i q_j \leftarrow q_k + H_{\land}(q_i, q_j, q_k; \delta_{i,j})
    \label{EQ_BoolRed1}
\end{equation}
where the compensation term is given by
\begin{equation}
    H_{\land}(q_i, q_j, q_k; \delta_{i,j}) = \delta_{i,j} \left( 3q_k +q_iq_j -2q_iq_k - 2q_jq_k \right)
    \label{EQ_BoolRed2}
\end{equation}
According to \cite{perdomo2008}, the value of $\delta_{i,j}$ should be chosen to be larger than the maximum value of its substituted function of energy, which in our case is $d_{ij}^2$. The substitution variable is also called an ancillary variable or simply called \emph{ancilla}, whereas the original one will be refered to as \emph{main variable}.

The input of a QAM or its simulator needs parameters (Ising coefficients) in $s$-domain, as indicated by Eq.(\ref{EQ_Hpot}), Eq.(\ref{EQ_Hkin}), and Eq.(\ref{EQ_HQA}). Therefore, from a general $k$-body interaction in $s$-domain energy function $E_k(s_i)$, we will transform it into $E_2(s_i)$ and eventually to its Hamiltonian $\hat{H}_2\left( \hat{\sigma}_i^z \right)$ by using steps given by the following Hamiltonian's \emph{construction diagram} 

\begin{equation}
  E_k(s_i) \rightarrow E_k(q_i) \rightarrow E_2(q_i) \rightarrow E_2(s_i)\rightarrow \hat{H}_2\left(\hat{\sigma}_i^z\right)
  \label{transdomain_diagram}
\end{equation}
%
In the following discussions, we will describe each of these transforms in the diagram and present examples to clarify the construction process.

First note that according to the transform given by Eq.(\ref{EQ_s2q}), the $q$-transformed energy from Eq.(\ref{E_ks}) into $E_k(q_i)$ will contains quartic terms $q_i q_j q_m q_n$. We observed that each term in $q_i q_j$ actually comes from a product of two column vectors $\sum_{r<M} q_{r,i}q_{r,j}$ (and so is $q_m q_n$). Therefore, it will be more convenient to arrange the substitution of $q_{r,i} q_{r,j} \leftarrow q_{r,t}$ (and so is  $q_m q_n$) column-wise. Then, we can make the arrangement of variables of the H-matrix and related ancillas as shown by the following table 

\begin{widetext}
  \begin{center} 
   \begin{tabular}{l l l l l| l l l l}
     \multicolumn{5}{c|}{main variables} &
     \multicolumn{4}{c}{ancillas} \\
     \hline
     $q_0$ & $q_M$ & $q_{2M}$ & $\cdots$ & $q_{(M-1)M}$ & $q_{M^2}$ & $q_{M^2+M}$ & $\cdots$ & $q_{M^2+M^2(M-1)/2-M+1}$\\
     $q_1$ & $q_{M+1}$ & $q_{2M+1}$ &$\cdots$ & $q_{(M-1)M+1}$ & $q_{M^2+1}$ & $q_{M^2+M+1}$ & $\cdots$ & $q_{M^2+M^2(M-1)/2-M+2}$\\
     $\cdots$ & $\cdots$ & $\cdots$ & $\cdots$ & $\cdots$ & $\cdots$ & $\cdots$ & $\cdots$  & $\cdots$\\
     $q_{M-1}$ & $q_{2M-1}$ & $q_{3M-1}$ & $\cdots$ & $q_{M^2-1}$ & $q_{M^2+M-1}$ & $q_{M^2+2M-1}$ & $\cdots$ & $q_{M^2+M^2(M-1)/2}$
   \end{tabular}
   \label{TBL_qHGEN}
 \end{center}
\end{widetext}
Left part of the table shows (main) variables of the matrix elements, whereas the right parts are ancillas. Using this arrangement, the substitution of a product of two binary variables by a single binary variable is done as follows
\begin{widetext}
 \begin{equation}
    \begin{matrix}
    q_0q_M \leftarrow q_{M^2} & q_0q_{2M} \leftarrow q_{M^2+M} & \cdots \\
    q_1q_{M+1} \leftarrow q_{M^2+1} & q_1q_{2M} \leftarrow q_{M^2+M+1} & \cdots \\
    \cdots & \cdots & \cdots \\
    q_{M-1}q_{2M-1} \leftarrow q_{M^2+M+1} & q_{M-1}q_{3M-1} \leftarrow q_{M^2+2M-1} & \cdots \\
    \end{matrix}
    \label{EQ_qSubs}
 \end{equation}
\end{widetext}

We can adopt similar conventions for the $s$-domain. The arrangement of variables is then given by the following table 

\begin{widetext}
\begin{center}
  \begin{tabular}{l l l l l| l l l l}
    \multicolumn{5}{c|}{main variables} &
    \multicolumn{4}{c}{ancillas} \\
    \hline
     $s_0$ & $s_M$ & $s_{2M}$ & $\cdots$ & $s_{(M-1)M}$ & $s_{M^2}$ & $s_{M^2+M}$ & $\cdots$ & $s_{M^2+M^2(M-1)/2-M+1}$\\
     $s_1$ & $s_{M+1}$ & $s_{2M+1}$ &$\cdots$ & $s_{(M-1)M+1}$ & $s_{M^2+1}$ & $s_{M^2+M+1}$ & $\cdots$ & $s_{M^2+M^2(M-1)/2-M+2}$\\
     $\cdots$ & $\cdots$ & $\cdots$ & $\cdots$ & $\cdots$ & $\cdots$ & $\cdots$ & $\cdots$  & $\cdots$\\
     $s_{M-1}$ & $s_{2M-1}$ & $s_{3M-1}$ & $\cdots$ & $s_{M^2-1}$ & $s_{M^2+M-1}$ & $s_{M^2+2M-1}$ & $\cdots$ & $s_{M^2+M^2(M-1)/2}$
  \end{tabular}
  \label{TBL_sHGEN}
 \end{center}
\end{widetext}
whereas the substitution scheme of a product of two-binary variables by a single variable will be conducted as follows
\begin{widetext}
 \begin{equation}
    \begin{matrix}
    s_0s_M \leftarrow s_{M^2} & s_0s_{2M} \leftarrow s_{M^2+M} & \cdots \\
    s_1s_{M+1} \leftarrow s_{M^2+1} & s_1s_{2M} \leftarrow s_{M^2+M+1} & \cdots \\
    \cdots & \cdots & \cdots \\
    s_{M-1}s_{2M-1} \leftarrow s_{M^2+M+1} & s_{M-1}s_{3M-1} \leftarrow s_{M^2+2M-1} & \cdots \\
    \end{matrix}
    \label{EQ_sSubs}
 \end{equation}
\end{widetext}
In practice, we do not perform the transform given by Eq.(\ref{EQ_sSubs}) directly since the substitution of a $k$-body to a 2-body interaction is always performed in the $q$-domain. The substitution in $s$-domain follows automatically when we transform the domain from $E_2(q_i)$ into $E_2(s_i)$ by substitution of variable $q_i \leftarrow s_i$. 
\subsection{Hamiltonian Formulation: Illustration by Low Order Case}

To clarify the method, we will explain the Hamiltonian formulation step-by-step for a low order case, which in this case is a H-matrix of order 2. The discussions follow the stages as illustrated by the Hamiltonian construction diagram depicted in Eq. (\ref{transdomain_diagram}).

\subsubsection{Formulation of $E_k(s_i)$.}

The formulation of energy function is started by an arrangement of variables, which for the finding H-matrix of order 2 problem is given by the followings 
\begin{equation*}
    \begin{pmatrix} 
     s_0 & s_2 \\
     s_1 & s_3
    \end{pmatrix}
    \label{EQ_sArray2}
\end{equation*}
The energy function is defined as the total sum of square of the off-diagonal elements of $D$-matrix, which in this case will only consist of a single term $d_{0,1}$. By using Eq.(\ref{EQ_dij2}) we obtain $E_k\left(s_i\right)=d_{0,1}^2=\left(s_0s_2+s_1s_3 \right)^2$ which leads to the following

\begin{equation}
 \begin{split}
    E_k\left(s_i\right) = \left(s_0^2s_2^2+s_1^2s_3^2\right) + 2\left(s_0s_2s_1s_3\right)
    \label{EQ_EH2_0}
  \end{split}
\end{equation}
By substitution of $s_i^2 \leftarrow 1$ into $E_k\left(s_i\right)$, we arrive to the following form
\begin{equation}
    E_k\left(s_i\right)= 2 + 2s_0s_1s_2s_3 
    \label{EQ_EH2} 
\end{equation}

The substitution $s_i^2 \leftarrow 1$ which is done in the last step simplifies greatly Eq.(\ref{EQ_EH2_0}) into Eq.(\ref{EQ_EH2}); this is one of the important steps to be highlighted in formulating the energy function. 

\subsubsection{Transformation $E_k(s_i) \rightarrow E_k(q_i)$.}

To obtain $E_k(q_i)$, we perform $q_i \leftarrow s_i $ substitution which is defined by Eq.(\ref{EQ_s2q}). Even for a two-term case of Eq.(\ref{EQ_EH2}), the number of terms starts to increase significantly into $16$, which is given by the following expression

\begin{widetext}
\begin{equation}
 \begin{split}
   E_k\left(q_i\right)= 4 - 4q_0 - 4q_1 - 4q_2 - 4q_3 + 8q_0q_1 + 8q_0q_2 + 8q_0q_3 + 8q_1q_2 + 8q_1q_3  +  8q_2q_3 \\
  - 16q_0q_1q_2 - 16q_0q_1q_3 - 16q_0q_2q_3 - 16q_1q_2q_3 +32q_0q_1q_2q_3     
 \end{split}
 \label{EQ_Hkq_orde2}
\end{equation}
\end{widetext}

By observing the terms in Eq.(\ref{EQ_Hkq_orde2}), we realize that the $q$-domain energy function $E_k\left(q_i\right)$ contains constants, quadratics,  cubics, and a quartic terms. The cubics and quartics terms should be converted into at most quadratics terms for implementation into a QAM. 

\subsubsection{Transformation $E_k(q_i) \rightarrow E_2(q_i)$.}

To reduce the degree of high order terms (cubics and quartics) into at most second order (quadratics), we employ the substitution by considering the following arrangement of variables as explained in the previous section

\begin{center}
  \begin{tabular}{ c c | c c }
     \multicolumn{2}{c|}{main variables} &
     \multicolumn{2}{c}{ancillas} \\
     \hline
      $q_0$ & $q_2$ & $q_4$ &\\
      $q_1$ & $q_3$ & $q_5$ & 
  \end{tabular}
  \label{TBL_H2}
\end{center}
%
Based on the arrangement, the substitutions to be done are $q_0q_2 \leftarrow q_4$ and $q_1q_3 \leftarrow q_5$, each of which is compensated by its corresponding $H_{\land}$. Then, based on Eq.(\ref{EQ_BoolRed1}) and Eq.(\ref{EQ_BoolRed2}), we should proceed as follows

\begin{equation}
  \begin{split}
    q_0q_2 \leftarrow q_4+\delta_{0,2} \left(3q_4 + q_0q_2 -2q_0q_4 -2q_2q_4 \right) \\
%
    q_1q_3 \leftarrow q_5+\delta_{1,3} \left(3q_5 + q_1q_3 -2q_1q_5 -2q_3q_5 \right)
    \end{split}
    \nonumber
\end{equation}

Since the substitution is done independently for each of the terms in $d_{i,j}^2$, the value of $\delta_{i,j}$ is determined by maximum value of $d_{i,j}^2=M^2$. In our case, we take $\delta_{0,2}=\delta_{1,3}\equiv \delta=4M^2=16$ for all terms undergoing the substitution. The result for finding $2$-order H-matrix problem is a $22$-terms $q$-domain energy function given as follows

\begin{widetext}
 \begin{equation}
  \begin{split}
    E_2\left(q_i\right) = 4 - 4q_0 -  4q_1 - 4q_2 - 4q_3 + 56q_4 + 56q_5
     + 8q_0q_1 + 16q_0q_2 + 8q_0q_3 - 32q_0q_4 - 16q_0q_5 \\
     + 8q_1q_2 + 16q_1q_3 - 16q_1q_4 - 32q_1q_5  + 8q_2q_3 - 32q_2q_4 - 16q_2q_5
     - 16q_3q_4 - 32q_3q_5  + 32q_4q_5 
  	\label{E_2k}
  \end{split}
 \end{equation}
\end{widetext}

\subsubsection{Transformation $E_2(q_i) \rightarrow E_2(s_i)$.}

After obtaining the $E_2(q_i)$ expression, based on the construction diagram, now we should transform it back to $s$-domain to obtain $E_2(s_i)$. The result is an $s$-domain energy function that also consists of $22$ terms given as follows

\begin{widetext}
 \begin{equation}
  \begin{split}
    E_2\left(s_i\right) = 28 + 6s_0 + 6s_1 + 6s_2 + 6s_3 - 12s_4 - 12s_5 
    + 2s_0s_1 + 4s_0s_2 + 2s_0s_3 - 8s_0s_4 - 4s_0s_5 \\
    + 2s_1s_2 + 4s_1s_3 - 4s_1s_4 - 8s_1s_5 + 2s_2s_3 - 8s_2s_4 - 4s_2s_5
    - 4s_3s_4 - 8s_3s_5 + 8s_4s_5
  	\label{E_2s}
  \end{split}
 \end{equation}
\end{widetext}

\subsubsection{Formulation of 2-Body Hamiltonian: $E_2(s_i) \rightarrow \hat{H}_2\left(\hat{\sigma}_i^z\right)$.}

The formulation of Hamiltonian for a given $E_2(s_i)$ is done by substitution of $s_i \leftarrow \hat{\sigma}_i^z$. Based on Eq.(\ref{E_2s}), we arrive to the following Hamiltonian of a 2-body interaction for H-SEARCH problem of order 2,

\begin{widetext}
\begin{equation}
 \begin{split}
    \hat{H}_2\left(\hat{\sigma}_i^z\right) =
    28 + 6\hat{\sigma}_0^z + 6\hat{\sigma}_1^z + 6\hat{\sigma}_2^z + 6\hat{\sigma}_3^z - 12\hat{\sigma}_4^z - 12\hat{\sigma}_5^z
    + 2\hat{\sigma}_0^z\hat{\sigma}_1^z + 2\hat{\sigma}_0^z\hat{\sigma}_2^z + 2\hat{\sigma}_0^z\hat{\sigma}_3^z - 8\hat{\sigma}_0^z\hat{\sigma}_4^z - 4\hat{\sigma}_0^z\hat{\sigma}_5^z\\ 
	+ 2\hat{\sigma}_1^z\hat{\sigma}_2^z + 4\hat{\sigma}_1^z\hat{\sigma}_3^z - 4\hat{\sigma}_1^z\hat{\sigma}_4^z - 8\hat{\sigma}_1^z\hat{\sigma}_5^z + 2\hat{\sigma}_2^z\hat{\sigma}_3^z - 8\hat{\sigma}_2^z\hat{\sigma}_4^z - 4\hat{\sigma}_2^z\hat{\sigma}_5^z 
	- 4\hat{\sigma}_3^z\hat{\sigma}_4^z - 8\hat{\sigma}_3^z\hat{\sigma}_5^z  + 8\hat{\sigma}_4^z\hat{\sigma}_5^z 
	\label{H_2sigma}
 \end{split}
\end{equation}
\end{widetext}
At this point, we can see that for implementation on a QAM, a simple two-terms $s$-domain energy function with only one quartic terms given by Eq.(\ref{E_ks}) transform into a $22$-terms Hamiltonian given by Eq.(\ref{H_2sigma}). Computation by hand for a higher order H-matrix problem surely will be not an easy task. This issue will be addressed in the following section.

\subsection{Higher Order Case: The Needs of Symbolic Computing}

The method to formulate Hamiltonian of finding 2-order H-matrix that has been described previously can be generalized to higher orders. It will be realized immediately that the problem start to occur due to the increasing number of variables and terms. An $M$ order H-SEARCH needs $M^2$ number of binary variables to represent the matrix and an additional of $M\times M(M-1)/2$ for the ancillas, giving $M^2 +M\times M(M-1)/2$ in total. Therefore, we have increased the number of variables (complexity) from $O(M^2)$ to $O(M^3)$. In the following discussion, when an expression of energy function or a Hamiltonian includes too many terms to write, we will only display partially. The complete expressions are provided separately in Appendix section.  

As an example, a problem of finding 4-order H-matrix needs $40$ binary variables, which consists of $16$ main variables and $24$ ancillas. We can arrange the $s$-variables as follows 

\begin{center}
 \begin{tabular}{c c c c | c c c c c c}
   \multicolumn{4}{c|}{main variables} &
   \multicolumn{6}{c}{ancillas} \\
   \hline
    $s_0$ &$s_4$ &$s_8$ &$s_{12}$ &$s_{16}$ &$s_{20}$ &$s_{24}$ &$s_{28}$ &$s_{32}$ &$s_{36}$\\
    $s_1$ &$s_5$ &$s_9$ &$s_{13}$ &$s_{17}$ &$s_{21}$ &$s_{25}$ &$s_{29}$ &$s_{33}$ &$s_{37}$\\
    $s_2$ &$s_6$ &$s_{10}$ &$s_{14}$ &$s_{18}$ &$s_{22}$ &$s_{26}$ &$s_{30}$ &$s_{34}$ &$s_{38}$\\
    $s_3$ &$s_7$ &$s_{11}$ &$s_{15}$ &$s_{19}$ &$s_{23}$ &$s_{27}$ &$s_{31}$ &$s_{35}$ &$s_{39}$\\
 \end{tabular}
 \label{TBL_H4s}
\end{center}
%
Similarly, this problem also needs 40 number of $q$-variables arranged as follows,
%
\begin{center}
 \begin{tabular}{c c c c | c c c c c c}
   \multicolumn{4}{c|}{main variables} &
   \multicolumn{6}{c}{ancillas} \\
   \hline
    $q_0$ &$q_4$ &$q_8$ &$q_{12}$ &$q_{16}$ &$q_{20}$ &$q_{24}$ &$q_{28}$ &$q_{32}$ &$q_{36}$\\
    $q_1$ &$q_5$ &$q_9$ &$q_{13}$ &$q_{17}$ &$q_{21}$ &$q_{25}$ &$q_{29}$ &$q_{33}$ &$q_{37}$\\
    $q_2$ &$q_6$ &$q_{10}$ &$q_{14}$ &$q_{18}$ &$q_{22}$ &$q_{26}$ &$q_{30}$ &$q_{34}$ &$q_{38}$\\
    $q_3$ &$q_7$ &$q_{11}$ &$q_{15}$ &$q_{19}$ &$q_{23}$ &$q_{27}$ &$q_{31}$ &$q_{35}$ &$q_{39}$\\
 \end{tabular}
 \label{TBL_H4q}
\end{center}
Although the formulation of $E_k(s_i)$ can be done similarly to the 2-order case, due to a large number of variables and terms, it will be better to automatize this process in a computer, i.e., we employ symbolic computing software to derive the energy function. We have formulated {\bf Algorithm \ref{ALG_Eks}} to calculate $E_k(s_i)$.
\begin{algorithm}[H]
  \caption{Construction of $E_k(s_i)$ by symbolic computation}
  \label{ALG_Eks}
  \begin{algorithmic}[1]
    \State Construct array of variables $\{s_i, s_j\}$ according to the order $M$ of the H-matrix
    \State Calculate inner product $d_{ij}=\langle s_i,s_j \rangle$ of symbols (variables) between every pairs of columns of the array 
    \State Calculate $E_k(s_i)= \sum d_{i,j}^2$
    \State Cleanup $s_i^2$ terms by substitution: $E_k(s_i)=E_k(s_i) \vert_{s_i^2 \leftarrow 1}$
  \end{algorithmic}
\end{algorithm}

The {\bf Algorithm \ref{ALG_Eks}} can be implemented into a programming language that has a symbolic computing capability. The energy function of finding $4$-order H-matrix problem is given as follows, 
\begin{equation}
  E_k(s_i)=\sum_{i<j<M} d_{ij}^2
  \nonumber
\end{equation}
Expanding the this energy function will generate a $37$-terms expression that can be written as follows,

\begin{widetext}
 \begin{equation}
    E_k(s_i) = 24 + 2s_0s_1s_4s_5 + \cdots + 2s_0s_1s_{12}s_{13} +  \cdots + 2s_{10}s_{11}s_{14}s_{15}
    \label{Ek_si4}
 \end{equation}
\end{widetext}
Likewise, the transformation of $E_k(s_i) \rightarrow E_k(q_i)$ can also be done automatically by using {\bf Algorithm \ref{ALG_Ekq}}.
\begin{algorithm}[H]
  \caption{Transform $E_k(s_i) \rightarrow E_k(q_i)$}
  \label{ALG_Ekq}
  \begin{algorithmic}[1]
    \State For all binary variables $s_i$:
    \State \hspace{20 pt} $E_k(s_i) \vert_{s_i \leftarrow q_i}$
  \end{algorithmic}
\end{algorithm}
The processing of the 4-order case yields an energy function with $317$ terms, which by setting $\delta=4\times H_{max}=64$, can be expressed as follows
\begin{widetext}
  \begin{equation}
    E_k(q_i) = 96 -36q_0 + \cdots -36q_{15} +\cdots
    + 24q_{14}q_{15} + \cdots - 16q_{11}q_{14}q_{15} + \cdots
    + 32q_{10}q_{11}q_{14}q_{15}
    \label{Ek_qi4}
 \end{equation}
\end{widetext}
The next stage of transforming $E_k(q_i) \rightarrow E_2(q_i)$ will yield an energy function with more number of terms. The processing for such transform is described in {\bf Algorithm \ref{ALG_E2q}}.
\begin{algorithm}[H]
  \caption{Transform $E_k(q_i) \rightarrow E_2(q_i)$}
  \label{ALG_E2q}
  \begin{algorithmic}[1]
     \State Construct a list of substitution pair $sPair [col_i, col_j, col_k]$
     \State $E_2(q_i) \leftarrow E_k(q_i)$
     \State For all high order terms in $E_2(q_i$) and based on $sPair$: 
     \State \hspace{20 pt} $E_2(q_i) \leftarrow E_2(q_i) \vert_{q_i\cdot q_j \leftarrow q_k} + H_{\land}(q_i,q_j,q_k,\delta)$
     \State Simplify $E_2(q_i)$   
  \end{algorithmic}
\end{algorithm}
In the finding $4$-order H-matrix case, the two-body $q$-domain energy function $E_2(q)$ will consist of $389$ terms, which are given as the followings,
\begin{widetext}
 \begin{equation}
    E_2(q_i) = 96 -36q_0 +\cdots +216q_{39} +24q_0q_1 + \cdots + 32q_{38}q_{39} 
    \label{E2_qi4}
 \end{equation}
\end{widetext}
%
The last stage of transformation $E_2(q_i) \rightarrow E_2(s_i)$ can be done by $q_i \leftarrow s_i$  substitution based on the {\bf Algorithm \ref{ALG_E2s} }.

\begin{algorithm}[H]
  \caption{Transform $E_2(q_i) \rightarrow E_2(s_i)$}
  \label{ALG_E2s}
  \begin{algorithmic}[1]
     \State For all variables $\{q_i\}$:
     \State \hspace{20 pt} $E_2(s_i) \leftarrow E_2(q_i) \vert_{q_i \leftarrow \left( \frac{1-s_i}{2}\right)}$
  \end{algorithmic}
\end{algorithm}
The final form of energy function $E_2(s_i)$ also consists of $389$ terms, which can be expressed as the followings,
\begin{widetext}
 \begin{equation}
    E_2(s_i) = 1,248 + 66s_0 + \cdots - 44s_{39} + 6s_0s_1 + \cdots + 8s_{38}s_{39}
     \label{E2_si4}
 \end{equation}
\end{widetext}

The Hamiltonian of the problem can be obtained directly from Eq.(\ref{E2_si4})  by replacing the binary variables by the corresponding operators  $s_i \leftarrow \hat{\sigma}_i^z$. The result is as follows
\begin{widetext}
 \begin{equation}
    \hat{H}_2\left(\hat{\sigma}_i^z\right) = 1,248 + 66\hat{\sigma}_0^z + \cdots -44\hat{\sigma}_{39}^z + 6 \hat{\sigma}_0^z\hat{\sigma}_1^z + \cdots + 8\hat{\sigma}_{38}^z\hat{\sigma}_{39}^z 
 \label{H2_sigma4}
 \end{equation}
\end{widetext}
which is the desired Hamiltonian of the 4-order H-SEARCH problem, which will become $\hat{H}_{pot}(\hat{\sigma})$ in the quantum annealing process given by Eq.(\ref{EQ_HQA}).

\subsection{Sub-Problem-1:Finding a Set of $N<M$ Orthogonal Binary Vectors}

In this sub-problem, we want to find $N$ number of $M$-length binary vectors, where $N<M$. The initial values of the binary variables of the $N$-vectors, which generally non-orthogonal to each other, can be set to either particular values or at random; therefore, this process of obtaining $N$-orthogonal vectors from the given initial vectors will also be called \emph{orthogonalization}. We arrange the variables similarly as before, but now with less number of variables. The number of ancillas is also reduced to $N\times N\times (N-1)/2$. We start with the following arrangement of $s$-variables

\begin{widetext}
\begin{center}
  \begin{tabular}{l l l l l| l l l l}
    \multicolumn{5}{c|}{main variables} &
    \multicolumn{4}{c}{ancillas} \\
    \hline
     $s_0$ & $s_M$ & $s_{2M}$ & $\cdots$ & $s_{(N-1)M}$ & $s_{NM}$ & $s_{M^2+M}$ & $\cdots$ & $s_{NM+NM(M-1)/2-M+1}$\\
     $s_1$ & $s_{M+1}$ & $s_{2M+1}$ &$\cdots$ & $s_{(N-1)M+1}$ & $s_{NM+1}$ & $s_{M^2+M+1}$ & $\cdots$ & $s_{NM+MN(N-1)/2-M+2}$\\
     $\cdots$ & $\cdots$ & $\cdots$ & $\cdots$ & $\cdots$ & $\cdots$ & $\cdots$ & $\cdots$  & $\cdots$\\
     $s_{M-1}$ & $s_{2M-1}$ & $s_{3M-1}$ & $\cdots$ & $s_{(N-1)(M-1)}$ & $s_{NM+M-1}$ & $s_{M^2+2M-1}$ & $\cdots$ & $s_{NM+MN(N-1)/2}$
  \end{tabular}
  \label{TBL_sSUBP1}
\end{center}
\end{widetext}

For a concrete illustration, consider $M=4$ and $N=3$, i.e., finding a set of $3$ binary ortho-vectors of order $4$. The arrangement of variables becomes as follows

\begin{center}
 \begin{tabular}{c c c | c c c}
   \multicolumn{3}{c|}{main variables} &
   \multicolumn{3}{c}{ancillas} \\
   \hline
    $s_0$ &$s_4$ &$s_8$     &$s_{12}$ &$s_{16}$ &$s_{20}$ \\
    $s_1$ &$s_5$ &$s_9$     &$s_{13}$ &$s_{17}$ &$s_{21}$ \\
    $s_2$ &$s_6$ &$s_{10}$  &$s_{14}$ &$s_{18}$ &$s_{22}$ \\
    $s_3$ &$s_7$ &$s_{11}$  &$s_{15}$ &$s_{19}$ &$s_{23}$ \\
 \end{tabular}
 \label{TBL_H4qSUB}
\end{center}
Compared to finding 4-order H-matrix problem, after performing the process described by the construction diagram, we found that the number of terms in $E_k(s_i)$ has been reduced to $19$, whereas there are $169$ number of terms in $E_k(q_i)$, and $205$ terms in each of $E_2(q_i)$ and  $E_2(s_i)$. The Hamiltonian of the system with $205$ terms has the following form
\begin{widetext}
 \begin{equation}
    \hat{H}_2\left(\hat{\sigma}_i^z\right) = 768 + 52\hat{\sigma}_0^z + \cdots -52\hat{\sigma}_{23}^z + 4\hat{\sigma}_0^z\hat{\sigma}_1^z + \cdots + 8\hat{\sigma}_{22}^z\hat{\sigma}_{23}^z
   \label{H2_SUB_N3_M4}
 \end{equation}
\end{widetext}
Some experiments to analyze a higher order case will be discused in more detail in the next Section.

\subsection{Sub-Problem-2: Hadamard-Matrix Completion}

In the H-matrix completion problem, the task is to find $N$-number of missing vectors of an $M$-order H-matrix. This means that $(M-N)$ number of (column) vectors are known. Construction of a H-matrix by random generation of $M$-order binary vector followed by orthogonality testing implies that finding the last vectors of a H-matrix become increasingly difficult, which is indicated by more number of iterations required in the later stages \cite{suksmono_arxiv2016}. It can be understood considering the orthogonality of a candidate vector should be tested to previously found vectors. Interestingly, in this sub-problem, we can use the known vectors as a constraint which further reduce the number of variables and therefore the number of qubits needed in the implementation of the problem in a QAM. 

Consider the problem of finding $1$ missing vector in a $2$-order H-matrix. When it is a seminormalized one, all elements in the first columns are $1$'s. Then, we have the following form of variable arrangements 

\begin{center}
  \begin{tabular}{ c c | c c}
     \multicolumn{2}{c|}{main variables} &
     \multicolumn{1}{c}{ancillas} \\
     \hline
      $1$ & $s_0$ & * \\
      $1$ & $s_1$ & *  
  \end{tabular}
  \label{TBL_qArraySub}
\end{center}
Note that in this case we do not need any ancilla, so that we put "*" to all of ancilla's positions the table. The expression of $E_k(s_i)$, after substitution $s_0^2 \leftarrow 1$, becomes

\begin{equation}
    \nonumber
    E_k(s_i) = (s_0+s_1)^2|_{s_i^2 \leftarrow 1} = 2 + 2s_0s_1=2(1+s_0s_1)
\end{equation}
It is easy to see that the minimum value of this energy function, which is $0$, will be achieved when $s_0s_1=-1$, i.e., either $s_0=-1$ and $s_1=1$ or $s_0=1$ and $s_1=-1$, which then gives the following solutions of the H-matrices
\begin{equation*}
    \begin{pmatrix} 
      & + & - \\
      & + & +  
    \end{pmatrix} \ and \ 
    \begin{pmatrix} 
      & + & + \\
      & + & -  
    \end{pmatrix}
    \label{EQ_qArray2SubSol}
\end{equation*}
Note that for conciseness, we have represented the elements by their signs, i.e, $-1$ is displayed as $-$, whereas $1$ is shown as $+$. 

Higher order cases can be treated similarly. Consider the problem of finding $2$ missing vectors in a $4$-order H-matrix. Instead of writing the known vectors at the first columns, we have written them in the last ones for convenience of indexing the variables and ancillas. The arrangement will become as follows,
\begin{center}
 \begin{tabular}{c c c c | c }
   \multicolumn{4}{c|}{main variables} &
   \multicolumn{1}{c}{ancillas} \\
   \hline
    $s_0$ &$s_4$ & + & + &$s_8$    \\
    $s_1$ &$s_5$ & - & + &$s_9$    \\
    $s_2$ &$s_6$ & + & + &$s_{10}$ \\
    $s_3$ &$s_7$ & - & + &$s_{11}$ \\
 \end{tabular}
 \label{TBL_sArray4_2known}
\end{center}
By following the previously explained symbolic computational procedures, we will obtain 11 terms in $E_k(s_i)$, 67 terms in $E_k(q_i)$, and 79 terms in each of $E_2(q_i)$ and $E_2(s_i)$. The Hamiltonian of the system, which also consists of $79$ terms, has the following form
\begin{widetext}
 \begin{equation}
    \hat{H}_2\left(\hat{\sigma}_i^z\right) = 128 + 14\hat{\sigma}_0^z + \cdots - 28\hat{\sigma}_{11}^z + 2\hat{\sigma}_0^z\hat{\sigma}_1^z + \cdots + 8\hat{\sigma}_{10}^z\hat{\sigma}_{11}^z 
   \label{H2_completion_M4}
 \end{equation}
\end{widetext}
Higher order case will be discussed and tested in the experiment section. Considering the current number of qubits and connection, we will try to find $1$-deleted vector of a 12-order H-matrix.


\section{Experiments and Analysis}

Experiments have been conducted to verify the proposed method, by both of simulation and actual implementation on a quantum annealer. 

In the simulation, a python-based simulated annealing package, the D-Wave's \emph{neal}, has been employed to find minimum energies and related configurations that yield solutions of the problem. Input of the simulator are Ising coefficients $\{h_i, J_{ij}\}$ of the problem's Hamiltonian or energy function. These coefficients can be extracted from either $\hat{H_2}(\hat{\sigma}_i^z)$ or $E_2(s_i)$, where its constant value is omitted which translates into the shift of the ground state energy to a negative value of the corresponding constant. Then, we normalize the coefficients by dividing them by the largest absolute values of the coefficients to simulate a real QAM input parameters.

We also have done experiments by using the D-Wave's DW2000Q quantum annealer. The "programming" of this quantum computer is performed by configuring the qubits which are connected by a Chimera graph, and assigning weight on each of the qubit and strength of the coupler that connect the qubits according to the Ising coefficients. A simple Hamiltonian can be implemented directly by manual configuration, whereas a more complex one needs an embedding tool. 

\subsection{Simulation on D-wave \emph{Neal} Simulator}

The input of the \emph{neal} simulated annealing software are Ising coefficients, which after scaling will simulate the input of the D-Wave quantum annealer, except that it is not necessary to take care of the restriction of the connection among the qubits imposed by the Chimera graph. 
%
\subsubsection{Finding 2-order and 4-order H-matrix}

To solve the problem of finding $2$-order H-matrix, we have used the energy function given by Eq.(\ref{E_2s}), which after normalization yields the following bias values
\begin{equation}
     h = (0.5,\ 0.5,\ 0.5,\ 0.5,\ -1.0,\ 1.0)^T
    \nonumber
\end{equation}
whereas the coupling coefficients between a pair of qubits are given as follows
\begin{equation*}
    J =
    \begin{pmatrix} 
        * & 0.167   & 0.333 & 0.167 & -0.667    & -0.333 \\
        * & *       & 0.167 & 0.333 & -0.333    & -0.667 \\
        * & *       & *     & 0.167 & -0.667    & -0.333 \\
        * & *       & *     & *     & -0.333    & -0.667 \\
        * & *       & *     & *     & *         &  0.667  \\
        * & *       & *     & *     & *         & *       
    \end{pmatrix}
    .
    \nonumber
\end{equation*}
Since the diagonal entries are not used and the $J$ matrix is symmetric, we only show the upper diagonal elements of the matrix.
We have set the number of sweeps in the simulator to $1,000$ and the number of configurations to $10$. Table \ref{TBL_H2conf} displays the obtained configurations with their corresponding energy values after the simulation has been finished
\begin{table}
  \centering
   \caption{\label{TBL_H2conf} Solutions of finding $2$-order H-matrix problem by D-Wave's  \emph{neal}. The final configurations of main variables and ancillas along with their associated energies are shown. The ground-state column indicates whether the lowest energy has been achieved, which is marked by "Y", or has not been achieved which is marked by "N".}
  \begin{tabular} {|c|c|c|c|} 
    \hline
     No & Configuration & Energy & Ground-State \\
    \hline
     1 & $(+, -, +, +, +, +)$ & -2.33 & Y \\
    \hline
     2 & $(+, -, -, -, +, -)$ & -2.33 & Y \\
    \hline
     3 & $(+, -, +, +, +, +)$ & -2.33 & Y \\
    \hline
     4 & $(-, +, +, +, +, +)$ & -2.33 & Y \\
    \hline
     5 & $(+, -, +, +, +, +)$ & -2.33 & Y \\
    \hline
     6 & $(+, -, -, -, +, -)$ & -2.33 & Y \\
    \hline
     7 & $(+, -, +, -, +, -)$ & -2.00 & N \\
    \hline
     8 & $(+, +, -, +, +, +)$ & -2.33 & Y \\
    \hline
     9 & $(+, -, +, -, +, -)$ & -2.00 & N \\
    \hline
     10 & $(-, +, +, +, +, +)$ & -2.33 & Y\\
    \hline
  \end{tabular}
\end{table}

Based on Eq.(\ref{E_2s}), we know that the value of the constant is $28.00$, whereas the largest (absolute value) of coefficients is $12.00$. By normalization, the constant becomes $2.33$, therefore the value of the lowest energy (the ground state) is $-2.33$, which is in agreement with the simulation result. We observed from the results that not all of the configurations achieved ground states. In the table, configurations achieving the ground states's are marked by "Y", whereas non-ground states are marked by "N". The elements of the obtained H-matrices are given by the first $4$ values of the configuration, such as $(s_0, s_1, s_2, s_3)=(+,\ -,\ +,\ +)$ for the first configuration, whereas the corresponding ancillas $(s_4,\ s_5)=(+,\ +)$ can be neglected. Reshaping the solutions into  $2\times 2$ matrices yields various $2\times 2$ orthogonal matrices, displayed subsequently as follows, 


\begin{equation}
    \begin{pmatrix}
        + & + \\
        - & + \\
    \end{pmatrix},\  
    \begin{pmatrix}
        + & - \\
        - & - \\
    \end{pmatrix}
         , \cdots
         , \ 
    \begin{pmatrix}
        - & + \\
        + & + \\
    \end{pmatrix}
   \nonumber
\end{equation}
It is easy to verify that the matrices that correspond to the ground state energy are indeed Hadamards.

In the second example, we consider the problem of finding $4$-order H-matrix. By taking $\delta=4\times H_{max}$, the energy function given by Eq.(\ref{E2_si4}). By setting the simulation parameters as before, we obtained the following set of energies (written to the second decimal places)

\begin{equation}
  \begin{split}
   -16.00,\ -15.62,\ -15.62,\ -15.71,\ -15.71,\\
   -15.71,\ -15.71,\ -15.71,\ -16.00,\ -15.62 \ 
   \nonumber
   \end{split}
\end{equation}

Our calculation shows that the ground state energy should have been $-16.00$, which only 2 out of 10 solutions have achieved. As an example, the first solution related to $E_2(s_i) = -16.00$ and the second one related to $E_2(s_i)=-15.62$ yields the following configurations 

\begin{equation}
  \begin{split}
    (+,+,+,+, +,-,-,+, -,-,+,+, -,+,-,+, +,+,+,+, \\
     +,+,+,+, +,+,+,+, +,-,+,+ ,+,+,-,+, -,+,+,+) 
   \end{split}
   \nonumber
\end{equation}
and
\begin{equation}
  \begin{split}
    (-,+,+,+, -,-,+,-, -,-,-,-, -,+,-,+, -,+,+,+, \\
     -,+,+,+, -,+,+,+, -,-,+,-, -,+,+,+, -,+,-,+)
   \end{split}
   \nonumber
\end{equation}
respectively. By taking the first $16$ elements of the solution vectors and reshaping them into $4\times 4$ matrices, we obtain the following results,


\begin{equation}
    \begin{pmatrix}
        + & + & + & + \\
        + & - & - & + \\
        - & - & + & + \\
        - & + & - & +
    \end{pmatrix}
        ,\  
    \begin{pmatrix}
       - & + & + & + \\
       - & - & + & - \\
       - & - & - & - \\
       - & + & - & +
    \end{pmatrix}
   \nonumber
\end{equation}
We can verify that the first solution with $E_2(s_i)=-16.00$ is actually an orthogonal matrix, whereas the second one related to $E_2(s_i)=-15.62$ is not. We also found that by increasing the number of sweeps, it is possible to obtain more  correct solutions.

\subsubsection{Finding a set of $N$-orthogonal $M$-order binary vectors}

In this experiment, our objective is to find a set of $3$-orthogonal binary vectors of length $12$. The number of binary variables that are required to do this task are $72$, whereas the number of $E_2(s_i)$ terms are $1,765$. The Hamiltonian obtained from $E_2(s_i)$ after symbolic computation yields the following expression 

\begin{widetext}
 \begin{equation}
    \hat{H}_2\left(\hat{\sigma}_i^z\right) = 19,872 + 404\hat{\sigma}_0^z + \cdots + 404\hat{\sigma}_{71}^z + 4\hat{\sigma}_0^z\hat{\sigma}_1^z + \cdots + 8\hat{\sigma}_{70}^z\hat{\sigma}_{71}^z 
    \label{H2_SUB_N3_M12}
 \end{equation}
\end{widetext}
Based on $E_2(s_i)$ and by using $\delta=5\times M^2$, the calculated ground-state energy is $-49.19$. Setting the number of sweep to $1000$ as in the previous case did not give a correct solution, therefore, we increased the number of sweeps to $500,000$ while keeping the number of configurations at $10$. We obtained the energies at each of the configuration in the solutions as follows
\begin{equation}
  \begin{split}
     -49.09,\ -49.11,\ -49.09,\ -49.13,\ -49.17, \\
     -49.11,\ -49.19,\ -49.19,\ -49.11,\ -49.11 \  
   \end{split}
   \nonumber
\end{equation}
Especially, the solution given by the ground-state with energy at $-49.19$ are as follows
\begin{equation}
  \begin{split}
      v_{g,1}=(+, -, -, -, +, +, -, -, +, -, +, -) \\    
      v_{g,2}=(-, -, -, -, -, -, +, -, +, -, -, +) \\ 
      v_{g,3}=(+, +, -, -, -, -, -, -, -, +, +, +)
      .
  \end{split}
  \nonumber
\end{equation}
We have verified that these three binary vectors are orthogonal to each other. On the other hand, the non-ground state vectors such as the solution with energy $-49.09$ given by the following set of vectors,
\begin{equation}
  \begin{split}
      v_{ng,1}=(+, -, -, +, -, +, -, -, +, +, +, +) \\    
      v_{ng,2}=(-, +, -, -, +, -, +, -, +, -, -, -) \\ 
      v_{ng,3}=(+, +, +, +, -, -, -, +, +, +, -, -)
  \end{split}
\end{equation}
are not 3 orthogonal set of binary vectors, and therefore not a correct solution.

\subsubsection{Finding a deleted vector in a $12$-order H-matrix}

For the completion problem, we have chosen a $12$-order H-matrix as a case, whose $1$ column vector has been deleted. The rests of $11$ known vectors are as follows, 
\begin{equation}
  \begin{split}
     v_0=( +, +, +, +, +, +, +, +, +, +, +, + ) \\
     v_1=( +, -, +, -, +, +, +, -, -, -, +, - )\\
     v_2=( +, -, -, +, -, +, +, +, -, -, -, + )\\
     v_3=( +, +, -, -, +, -, +, +, +, -, -, - )\\
     v_4=( +, -, +, -, -, +, -, +, +, +, -, - )\\
     v_5=( +, -, -, +, -, -, +, -, +, +, +, - )\\
     v_6=( +, -, -, -, +, -, -, +, -, +, +, + )\\
     v_7=( +, +, -, -, -, +, -, -, +, -, +, + )\\
     v_8=( +, +, +, +, -, -, -, +, -, -, +, - )\\
     v_9=( +, +, +, -, -, -, +, -, -, +, -, + )\\
  v_{10}=( +, +, -, +, +, +, -, -, -, +, -,- ) \\ 
  \end{split}
  \label{H12_Compl}
\end{equation}
%
Since all of the elements of $v_{0}$ are $1$, it is a seminormalized H-matrix. Our symbolic computation yields the number of terms in $E_k(s_i) $ is $379$, $E_k(q_i) $ is $407$, $E_2(q_i)$ is $407$ and $E_2(s_i)$ is $379$. The Hamiltonian obtained from $E_2(s_i)$, after symbolic computation, is as follows 

\begin{equation}
    \hat{H}_2\left(\hat{\sigma}_i^z\right) = 756 + 2\hat{\sigma}_0^z\hat{\sigma}_1^z + \cdots + 2\hat{\sigma}_{0}^z\hat{\sigma}_{27}^z +  \cdots - 2\hat{\sigma}_{26}^z\hat{\sigma}_{27}^z
    \label{H2_COMPLETION_M12}
\end{equation}

By setting the number of sweeps to $1,000$ we obtained the energy equal to $-66.00$, which are identical for all of 10 configurations in the solution. This result shows that all of the configuration achieved lowest energy, they consist of two binary vectors as follows
\begin{equation}
  \begin{split}
    v_{11,1}=( +,-, +, +, +,-,-,-, +,-,-, +) \\
    v_{11,2}=(-, +,-,-,-, +, +, +,-, +, +,-)
  \end{split}
  \label{H12_Compl_sol}
\end{equation}
By inspection, we can see that $v_{11,2} = -v_{11,1}$ and therefore both of them are correct solutions that completes the $12$-order H-matrix.

\subsection{Experiments on D-Wave Quantum Annealer}

We also have implemented the Hamiltonian of H-SEARCH problems (for order 2 and 4), finding a set of $N<M$ orthogonal binary vectors of order $M$, and H-matrix completion problems into DW2000Q quantum annealer. The DW2000Q has up to 2048 qubits and 6016 couplers, where the qubits are connected by a C16 Chimera graph, which means that its 2048 qubits are logically map into a $16 \times 16$ matrix of unit cell, whose each cell consists of 8 qubits \cite{dwave2018}. The layout of the cell can be represented  either by a \emph{column} or by a \emph{cross}. In this paper, we use the \emph{cross} layout to show the connection among the qubits in each of the presented problem.

The schedule of quantum annealing process in DW2000Q can be adjusted by the user. However, in the following experiments, we have used the default schedule defined in \cite{dwave2018qpu}; where the kinetic energy $E_{kin}/h$ (with $h$ is the Planck constant) has been set at around $6$ GHz at the beginning; which is decreased exponentially to around zero at the end of the annealing process. Meanwhile, the potential energy $E_{pot}/h$ is started from zero at the beginning and then increased exponentially to around $12$ GHz at the end of the annealing.

\subsubsection{Finding 2-order and 4-order H-matrix}

\begin{figure}
	\includegraphics[width=1.0\columnwidth]{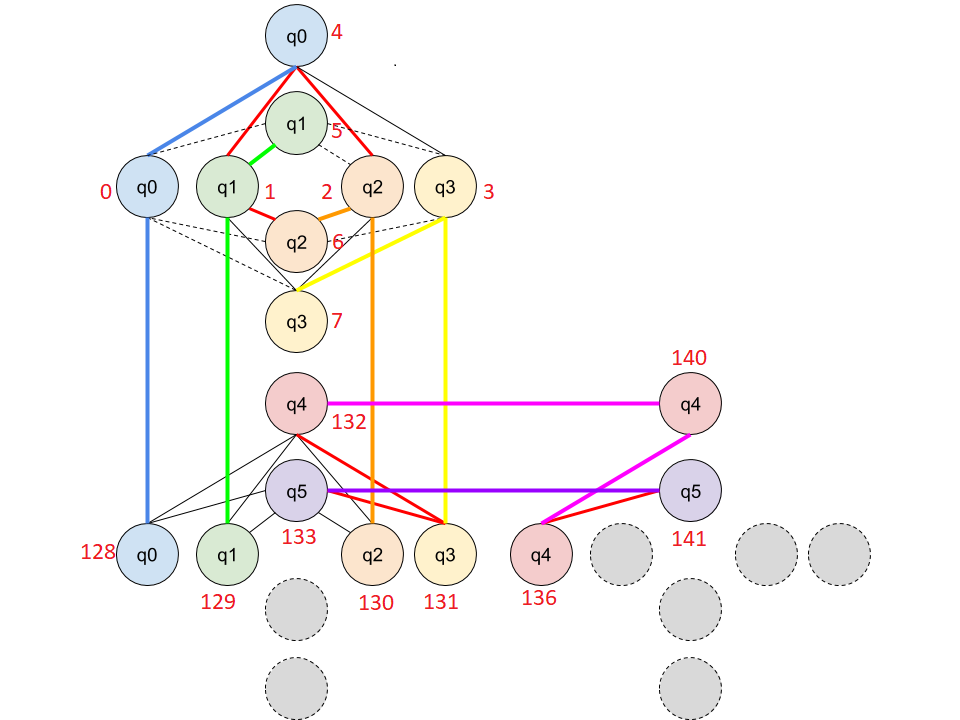}\\
    (a)\\
	\includegraphics[width=1.0\columnwidth]{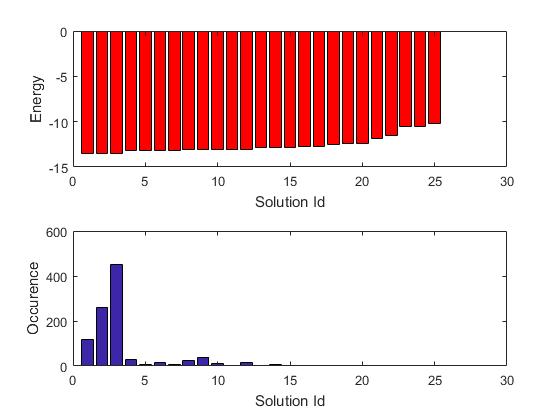}\\
    (b)\\
	\protect\caption{\label{FIG_DWCON_H2} Implementation of finding a 2-order H-matrix problem's Hamiltonian into a quantum annealer: (a) Embedding 6 logical qubits into 17 physical qubits in a Chimera graph's of the DW2000Q, (b) Obtained results after quantum annealing: the distribution of energy related to each solution (top) and distribution of the solutions obtained by 1000 reads (bottom).
	}
\end{figure}

The Hamiltonian of the finding 2-order H-matrix problem given by Eq.(\ref{H_2sigma}) indicates that $6$ (logical) qubits are required. However, implementation on the Chimera graph increases the number into $17$ (physical) qubits which are located in the neighbouring blocks (unit cells). We have manually designed the qubit's connection, whose configuration result is shown in Fig.\ref{FIG_DWCON_H2}(a).

We have used a default annealing schedule, whereas the number of reads is set to $1000$. Energy distribution of the result and its related occurrence number of each solution are shown in the top and bottom parts of Fig.\ref{FIG_DWCON_H2}(b), respectively. We obtained a minimum energy of $-13.52$, which corresponds to solution vector $(1,-,-,-)^T$ for the first four qubits while values of the ancilla qubits can be ignored.  The solution can be rearranged into a $2\times2$ arrays as follows 

\begin{equation}
    \begin{pmatrix}
        + & - \\
        - & - 
    \end{pmatrix}
   \nonumber
\end{equation}
which actually is a 2-order H-matrix.

For the $4$-order H-SEARCH problem, the Hamiltonian expressed in Eq.(\ref{H2_sigma4}) indicates that $40$ (logical) qubits are required. This number increases when it is implemented on the set of qubits with Chimera graph connection. We have employed SAPI (Solver Application Programming Interface) embedding tool which is provided by the D-Wave to construct the connection among the qubits automatically. After optimization, the SAPI indicates that $344$ (physical) qubits are required. 
%
\begin{figure}[H]
    \begin{center}
    \includegraphics[width=1.0\columnwidth]{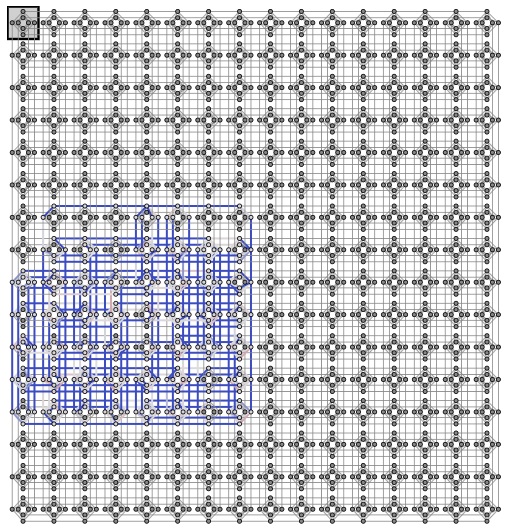} \\
    (a)\\
	\includegraphics[width=1.0\columnwidth]{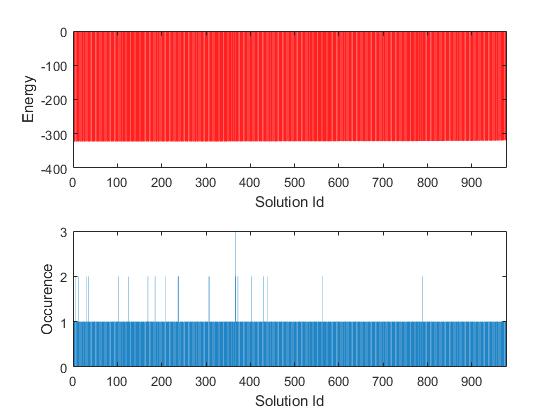}\\
    (b)\\
	\protect\caption{\label{FIG_DWCON_H4} Implementation of finding a 4-order H-matrix problem's Hamiltonian into a quantum annealer: (a) Embedding diagram of 40 logical qubits into 344 physical qubits in DW2000Q which is obtained by an embedding tool, (b) Results after finishing the quantum annealing evolution: the distribution of energy related to each solution (top) and distribution of the solutions obtained by 1000 reads (bottom).
	}
	\end{center}
\end{figure}

Sketched of the qubits connection is displayed in Fig.\ref{FIG_DWCON_H4}(a), whereas the distribution of energy and its related population are depicted in top and bottom part of Fig.\ref{FIG_DWCON_H4}(b) respectively. In contrast to the 2-order case, the figure shows an almost uniform distribution, except for a few number of solutions. Connection diagram displayed in Fig.\ref{FIG_DWCON_H4}(a) shows that a 4-order H-SEARCH problem already occupied a significant number of available qubits and couplers of the DW2000Q quantum processor. 

Default annealing schedule has been used and we also set the number of reads to $1000$. The achieved lowest energy for the given configuration is $-322.91$. The corresponding solution, after neglecting the ancillas and reformatting it into a $4 \times 4$ matrix, is as follows

\begin{equation}
    \begin{pmatrix}
        - & - & + & + \\
        + & - & + & - \\
        - & - & - & - \\
        - & + & + & -
    \end{pmatrix}
   \nonumber
\end{equation}
We can verify that the solution is indeed a H-matrix, therefore the D-Wave has successfully found the H-matrix of order-4.

\subsubsection{Finding a set of $3$-orthogonal $12$-order binary vectors}

\begin{figure}%
	\includegraphics[width=1.0\columnwidth]{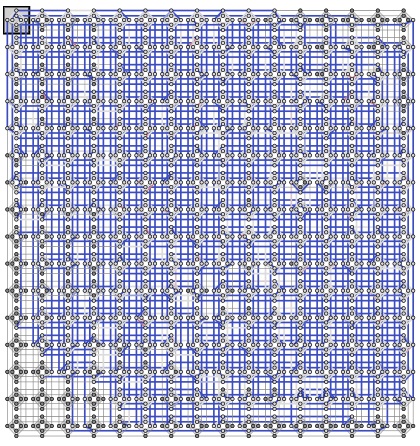}\\
    (a)\\
	\includegraphics[width=1.0\columnwidth]{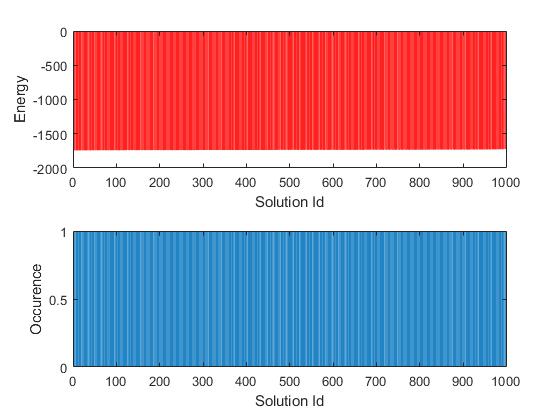}\\
    (b)\\
	\protect\caption{\label{FIG_DW_ortho12_3} Realization of finding a set of $3$-orthogonal binary vectors of order (length) $12$ into a quantum annealer: (a) Embedding $72$ logical qubits into $1,766$ physical qubits in a Chimera-connected qubits of the DW2000Q, (b) Obtained results after quantum annealing: the distribution of energy related to each solution (top) and distribution of the solutions obtained by 1000 reads (bottom). }
\end{figure}

In this experiment, we configured the D-Wave to find a set of 3 orthogonal binary vectors of order $12$. The Hamiltonian given by Eq.(\ref{H2_SUB_N3_M12}) indicates that $72$ (logical) qubits is necessary. We also used SAPI embedding tool to configure the Chimera graph to obtain the qubits connection. After several steps of optimizations, the SAPI shows that $1,766$ (physical) qubits are required. The sketch of configuration in the Chimera is displayed in Fig.\ref{FIG_DW_ortho12_3}(a).

We have set the annealing schedule to the default and also set the number of reads to $1,000$ as before. The the distribution of energy and population of each configurations are shown in Fig.\ref{FIG_DW_ortho12_3}(b). The achieved minimum energy with this configuration is $-1746.26$ which is corresponding to the following vectors as the solution

\begin{equation}
  \begin{split}
     v_0=( +, -, +, -,  -, +, +, -,  +, +, -, - ) \\
     v_1=( +, -, -, -,  +, +, -, +,  +, -, +, - ) \\
     v_2=( +, +, -, +,  +, +, +, +,  +, +, -, + ) 
  \end{split}
  \label{H12_subortho}
\end{equation}
We can verify that these set of three binary vectors are orthogonal to each others. The distribution of the solution shown in Fig.\ref{FIG_DW_ortho12_3}(b) is uniform, which means that every solution achieved minimum energy level. The connection diagram in Fig.\ref{FIG_DW_ortho12_3}(a) shows that for order-$12$, problem of finding three orthogonal binary vectors already occupied most of the qubits and connections of the processor.

\subsubsection{Finding a deleted vector of $12$-order H-matrix}

\begin{figure}
 \includegraphics[width=0.95\columnwidth]{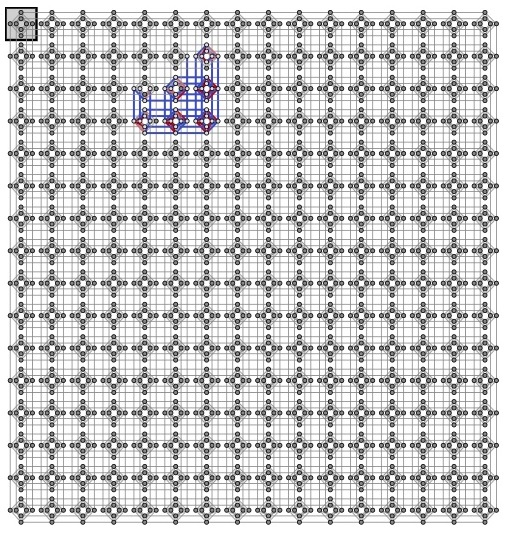}\\
  (a)\\
  \includegraphics[width=0.95\columnwidth]{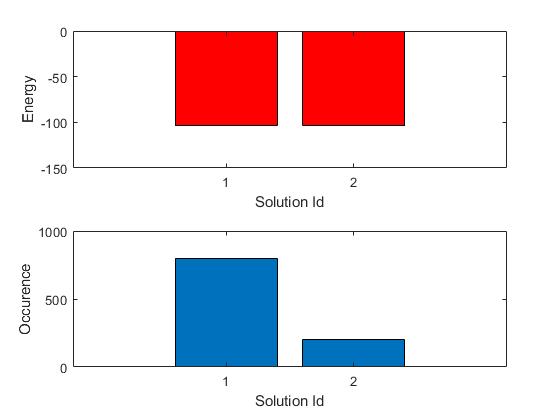}\\
  (b)\\
  \protect\caption{\label{FIG_H12Completion} Implementation of finding $1$-deleted vector of 12-order H-matrix's Hamiltonian into a quantum annealer: (a) Embedding diagram of $28$ logical qubits into $50$ physical qubits in a Chimera-connected qubits of the DW2000Q, (b) Obtained results after quantum annealing: the distribution of energy related to each solution (top) and distribution of the solutions obtained by $1,000$ reads (bottom).
	}
\end{figure}

In this experiment, the D-Wave is programmed to find one vector missing in an 12-order H-matrix. The known $11$ vectors are identical to the simulation case given by Eq.(\ref{H12_Compl}). Based on the Hamiltonian given by Eq.(\ref{H2_COMPLETION_M12}), we realized that $28$ logical qubits are needed. We rely on SAPI embedding module to configure the Chimera-connection of the qubits, which shows that $50$ physical qubits are required. Fig.\ref{FIG_H12Completion}(a) shows the realization of qubits connection in the Chimera graph. Although the order of the matrix is sufficiently high, since the required qubits and couplers for this problem are small, it only occupies a small area in the processor. 

By using the default annealing schedule with $1000$ reads as before, we have obtained the minimum energy of $-104.00$ and the following binary vector as a solution,
\begin{equation}
    v_{11}=
    \begin{pmatrix}
       -, +, -, -, -, +, +, +, -, +, +, - 
    \end{pmatrix}^T
   \nonumber
\end{equation}
which can be verified to be a correct one; i.e., along with $11$ vectors in Eq.(\ref{H12_Compl}), this vector constructs a $12$-order H-matrix. Fig.\ref{FIG_H12Completion}(b) shows the distribution of energy and occurence of the solutions. We see that only two kind of solutions are exists, both of them are at the identical minimum energy level. 

\section{Conclusions}

We have investigated the possibility of quantum computing to solve the problem of finding H-matrix among possible binary matrices of the same order, which is a hard problem. The QAM or quantum annealer has been considered for its realization, which requires the problem to be translated into a Hamiltonian. We have proposed a method to formulate the Hamiltonian's of finding H-matrix and its related problems. 

Existing quantum annealer permits only up to quadratic terms for realization. Since the problem naturally induces higher order terms, we have to perform boolean reduction to obtain realizable Hamiltonians. Manipulation of large number of terms implied by both of growing number of variables with order and the boolean reduction procedure requires a computer-assisted process in constructing the Hamiltonians. The proposed method consists of a set of symbolic computing algorithms to formulate the energy function that lead to the Hamiltonian of the problems. The obtained Hamiltonians are then evaluated by both of simulation and implementation in a 2048 qubits DW2000Q quantum annealer.

For the H-SEARCH problem, existing quantum annealer achieved up to finding 4-order H-matrix. We also have successfully solved the problem of finding $3$ orthogonal binary vectors of length $12$ and the problem of finding $1$ missing vector in a $12$-order H-matrix. In the future, it is expected that higher order H-matrix searching problem can be solved when the device allows more than 2-body interaction or a better qubits connection beyond the Chimera graph is available. 

\section*{Acknowledgments}
This work has been supported by P3MI ITB Grant of Research 2018 and MDR Inc, Tokyo.

\clearpage


\bibliography{5_References.bib}

\begin{thebibliography}{44}%
\makeatletter
\providecommand \@ifxundefined [1]{%
 \@ifx{#1\undefined}
}%
\providecommand \@ifnum [1]{%
 \ifnum #1\expandafter \@firstoftwo
 \else \expandafter \@secondoftwo
 \fi
}%
\providecommand \@ifx [1]{%
 \ifx #1\expandafter \@firstoftwo
 \else \expandafter \@secondoftwo
 \fi
}%
\providecommand \natexlab [1]{#1}%
\providecommand \enquote  [1]{``#1''}%
\providecommand \bibnamefont  [1]{#1}%
\providecommand \bibfnamefont [1]{#1}%
\providecommand \citenamefont [1]{#1}%
\providecommand \href@noop [0]{\@secondoftwo}%
\providecommand \href [0]{\begingroup \@sanitize@url \@href}%
\providecommand \@href[1]{\@@startlink{#1}\@@href}%
\providecommand \@@href[1]{\endgroup#1\@@endlink}%
\providecommand \@sanitize@url [0]{\catcode `\\12\catcode `\$12\catcode
  `\&12\catcode `\#12\catcode `\^12\catcode `\_12\catcode `\%12\relax}%
\providecommand \@@startlink[1]{}%
\providecommand \@@endlink[0]{}%
\providecommand \url  [0]{\begingroup\@sanitize@url \@url }%
\providecommand \@url [1]{\endgroup\@href {#1}{\urlprefix }}%
\providecommand \urlprefix  [0]{URL }%
\providecommand \Eprint [0]{\href }%
\providecommand \doibase [0]{http://dx.doi.org/}%
\providecommand \selectlanguage [0]{\@gobble}%
\providecommand \bibinfo  [0]{\@secondoftwo}%
\providecommand \bibfield  [0]{\@secondoftwo}%
\providecommand \translation [1]{[#1]}%
\providecommand \BibitemOpen [0]{}%
\providecommand \bibitemStop [0]{}%
\providecommand \bibitemNoStop [0]{.\EOS\space}%
\providecommand \EOS [0]{\spacefactor3000\relax}%
\providecommand \BibitemShut  [1]{\csname bibitem#1\endcsname}%
\let\auto@bib@innerbib\@empty
\bibitem [{\citenamefont {Feynman}(1982)}]{feynman1982}%
  \BibitemOpen
  \bibfield  {author} {\bibinfo {author} {\bibfnamefont {R.P.}\ \bibnamefont
  {Feynman}},\ }\bibfield  {title} {\enquote {\bibinfo {title} {Simulating
  physics with computers},}\ }\href@noop {} {\bibfield  {journal} {\bibinfo
  {journal} {Int. J. Theor. Phys.}\ }\textbf {\bibinfo {volume} {21}},\
  \bibinfo {pages} {467--488} (\bibinfo {year} {1982})}\BibitemShut {NoStop}%
\bibitem [{\citenamefont {Deutsch}(1985)}]{Deutsch1985}%
  \BibitemOpen
  \bibfield  {author} {\bibinfo {author} {\bibfnamefont {D.}~\bibnamefont
  {Deutsch}},\ }\bibfield  {title} {\enquote {\bibinfo {title} {Quantum theory,
  the church-turing principle and the universal quantum computer},}\
  }\href@noop {} {\bibfield  {journal} {\bibinfo  {journal} {Proceedings of the
  Royal Society A}\ }\textbf {\bibinfo {volume} {400}},\ \bibinfo {pages}
  {97--117} (\bibinfo {year} {1985})}\BibitemShut {NoStop}%
\bibitem [{\citenamefont {Shor}(1997)}]{Shor1997}%
  \BibitemOpen
  \bibfield  {author} {\bibinfo {author} {\bibfnamefont {P.W.}\ \bibnamefont
  {Shor}},\ }\bibfield  {title} {\enquote {\bibinfo {title} {Polynomial-time
  algorithms for prime factorization and discrete logarithms on a quantum
  computer},}\ }\href@noop {} {\bibfield  {journal} {\bibinfo  {journal} {SIAM
  Journal on Computing}\ }\textbf {\bibinfo {volume} {26}},\ \bibinfo {pages}
  {1484--1509} (\bibinfo {year} {1997})}\BibitemShut {NoStop}%
\bibitem [{\citenamefont {Vandersypen}\ \emph {et~al.}(2001)\citenamefont
  {Vandersypen}, \citenamefont {Breyta}, \citenamefont {Steffen}, \citenamefont
  {Yannoni}, \citenamefont {Sherwood},\ and\ \citenamefont
  {Chuang}}]{Chuang2001}%
  \BibitemOpen
  \bibfield  {author} {\bibinfo {author} {\bibfnamefont {L.M.K.}\ \bibnamefont
  {Vandersypen}}, \bibinfo {author} {\bibfnamefont {G.}~\bibnamefont {Breyta}},
  \bibinfo {author} {\bibfnamefont {M.}~\bibnamefont {Steffen}}, \bibinfo
  {author} {\bibfnamefont {C.S.}\ \bibnamefont {Yannoni}}, \bibinfo {author}
  {\bibfnamefont {M.H.}\ \bibnamefont {Sherwood}}, \ and\ \bibinfo {author}
  {\bibfnamefont {I.L.}\ \bibnamefont {Chuang}},\ }\bibfield  {title} {\enquote
  {\bibinfo {title} {Experimental realization of shor's quantum factoring
  algorithm using nuclear magnetic resonance},}\ }\href@noop {} {\bibfield
  {journal} {\bibinfo  {journal} {Nature}\ }\textbf {\bibinfo {volume} {414}},\
  \bibinfo {pages} {883--887} (\bibinfo {year} {2001})}\BibitemShut {NoStop}%
\bibitem [{\citenamefont {Johnson}\ \emph {et~al.}(2011)\citenamefont
  {Johnson}, \citenamefont {Amin}, \citenamefont {Gildert}, \citenamefont
  {Lanting}, \citenamefont {Hamze}, \citenamefont {Dickson}, \citenamefont
  {Harris}, \citenamefont {Johansson}, \citenamefont {Bunyk} \emph
  {et~al.}}]{johnson2011}%
  \BibitemOpen
  \bibfield  {author} {\bibinfo {author} {\bibfnamefont {M.W.}\ \bibnamefont
  {Johnson}}, \bibinfo {author} {\bibfnamefont {M.H.S.}\ \bibnamefont {Amin}},
  \bibinfo {author} {\bibfnamefont {S.}~\bibnamefont {Gildert}}, \bibinfo
  {author} {\bibfnamefont {T.}~\bibnamefont {Lanting}}, \bibinfo {author}
  {\bibfnamefont {F.}~\bibnamefont {Hamze}}, \bibinfo {author} {\bibfnamefont
  {N.}~\bibnamefont {Dickson}}, \bibinfo {author} {\bibfnamefont {A.J.}\
  \bibnamefont {Harris}, \bibfnamefont {R.and~Berkley}}, \bibinfo {author}
  {\bibfnamefont {J.}~\bibnamefont {Johansson}}, \bibinfo {author}
  {\bibfnamefont {P.}~\bibnamefont {Bunyk}},  \emph {et~al.},\ }\bibfield
  {title} {\enquote {\bibinfo {title} {Quantum annealing with manufactured
  spins},}\ }\href@noop {} {\bibfield  {journal} {\bibinfo  {journal} {Nature}\
  }\textbf {\bibinfo {volume} {473}},\ \bibinfo {pages} {194--198} (\bibinfo
  {year} {2011})}\BibitemShut {NoStop}%
\bibitem [{\citenamefont {Suksmono}(2018)}]{suksmono2018}%
  \BibitemOpen
  \bibfield  {author} {\bibinfo {author} {\bibfnamefont {A.B.}\ \bibnamefont
  {Suksmono}},\ }\bibfield  {title} {\enquote {\bibinfo {title} {Finding a
  {H}adamard matrix by simulated quantum annealing},}\ }\href@noop {}
  {\bibfield  {journal} {\bibinfo  {journal} {Entropy}\ }\textbf {\bibinfo
  {volume} {20}},\ \bibinfo {pages} {141} (\bibinfo {year} {2018})}\BibitemShut
  {NoStop}%
\bibitem [{\citenamefont {Suksmono}(2017)}]{suksmono2012}%
  \BibitemOpen
  \bibfield  {author} {\bibinfo {author} {\bibfnamefont {A.B.}\ \bibnamefont
  {Suksmono}},\ }\bibfield  {title} {\enquote {\bibinfo {title} {Finding a
  hadamard matrix by simulated annealing of spin vectors},}\ }\href@noop {}
  {\bibfield  {journal} {\bibinfo  {journal} {J. Phys.: Conf. Ser.}\ }\textbf
  {\bibinfo {volume} {856}},\ \bibinfo {pages} {012012} (\bibinfo {year}
  {2017})}\BibitemShut {NoStop}%
\bibitem [{\citenamefont {Sylvester}(1867)}]{sylvester1867}%
  \BibitemOpen
  \bibfield  {author} {\bibinfo {author} {\bibfnamefont {J.J.}\ \bibnamefont
  {Sylvester}},\ }\bibfield  {title} {\enquote {\bibinfo {title} {Thoughts on
  inverse orthogonal matrices, simultaneous sign successions, and tessellated
  pavements in two or more colours, with applications to newton's rule,
  ornamental tile-work, and the theory of numbers},}\ }\href@noop {} {\bibfield
   {journal} {\bibinfo  {journal} {Philos. Mag.}\ }\textbf {\bibinfo {volume}
  {34}},\ \bibinfo {pages} {461--475} (\bibinfo {year} {1867})}\BibitemShut
  {NoStop}%
\bibitem [{\citenamefont {Hadamard}(1893)}]{hadamard1893}%
  \BibitemOpen
  \bibfield  {author} {\bibinfo {author} {\bibfnamefont {J.}~\bibnamefont
  {Hadamard}},\ }\bibfield  {title} {\enquote {\bibinfo {title} {Resolution
  d'une question relative aux determinants},}\ }\href@noop {} {\bibfield
  {journal} {\bibinfo  {journal} {Bull. des sciences math.}\ }\textbf {\bibinfo
  {volume} {2}},\ \bibinfo {pages} {240--246} (\bibinfo {year}
  {1893})}\BibitemShut {NoStop}%
\bibitem [{\citenamefont {Hedayat}\ and\ \citenamefont
  {Wallis}(1978)}]{hedayat1978}%
  \BibitemOpen
  \bibfield  {author} {\bibinfo {author} {\bibfnamefont {A.}~\bibnamefont
  {Hedayat}}\ and\ \bibinfo {author} {\bibfnamefont {W.D.}\ \bibnamefont
  {Wallis}},\ }\bibfield  {title} {\enquote {\bibinfo {title} {Hadamard
  matrices and their applications},}\ }\href@noop {} {\bibfield  {journal}
  {\bibinfo  {journal} {Ann. Stat.}\ }\textbf {\bibinfo {volume} {6}},\
  \bibinfo {pages} {1184--1238} (\bibinfo {year} {1978})}\BibitemShut {NoStop}%
\bibitem [{\citenamefont {Horadam}(2007)}]{horadam2007}%
  \BibitemOpen
  \bibfield  {author} {\bibinfo {author} {\bibfnamefont {K.J.}\ \bibnamefont
  {Horadam}},\ }\href@noop {} {\emph {\bibinfo {title} {Hadamard Matrices and
  Their Applications}}}\ (\bibinfo  {publisher} {Princeton University Press},\
  \bibinfo {year} {2007})\BibitemShut {NoStop}%
\bibitem [{\citenamefont {Seberry}\ \emph {et~al.}(2005)\citenamefont
  {Seberry}, \citenamefont {Wysocki},\ and\ \citenamefont
  {Wysocki}}]{seberry2005}%
  \BibitemOpen
  \bibfield  {author} {\bibinfo {author} {\bibfnamefont {J.}~\bibnamefont
  {Seberry}}, \bibinfo {author} {\bibfnamefont {B.J.}\ \bibnamefont {Wysocki}},
  \ and\ \bibinfo {author} {\bibfnamefont {T.A.}\ \bibnamefont {Wysocki}},\
  }\bibfield  {title} {\enquote {\bibinfo {title} {On some applications of
  {H}adamard matrices},}\ }\href@noop {} {\bibfield  {journal} {\bibinfo
  {journal} {Metrika}\ }\textbf {\bibinfo {volume} {62}},\ \bibinfo {pages}
  {221--239} (\bibinfo {year} {2005})}\BibitemShut {NoStop}%
\bibitem [{\citenamefont {Garg}(2007)}]{garg2007}%
  \BibitemOpen
  \bibfield  {author} {\bibinfo {author} {\bibfnamefont {V.}~\bibnamefont
  {Garg}},\ }\href@noop {} {\emph {\bibinfo {title} {Wireless Communications \&
  Networking}}}\ (\bibinfo  {publisher} {Morgan-Kaufman},\ \bibinfo {year}
  {2007})\BibitemShut {NoStop}%
\bibitem [{\citenamefont {Dade}\ and\ \citenamefont
  {Goldberg}(1959)}]{dadegoldberg1959}%
  \BibitemOpen
  \bibfield  {author} {\bibinfo {author} {\bibfnamefont {E.C.}\ \bibnamefont
  {Dade}}\ and\ \bibinfo {author} {\bibfnamefont {K.}~\bibnamefont
  {Goldberg}},\ }\bibfield  {title} {\enquote {\bibinfo {title} {The
  construction of {H}adamard matrices},}\ }\href@noop {} {\bibfield  {journal}
  {\bibinfo  {journal} {Michigan Math. J.}\ }\textbf {\bibinfo {volume} {6}},\
  \bibinfo {pages} {247--250} (\bibinfo {year} {1959})}\BibitemShut {NoStop}%
\bibitem [{\citenamefont {Williamson}\ \emph {et~al.}(1944)\citenamefont
  {Williamson} \emph {et~al.}}]{williamson1944}%
  \BibitemOpen
  \bibfield  {author} {\bibinfo {author} {\bibfnamefont {J.}~\bibnamefont
  {Williamson}} \emph {et~al.},\ }\bibfield  {title} {\enquote {\bibinfo
  {title} {Hadamard's determinant theorem and the sum of four squares},}\
  }\href@noop {} {\bibfield  {journal} {\bibinfo  {journal} {Duke Math. J.}\
  }\textbf {\bibinfo {volume} {11}},\ \bibinfo {pages} {65--81} (\bibinfo
  {year} {1944})}\BibitemShut {NoStop}%
\bibitem [{\citenamefont {Bush}(1971{\natexlab{a}})}]{bush1971a}%
  \BibitemOpen
  \bibfield  {author} {\bibinfo {author} {\bibfnamefont {K.A.}\ \bibnamefont
  {Bush}},\ }\bibfield  {title} {\enquote {\bibinfo {title} {Unbalanced
  {H}adamard matrices and finite projective planes of even order},}\
  }\href@noop {} {\bibfield  {journal} {\bibinfo  {journal} {J. Combin. Theory
  A}\ }\textbf {\bibinfo {volume} {11}},\ \bibinfo {pages} {38--44} (\bibinfo
  {year} {1971}{\natexlab{a}})}\BibitemShut {NoStop}%
\bibitem [{\citenamefont {Bush}(1971{\natexlab{b}})}]{bush1971b}%
  \BibitemOpen
  \bibfield  {author} {\bibinfo {author} {\bibfnamefont {K.A.}\ \bibnamefont
  {Bush}},\ }\href@noop {} {\bibfield  {journal} {\bibinfo  {journal} {Atti del
  Convegno di Geometria Combinatoria e sue Applicazioni}\ }\textbf {\bibinfo
  {volume} {131}} (\bibinfo {year} {1971}{\natexlab{b}})}\BibitemShut {NoStop}%
\bibitem [{\citenamefont {Paley}(1933)}]{paley1933}%
  \BibitemOpen
  \bibfield  {author} {\bibinfo {author} {\bibfnamefont {R.E.A.C.}\
  \bibnamefont {Paley}},\ }\bibfield  {title} {\enquote {\bibinfo {title} {On
  orthogonal matrices},}\ }\href@noop {} {\bibfield  {journal} {\bibinfo
  {journal} {J. Math. Phys.}\ }\textbf {\bibinfo {volume} {12}},\ \bibinfo
  {pages} {311--320} (\bibinfo {year} {1933})}\BibitemShut {NoStop}%
\bibitem [{\citenamefont {Wallis}(1976)}]{wallis1976}%
  \BibitemOpen
  \bibfield  {author} {\bibinfo {author} {\bibfnamefont {J.S.}\ \bibnamefont
  {Wallis}},\ }\bibfield  {title} {\enquote {\bibinfo {title} {On the existence
  of {H}adamard matrices},}\ }\href@noop {} {\bibfield  {journal} {\bibinfo
  {journal} {J. Combin. Theory A}\ }\textbf {\bibinfo {volume} {21}},\ \bibinfo
  {pages} {188--195} (\bibinfo {year} {1976})}\BibitemShut {NoStop}%
\bibitem [{\citenamefont {Metropolis}\ \emph {et~al.}(1953)\citenamefont
  {Metropolis}, \citenamefont {Rosenbluth}, \citenamefont {Rosenbluth},
  \citenamefont {Teller},\ and\ \citenamefont {Teller}}]{metropolis1953}%
  \BibitemOpen
  \bibfield  {author} {\bibinfo {author} {\bibfnamefont {N.}~\bibnamefont
  {Metropolis}}, \bibinfo {author} {\bibfnamefont {A.W.}\ \bibnamefont
  {Rosenbluth}}, \bibinfo {author} {\bibfnamefont {M.N.}\ \bibnamefont
  {Rosenbluth}}, \bibinfo {author} {\bibfnamefont {A.H.}\ \bibnamefont
  {Teller}}, \ and\ \bibinfo {author} {\bibfnamefont {E.}~\bibnamefont
  {Teller}},\ }\bibfield  {title} {\enquote {\bibinfo {title} {Equation of
  state calculations by fast computing machines},}\ }\href@noop {} {\bibfield
  {journal} {\bibinfo  {journal} {J. Chem. Phys.}\ }\textbf {\bibinfo {volume}
  {21}},\ \bibinfo {pages} {1087--1092} (\bibinfo {year} {1953})}\BibitemShut
  {NoStop}%
\bibitem [{\citenamefont {Kirkpatrick}\ \emph {et~al.}(1983)\citenamefont
  {Kirkpatrick}, \citenamefont {Gelatt},\ and\ \citenamefont
  {Vecchi}}]{kirkpatrick1983}%
  \BibitemOpen
  \bibfield  {author} {\bibinfo {author} {\bibfnamefont {S.}~\bibnamefont
  {Kirkpatrick}}, \bibinfo {author} {\bibfnamefont {C.D.}\ \bibnamefont
  {Gelatt}}, \ and\ \bibinfo {author} {\bibfnamefont {M.P.}\ \bibnamefont
  {Vecchi}},\ }\bibfield  {title} {\enquote {\bibinfo {title} {Optimization by
  simulated annealing},}\ }\href@noop {} {\bibfield  {journal} {\bibinfo
  {journal} {Science}\ }\textbf {\bibinfo {volume} {220}},\ \bibinfo {pages}
  {671--680} (\bibinfo {year} {1983})}\BibitemShut {NoStop}%
\bibitem [{\citenamefont {Cerny}(1985)}]{cerny1985}%
  \BibitemOpen
  \bibfield  {author} {\bibinfo {author} {\bibfnamefont {V.}~\bibnamefont
  {Cerny}},\ }\bibfield  {title} {\enquote {\bibinfo {title} {Thermodynamical
  approach to the traveling salesman problem: An efficient simulation
  algorithm},}\ }\href@noop {} {\bibfield  {journal} {\bibinfo  {journal} {J.
  Optim. Theory Appl.}\ }\textbf {\bibinfo {volume} {45}},\ \bibinfo {pages}
  {41--51} (\bibinfo {year} {1985})}\BibitemShut {NoStop}%
\bibitem [{\citenamefont {Battaglia}\ \emph {et~al.}(2005)\citenamefont
  {Battaglia}, \citenamefont {Santoro},\ and\ \citenamefont
  {Tosatti}}]{battaglia2005}%
  \BibitemOpen
  \bibfield  {author} {\bibinfo {author} {\bibfnamefont {D.A.}\ \bibnamefont
  {Battaglia}}, \bibinfo {author} {\bibfnamefont {G.E.}\ \bibnamefont
  {Santoro}}, \ and\ \bibinfo {author} {\bibfnamefont {E.}~\bibnamefont
  {Tosatti}},\ }\bibfield  {title} {\enquote {\bibinfo {title} {Optimization by
  quantum annealing: Lessons from hard satisfiability problems},}\ }\href@noop
  {} {\bibfield  {journal} {\bibinfo  {journal} {Phys. Rev. E}\ }\textbf
  {\bibinfo {volume} {71}},\ \bibinfo {pages} {066707} (\bibinfo {year}
  {2005})}\BibitemShut {NoStop}%
\bibitem [{\citenamefont {Kadowaki}\ and\ \citenamefont
  {Nishimori}(1998)}]{kadowaki1988}%
  \BibitemOpen
  \bibfield  {author} {\bibinfo {author} {\bibfnamefont {T.}~\bibnamefont
  {Kadowaki}}\ and\ \bibinfo {author} {\bibfnamefont {H.}~\bibnamefont
  {Nishimori}},\ }\bibfield  {title} {\enquote {\bibinfo {title} {Quantum
  annealing in the transverse ising model},}\ }\href@noop {} {\bibfield
  {journal} {\bibinfo  {journal} {Phys. Rev. E}\ }\textbf {\bibinfo {volume}
  {58}},\ \bibinfo {pages} {5355} (\bibinfo {year} {1998})}\BibitemShut
  {NoStop}%
\bibitem [{\citenamefont {Santoro}\ \emph {et~al.}(2002)\citenamefont
  {Santoro}, \citenamefont {Marto{\v{n}}{\'a}k}, \citenamefont {Tosatti},\ and\
  \citenamefont {Car}}]{santoro2002}%
  \BibitemOpen
  \bibfield  {author} {\bibinfo {author} {\bibfnamefont {G.E.}\ \bibnamefont
  {Santoro}}, \bibinfo {author} {\bibfnamefont {R.}~\bibnamefont
  {Marto{\v{n}}{\'a}k}}, \bibinfo {author} {\bibfnamefont {E.}~\bibnamefont
  {Tosatti}}, \ and\ \bibinfo {author} {\bibfnamefont {R.}~\bibnamefont
  {Car}},\ }\bibfield  {title} {\enquote {\bibinfo {title} {Theory of quantum
  annealing of an {I}sing spin glass},}\ }\href@noop {} {\bibfield  {journal}
  {\bibinfo  {journal} {Science}\ }\textbf {\bibinfo {volume} {295}},\ \bibinfo
  {pages} {2427--2430} (\bibinfo {year} {2002})}\BibitemShut {NoStop}%
\bibitem [{\citenamefont {Biamonte}(2008)}]{biamonte2008}%
  \BibitemOpen
  \bibfield  {author} {\bibinfo {author} {\bibfnamefont {J.~D.}\ \bibnamefont
  {Biamonte}},\ }\bibfield  {title} {\enquote {\bibinfo {title}
  {Nonperturbative k-body to two-body commuting conversion {H}amiltonians and
  embedding problem instances into ising spins},}\ }\href@noop {} {\bibfield
  {journal} {\bibinfo  {journal} {Phys. Rev. A}\ }\textbf {\bibinfo {volume}
  {77}},\ \bibinfo {pages} {052331} (\bibinfo {year} {2008})}\BibitemShut
  {NoStop}%
\bibitem [{\citenamefont {Perdomo}\ \emph {et~al.}(2008)\citenamefont
  {Perdomo}, \citenamefont {Truncik}, \citenamefont {Tubert-Brohman},
  \citenamefont {Rose},\ and\ \citenamefont {Aspuru-Guzik}}]{perdomo2008}%
  \BibitemOpen
  \bibfield  {author} {\bibinfo {author} {\bibfnamefont {A.}~\bibnamefont
  {Perdomo}}, \bibinfo {author} {\bibfnamefont {C.}~\bibnamefont {Truncik}},
  \bibinfo {author} {\bibfnamefont {I.}~\bibnamefont {Tubert-Brohman}},
  \bibinfo {author} {\bibfnamefont {G.}~\bibnamefont {Rose}}, \ and\ \bibinfo
  {author} {\bibfnamefont {A.}~\bibnamefont {Aspuru-Guzik}},\ }\bibfield
  {title} {\enquote {\bibinfo {title} {Construction of model {H}amiltonians for
  adiabatic quantum computation and its application to finding low-energy
  conformations of lattice protein models},}\ }\href@noop {} {\bibfield
  {journal} {\bibinfo  {journal} {Phys. Rev. A}\ }\textbf {\bibinfo {volume}
  {78}},\ \bibinfo {pages} {012320} (\bibinfo {year} {2008})}\BibitemShut
  {NoStop}%
\bibitem [{\citenamefont {Jiang}\ \emph {et~al.}(2017)\citenamefont {Jiang},
  \citenamefont {Smelyanskiy}, \citenamefont {Isakov}, \citenamefont {Boixo},
  \citenamefont {Mazzola}, \citenamefont {Troyer},\ and\ \citenamefont
  {Neven}}]{jiang2017}%
  \BibitemOpen
  \bibfield  {author} {\bibinfo {author} {\bibfnamefont {Z.}~\bibnamefont
  {Jiang}}, \bibinfo {author} {\bibfnamefont {V.N.}\ \bibnamefont
  {Smelyanskiy}}, \bibinfo {author} {\bibfnamefont {S.V.}\ \bibnamefont
  {Isakov}}, \bibinfo {author} {\bibfnamefont {S.}~\bibnamefont {Boixo}},
  \bibinfo {author} {\bibfnamefont {G.}~\bibnamefont {Mazzola}}, \bibinfo
  {author} {\bibfnamefont {M.}~\bibnamefont {Troyer}}, \ and\ \bibinfo {author}
  {\bibfnamefont {H.}~\bibnamefont {Neven}},\ }\bibfield  {title} {\enquote
  {\bibinfo {title} {Scaling analysis and instantons for thermally assisted
  tunneling and quantum simulations},}\ }\href@noop {} {\bibfield  {journal}
  {\bibinfo  {journal} {Phys. Rev. A}\ }\textbf {\bibinfo {volume} {95}},\
  \bibinfo {pages} {012322} (\bibinfo {year} {2017})}\BibitemShut {NoStop}%
\bibitem [{\citenamefont {Hormozi}\ \emph {et~al.}(2017)\citenamefont
  {Hormozi}, \citenamefont {Brown}, \citenamefont {Carleo},\ and\ \citenamefont
  {Troyer}}]{hormozi2017}%
  \BibitemOpen
  \bibfield  {author} {\bibinfo {author} {\bibfnamefont {L.}~\bibnamefont
  {Hormozi}}, \bibinfo {author} {\bibfnamefont {E.W.}\ \bibnamefont {Brown}},
  \bibinfo {author} {\bibfnamefont {G.}~\bibnamefont {Carleo}}, \ and\ \bibinfo
  {author} {\bibfnamefont {M.}~\bibnamefont {Troyer}},\ }\bibfield  {title}
  {\enquote {\bibinfo {title} {Nonstoquastic hamiltonians and quantum annealing
  of an ising spin glass},}\ }\href@noop {} {\bibfield  {journal} {\bibinfo
  {journal} {Phys. Rev. B}\ }\textbf {\bibinfo {volume} {95}},\ \bibinfo
  {pages} {184416} (\bibinfo {year} {2017})}\BibitemShut {NoStop}%
\bibitem [{\citenamefont {Boixo}\ \emph {et~al.}(2014)\citenamefont {Boixo},
  \citenamefont {R{\o}nnow}, \citenamefont {Isakov}, \citenamefont {Wang},
  \citenamefont {Martinis},\ and\ \citenamefont {Troyer}}]{boixo2014}%
  \BibitemOpen
  \bibfield  {author} {\bibinfo {author} {\bibfnamefont {S}~\bibnamefont
  {Boixo}}, \bibinfo {author} {\bibfnamefont {T.F.}\ \bibnamefont {R{\o}nnow}},
  \bibinfo {author} {\bibfnamefont {S.V.}\ \bibnamefont {Isakov}}, \bibinfo
  {author} {\bibfnamefont {Z.}~\bibnamefont {Wang}}, \bibinfo {author}
  {\bibfnamefont {J.M.}\ \bibnamefont {Martinis}}, \ and\ \bibinfo {author}
  {\bibfnamefont {M.}~\bibnamefont {Troyer}},\ }\bibfield  {title} {\enquote
  {\bibinfo {title} {Evidence for quantum annealing with more than one hundred
  qubits},}\ }\href@noop {} {\bibfield  {journal} {\bibinfo  {journal} {Nature
  Physics}\ }\textbf {\bibinfo {volume} {10}},\ \bibinfo {pages} {218--224}
  (\bibinfo {year} {2014})}\BibitemShut {NoStop}%
\bibitem [{\citenamefont {Heim}\ \emph {et~al.}(2015)\citenamefont {Heim},
  \citenamefont {R{\o}nnow}, \citenamefont {Isakov},\ and\ \citenamefont
  {Troyer}}]{heim2015}%
  \BibitemOpen
  \bibfield  {author} {\bibinfo {author} {\bibfnamefont {B.}~\bibnamefont
  {Heim}}, \bibinfo {author} {\bibfnamefont {T.F.}\ \bibnamefont {R{\o}nnow}},
  \bibinfo {author} {\bibfnamefont {S.V.}\ \bibnamefont {Isakov}}, \ and\
  \bibinfo {author} {\bibfnamefont {M.}~\bibnamefont {Troyer}},\ }\bibfield
  {title} {\enquote {\bibinfo {title} {Quantum versus classical annealing of
  {I}sing spin glasses},}\ }\href@noop {} {\bibfield  {journal} {\bibinfo
  {journal} {Science}\ }\textbf {\bibinfo {volume} {348}},\ \bibinfo {pages}
  {215--217} (\bibinfo {year} {2015})}\BibitemShut {NoStop}%
\bibitem [{\citenamefont {Isakov}\ \emph {et~al.}(2016)\citenamefont {Isakov},
  \citenamefont {Mazzola}, \citenamefont {Smelyanskiy}, \citenamefont {Jiang},
  \citenamefont {Boixo}, \citenamefont {Neven},\ and\ \citenamefont
  {Troyer}}]{isakov2016}%
  \BibitemOpen
  \bibfield  {author} {\bibinfo {author} {\bibfnamefont {S.V.}\ \bibnamefont
  {Isakov}}, \bibinfo {author} {\bibfnamefont {G.}~\bibnamefont {Mazzola}},
  \bibinfo {author} {\bibfnamefont {V.N.}\ \bibnamefont {Smelyanskiy}},
  \bibinfo {author} {\bibfnamefont {Z.}~\bibnamefont {Jiang}}, \bibinfo
  {author} {\bibfnamefont {S.}~\bibnamefont {Boixo}}, \bibinfo {author}
  {\bibfnamefont {H.}~\bibnamefont {Neven}}, \ and\ \bibinfo {author}
  {\bibfnamefont {M.}~\bibnamefont {Troyer}},\ }\bibfield  {title} {\enquote
  {\bibinfo {title} {Understanding quantum tunneling through quantum {M}onte
  {C}arlo simulations},}\ }\href@noop {} {\bibfield  {journal} {\bibinfo
  {journal} {Phys. Rev. Lett.}\ }\textbf {\bibinfo {volume} {117}},\ \bibinfo
  {pages} {180402} (\bibinfo {year} {2016})}\BibitemShut {NoStop}%
\bibitem [{\citenamefont {R{\o}nnow}\ \emph {et~al.}(2014)\citenamefont
  {R{\o}nnow}, \citenamefont {Wang}, \citenamefont {Job}, \citenamefont
  {Boixo}, \citenamefont {Isakov}, \citenamefont {Wecker}, \citenamefont
  {Martinis}, \citenamefont {Lidar},\ and\ \citenamefont
  {Troyer}}]{ronnow2014}%
  \BibitemOpen
  \bibfield  {author} {\bibinfo {author} {\bibfnamefont {T.F.}\ \bibnamefont
  {R{\o}nnow}}, \bibinfo {author} {\bibfnamefont {Z.}~\bibnamefont {Wang}},
  \bibinfo {author} {\bibfnamefont {J.}~\bibnamefont {Job}}, \bibinfo {author}
  {\bibfnamefont {S.}~\bibnamefont {Boixo}}, \bibinfo {author} {\bibfnamefont
  {S.V.}\ \bibnamefont {Isakov}}, \bibinfo {author} {\bibfnamefont
  {D.}~\bibnamefont {Wecker}}, \bibinfo {author} {\bibfnamefont {J.M.}\
  \bibnamefont {Martinis}}, \bibinfo {author} {\bibfnamefont {D.A.}\
  \bibnamefont {Lidar}}, \ and\ \bibinfo {author} {\bibfnamefont
  {M.}~\bibnamefont {Troyer}},\ }\bibfield  {title} {\enquote {\bibinfo {title}
  {Defining and detecting quantum speedup},}\ }\href@noop {} {\bibfield
  {journal} {\bibinfo  {journal} {Science}\ }\textbf {\bibinfo {volume}
  {345}},\ \bibinfo {pages} {420--424} (\bibinfo {year} {2014})}\BibitemShut
  {NoStop}%
\bibitem [{\citenamefont {Mazzola}\ \emph {et~al.}(2017)\citenamefont
  {Mazzola}, \citenamefont {Smelyanskiy},\ and\ \citenamefont
  {Troyer}}]{mazzola2017a}%
  \BibitemOpen
  \bibfield  {author} {\bibinfo {author} {\bibfnamefont {G.}~\bibnamefont
  {Mazzola}}, \bibinfo {author} {\bibfnamefont {V.N.}\ \bibnamefont
  {Smelyanskiy}}, \ and\ \bibinfo {author} {\bibfnamefont {M.}~\bibnamefont
  {Troyer}},\ }\bibfield  {title} {\enquote {\bibinfo {title} {Quantum monte
  carlo tunneling from quantum chemistry to quantum annealing},}\ }\href@noop
  {} {\bibfield  {journal} {\bibinfo  {journal} {Phys. Rev. B}\ }\textbf
  {\bibinfo {volume} {96}},\ \bibinfo {pages} {134305} (\bibinfo {year}
  {2017})}\BibitemShut {NoStop}%
\bibitem [{\citenamefont {Marto{\v{n}}{\'a}k}\ \emph
  {et~al.}(2004)\citenamefont {Marto{\v{n}}{\'a}k}, \citenamefont {Santoro},\
  and\ \citenamefont {Tosatti}}]{martonak2004}%
  \BibitemOpen
  \bibfield  {author} {\bibinfo {author} {\bibfnamefont {R.}~\bibnamefont
  {Marto{\v{n}}{\'a}k}}, \bibinfo {author} {\bibfnamefont {G.E.}\ \bibnamefont
  {Santoro}}, \ and\ \bibinfo {author} {\bibfnamefont {E.}~\bibnamefont
  {Tosatti}},\ }\bibfield  {title} {\enquote {\bibinfo {title} {Quantum
  annealing of the traveling-salesman problem},}\ }\href@noop {} {\bibfield
  {journal} {\bibinfo  {journal} {Phys. Rev. E}\ }\textbf {\bibinfo {volume}
  {70}},\ \bibinfo {pages} {057701} (\bibinfo {year} {2004})}\BibitemShut
  {NoStop}%
\bibitem [{\citenamefont {Titiloye}\ and\ \citenamefont
  {Crispin}(2011)}]{titiloye2011}%
  \BibitemOpen
  \bibfield  {author} {\bibinfo {author} {\bibfnamefont {O.}~\bibnamefont
  {Titiloye}}\ and\ \bibinfo {author} {\bibfnamefont {A.}~\bibnamefont
  {Crispin}},\ }\bibfield  {title} {\enquote {\bibinfo {title} {Quantum
  annealing of the graph coloring problem},}\ }\href@noop {} {\bibfield
  {journal} {\bibinfo  {journal} {Discrete Optimization}\ }\textbf {\bibinfo
  {volume} {8}},\ \bibinfo {pages} {376--384} (\bibinfo {year}
  {2011})}\BibitemShut {NoStop}%
\bibitem [{\citenamefont {Zick}\ \emph {et~al.}(2015)\citenamefont {Zick},
  \citenamefont {Shehab},\ and\ \citenamefont {French}}]{zick2015}%
  \BibitemOpen
  \bibfield  {author} {\bibinfo {author} {\bibfnamefont {K.M.}\ \bibnamefont
  {Zick}}, \bibinfo {author} {\bibfnamefont {O.}~\bibnamefont {Shehab}}, \ and\
  \bibinfo {author} {\bibfnamefont {M.}~\bibnamefont {French}},\ }\bibfield
  {title} {\enquote {\bibinfo {title} {Experimental quantum annealing: case
  study involving the graph isomorphism problem},}\ }\href@noop {} {\bibfield
  {journal} {\bibinfo  {journal} {Sci. Rep.}\ }\textbf {\bibinfo {volume}
  {5}},\ \bibinfo {pages} {11168} (\bibinfo {year} {2015})}\BibitemShut
  {NoStop}%
\bibitem [{\citenamefont {Jiang}\ \emph {et~al.}(2018)\citenamefont {Jiang},
  \citenamefont {Britt}, \citenamefont {McCaskey}, \citenamefont {Humble},\
  and\ \citenamefont {Kais}}]{Jiang2018}%
  \BibitemOpen
  \bibfield  {author} {\bibinfo {author} {\bibfnamefont {S.}~\bibnamefont
  {Jiang}}, \bibinfo {author} {\bibfnamefont {K.A.}\ \bibnamefont {Britt}},
  \bibinfo {author} {\bibfnamefont {A.J.}\ \bibnamefont {McCaskey}}, \bibinfo
  {author} {\bibfnamefont {T.S.}\ \bibnamefont {Humble}}, \ and\ \bibinfo
  {author} {\bibfnamefont {S.}~\bibnamefont {Kais}},\ }\bibfield  {title}
  {\enquote {\bibinfo {title} {Quantum annealing for prime factorization},}\
  }\href@noop {} {\bibfield  {journal} {\bibinfo  {journal} {Scientific
  Reports}\ }\textbf {\bibinfo {volume} {8}} (\bibinfo {year}
  {2018})}\BibitemShut {NoStop}%
\bibitem [{\citenamefont {Benedetti}\ \emph {et~al.}(2017)\citenamefont
  {Benedetti}, \citenamefont {Realpe-G\'omez}, \citenamefont {Biswas},\ and\
  \citenamefont {Perdomo-Ortiz}}]{Benedetti2017}%
  \BibitemOpen
  \bibfield  {author} {\bibinfo {author} {\bibfnamefont {M.}~\bibnamefont
  {Benedetti}}, \bibinfo {author} {\bibfnamefont {J.}~\bibnamefont
  {Realpe-G\'omez}}, \bibinfo {author} {\bibfnamefont {R.}~\bibnamefont
  {Biswas}}, \ and\ \bibinfo {author} {\bibfnamefont {A.}~\bibnamefont
  {Perdomo-Ortiz}},\ }\bibfield  {title} {\enquote {\bibinfo {title}
  {Quantum-assisted learning of hardware-embedded probabilistic graphical
  models},}\ }\href {\doibase 10.1103/PhysRevX.7.041052} {\bibfield  {journal}
  {\bibinfo  {journal} {Phys. Rev. X}\ }\textbf {\bibinfo {volume} {7}},\
  \bibinfo {pages} {041052} (\bibinfo {year} {2017})}\BibitemShut {NoStop}%
\bibitem [{\citenamefont {Li}\ \emph {et~al.}(2018)\citenamefont {Li},
  \citenamefont {Felice}, \citenamefont {Rohs},\ and\ \citenamefont
  {Lidar}}]{Li2018}%
  \BibitemOpen
  \bibfield  {author} {\bibinfo {author} {\bibfnamefont {R.Y.}\ \bibnamefont
  {Li}}, \bibinfo {author} {\bibfnamefont {R.D.}\ \bibnamefont {Felice}},
  \bibinfo {author} {\bibfnamefont {R.}~\bibnamefont {Rohs}}, \ and\ \bibinfo
  {author} {\bibfnamefont {D.A.}\ \bibnamefont {Lidar}},\ }\bibfield  {title}
  {\enquote {\bibinfo {title} {Quantum annealing versus classical machine
  learning applied to a simplified computational biology problem},}\
  }\href@noop {} {\bibfield  {journal} {\bibinfo  {journal} {npj Quantum
  Information}\ }\textbf {\bibinfo {volume} {4}} (\bibinfo {year}
  {2018})}\BibitemShut {NoStop}%
\bibitem [{\citenamefont {O'Malley}(2018)}]{OMalley2018}%
  \BibitemOpen
  \bibfield  {author} {\bibinfo {author} {\bibfnamefont {D.}~\bibnamefont
  {O'Malley}},\ }\bibfield  {title} {\enquote {\bibinfo {title} {An approach to
  quantum-computational hydrologic inverse analysis},}\ }\href@noop {}
  {\bibfield  {journal} {\bibinfo  {journal} {Scientific Reports}\ }\textbf
  {\bibinfo {volume} {8}} (\bibinfo {year} {2018})}\BibitemShut {NoStop}%
\bibitem [{\citenamefont {Suksmono}(2016)}]{suksmono_arxiv2016}%
  \BibitemOpen
  \bibfield  {author} {\bibinfo {author} {\bibfnamefont {A.B.}\ \bibnamefont
  {Suksmono}},\ }\bibfield  {title} {\enquote {\bibinfo {title} {Probabilistic
  construction and analysis of seminormalized {H}adamard matrices},}\
  }\href@noop {} {\bibfield  {journal} {\bibinfo  {journal} {eprint
  arXiv:1606.09368v1}\ } (\bibinfo {year} {2016})}\BibitemShut {NoStop}%
\bibitem [{\citenamefont {Inc.}(2018{\natexlab{a}})}]{dwave2018}%
  \BibitemOpen
  \bibfield  {author} {\bibinfo {author} {\bibfnamefont {D-Wave~System}\
  \bibnamefont {Inc.}},\ }\href@noop {} {\emph {\bibinfo {title} {Getting
  Started with the D-Wave System: User Manual}}}\ (\bibinfo  {publisher}
  {D-Wave System Inc.},\ \bibinfo {year} {2018})\BibitemShut {NoStop}%
\bibitem [{\citenamefont {Inc.}(2018{\natexlab{b}})}]{dwave2018qpu}%
  \BibitemOpen
  \bibfield  {author} {\bibinfo {author} {\bibfnamefont {D-Wave~System}\
  \bibnamefont {Inc.}},\ }\href@noop {} {\emph {\bibinfo {title} {Technical
  Description of the D-Wave Quantum Processing Unit: User Manual}}}\ (\bibinfo
  {publisher} {D-Wave System Inc.},\ \bibinfo {year} {2018})\BibitemShut
  {NoStop}%
\end{thebibliography}%

\section*{Appendix}


\subsection*{A complete expression of Eq.(\ref{Ek_si4})}

$E_k(s_i) = 2s_{0}s_{1}s_{12}s_{13} + 2s_{0}s_{1}s_{4}s_{5} + 2s_{0}s_{1}s_{8}s_{9} + 2s_{0}s_{10}s_{2}s_{8} + 2s_{0}s_{11}s_{3}s_{8} + 2s_{0}s_{12}s_{14}s_{2} + 2s_{0}s_{12}s_{15}s_{3} + 2s_{0}s_{2}s_{4}s_{6} + 2s_{0}s_{3}s_{4}s_{7} + 2s_{1}s_{10}s_{2}s_{9} + 2s_{1}s_{11}s_{3}s_{9} + 2s_{1}s_{13}s_{14}s_{2} + 2s_{1}s_{13}s_{15}s_{3} + 2s_{1}s_{2}s_{5}s_{6} + 2s_{1}s_{3}s_{5}s_{7} + 2s_{10}s_{11}s_{14}s_{15} + 2s_{10}s_{11}s_{2}s_{3} + 2s_{10}s_{11}s_{6}s_{7} + 2s_{10}s_{12}s_{14}s_{8} + 2s_{10}s_{13}s_{14}s_{9} + 2s_{10}s_{4}s_{6}s_{8} + 2s_{10}s_{5}s_{6}s_{9} + 2s_{11}s_{12}s_{15}s_{8} + 2s_{11}s_{13}s_{15}s_{9} + 2s_{11}s_{4}s_{7}s_{8} + 2s_{11}s_{5}s_{7}s_{9} + 2s_{12}s_{13}s_{4}s_{5} + 2s_{12}s_{13}s_{8}s_{9} + 2s_{12}s_{14}s_{4}s_{6} + 2s_{12}s_{15}s_{4}s_{7} + 2s_{13}s_{14}s_{5}s_{6} + 2s_{13}s_{15}s_{5}s_{7} + 2s_{14}s_{15}s_{2}s_{3} + 2s_{14}s_{15}s_{6}s_{7} + 2s_{2}s_{3}s_{6}s_{7} + 2s_{4}s_{5}s_{8}s_{9} + 24$


\subsection*{A complete expression of Eq.(\ref{Ek_qi4})}

$ E_k(q_i) = 32q_{0}q_1q_{12}q_{13} - 16q_{0}q_1q_{12} - 16q_{0}q_1q_{13} + 32q_{0}q_1q_4q_5 - 16q_{0}q_1q_4 - 16q_{0}q_1q_5 + 32q_{0}q_1q_8q_9 - 16q_{0}q_1q_8 - 16q_{0}q_1q_9 + 24q_{0}q_1 + 32q_{0}q_{10}q_2q_8 - 16q_{0}q_{10}q_2 - 16q_{0}q_{10}q_8 + 8q_{0}q_{10} + 32q_{0}q_{11}q_3q_8 - 16q_{0}q_{11}q_3 - 16q_{0}q_{11}q_8 + 8q_{0}q_{11} - 16q_{0}q_{12}q_{13} + 32q_{0}q_{12}q_{14}q_2 - 16q_{0}q_{12}q_{14} + 32q_{0}q_{12}q_{15}q_3 - 16q_{0}q_{12}q_{15} - 16q_{0}q_{12}q_2 - 16q_{0}q_{12}q_3 + 24q_{0}q_{12} + 8q_{0}q_{13} - 16q_{0}q_{14}q_2 + 8q_{0}q_{14} - 16q_{0}q_{15}q_3 + 8q_{0}q_{15} + 32q_{0}q_2q_4q_6 - 16q_{0}q_2q_4 - 16q_{0}q_2q_6 - 16q_{0}q_2q_8 + 24q_{0}q_2 + 32q_{0}q_3q_4q_7 - 16q_{0}q_3q_4 - 16q_{0}q_3q_7 - 16q_{0}q_3q_8 + 24q_{0}q_3 - 16q_{0}q_4q_5 - 16q_{0}q_4q_6 - 16q_{0}q_4q_7 + 24q_{0}q_4 + 8q_{0}q_5 + 8q_{0}q_6 + 8q_{0}q_7 - 16q_{0}q_8q_9 + 24q_{0}q_8 + 8q_{0}q_9 - 36q_{0} + 32q_1q_{10}q_2q_9 - 16q_1q_{10}q_2 - 16q_1q_{10}q_9 + 8q_1q_{10} + 32q_1q_{11}q_3q_9 - 16q_1q_{11}q_3 - 16q_1q_{11}q_9 + 8q_1q_{11} - 16q_1q_{12}q_{13} + 8q_1q_{12} + 32q_1q_{13}q_{14}q_2 - 16q_1q_{13}q_{14} + 32q_1q_{13}q_{15}q_3 - 16q_1q_{13}q_{15} - 16q_1q_{13}q_2 - 16q_1q_{13}q_3 + 24q_1q_{13} - 16q_1q_{14}q_2 + 8q_1q_{14} - 16q_1q_{15}q_3 + 8q_1q_{15} + 32q_1q_2q_5q_6 - 16q_1q_2q_5 - 16q_1q_2q_6 - 16q_1q_2q_9 + 24q_1q_2 + 32q_1q_3q_5q_7 - 16q_1q_3q_5 - 16q_1q_3q_7 - 16q_1q_3q_9 + 24q_1q_3 - 16q_1q_4q_5 + 8q_1q_4 - 16q_1q_5q_6 - 16q_1q_5q_7 + 24q_1q_5 + 8q_1q_6 + 8q_1q_7 - 16q_1q_8q_9 + 8q_1q_8 + 24q_1q_9 - 36q_1 + 32q_{10}q_{11}q_{14}q_{15} - 16q_{10}q_{11}q_{14} - 16q_{10}q_{11}q_{15} + 32q_{10}q_{11}q_2q_3 - 16q_{10}q_{11}q_2 - 16q_{10}q_{11}q_3 + 32q_{10}q_{11}q_6q_7 - 16q_{10}q_{11}q_6 - 16q_{10}q_{11}q_7 + 24q_{10}q_{11} + 32q_{10}q_{12}q_{14}q_8 - 16q_{10}q_{12}q_{14} - 16q_{10}q_{12}q_8 + 8q_{10}q_{12} + 32q_{10}q_{13}q_{14}q_9 - 16q_{10}q_{13}q_{14} - 16q_{10}q_{13}q_9 + 8q_{10}q_{13} - 16q_{10}q_{14}q_{15} - 16q_{10}q_{14}q_8 - 16q_{10}q_{14}q_9 + 24q_{10}q_{14} + 8q_{10}q_{15} - 16q_{10}q_2q_3 - 16q_{10}q_2q_8 - 16q_{10}q_2q_9 + 24q_{10}q_2 + 8q_{10}q_3 + 32q_{10}q_4q_6q_8 - 16q_{10}q_4q_6 - 16q_{10}q_4q_8 + 8q_{10}q_4 + 32q_{10}q_5q_6q_9 - 16q_{10}q_5q_6 - 16q_{10}q_5q_9 + 8q_{10}q_5 - 16q_{10}q_6q_7 - 16q_{10}q_6q_8 - 16q_{10}q_6q_9 + 24q_{10}q_6 + 8q_{10}q_7 + 24q_{10}q_8 + 24q_{10}q_9 - 36q_{10} + 32q_{11}q_{12}q_{15}q_8 - 16q_{11}q_{12}q_{15} - 16q_{11}q_{12}q_8 + 8q_{11}q_{12} + 32q_{11}q_{13}q_{15}q_9 - 16q_{11}q_{13}q_{15} - 16q_{11}q_{13}q_9 + 8q_{11}q_{13} - 16q_{11}q_{14}q_{15} + 8q_{11}q_{14} - 16q_{11}q_{15}q_8 - 16q_{11}q_{15}q_9 + 24q_{11}q_{15} - 16q_{11}q_2q_3 + 8q_{11}q_2 - 16q_{11}q_3q_8 - 16q_{11}q_3q_9 + 24q_{11}q_3 + 32q_{11}q_4q_7q_8 - 16q_{11}q_4q_7 - 16q_{11}q_4q_8 + 8q_{11}q_4 + 32q_{11}q_5q_7q_9 - 16q_{11}q_5q_7 - 16q_{11}q_5q_9 + 8q_{11}q_5 - 16q_{11}q_6q_7 + 8q_{11}q_6 - 16q_{11}q_7q_8 - 16q_{11}q_7q_9 + 24q_{11}q_7 + 24q_{11}q_8 + 24q_{11}q_9 - 36q_{11} + 32q_{12}q_{13}q_4q_5 - 16q_{12}q_{13}q_4 - 16q_{12}q_{13}q_5 + 32q_{12}q_{13}q_8q_9 - 16q_{12}q_{13}q_8 - 16q_{12}q_{13}q_9 + 24q_{12}q_{13} - 16q_{12}q_{14}q_2 + 32q_{12}q_{14}q_4q_6 - 16q_{12}q_{14}q_4 - 16q_{12}q_{14}q_6 - 16q_{12}q_{14}q_8 + 24q_{12}q_{14} - 16q_{12}q_{15}q_3 + 32q_{12}q_{15}q_4q_7 - 16q_{12}q_{15}q_4 - 16q_{12}q_{15}q_7 - 16q_{12}q_{15}q_8 + 24q_{12}q_{15} + 8q_{12}q_2 + 8q_{12}q_3 - 16q_{12}q_4q_5 - 16q_{12}q_4q_6 - 16q_{12}q_4q_7 + 24q_{12}q_4 + 8q_{12}q_5 + 8q_{12}q_6 + 8q_{12}q_7 - 16q_{12}q_8q_9 + 24q_{12}q_8 + 8q_{12}q_9 - 36q_{12} - 16q_{13}q_{14}q_2 + 32q_{13}q_{14}q_5q_6 - 16q_{13}q_{14}q_5 - 16q_{13}q_{14}q_6 - 16q_{13}q_{14}q_9 + 24q_{13}q_{14} - 16q_{13}q_{15}q_3 + 32q_{13}q_{15}q_5q_7 - 16q_{13}q_{15}q_5 - 16q_{13}q_{15}q_7 - 16q_{13}q_{15}q_9 + 24q_{13}q_{15} + 8q_{13}q_2 + 8q_{13}q_3 - 16q_{13}q_4q_5 + 8q_{13}q_4 - 16q_{13}q_5q_6 - 16q_{13}q_5q_7 + 24q_{13}q_5 + 8q_{13}q_6 + 8q_{13}q_7 - 16q_{13}q_8q_9 + 8q_{13}q_8 + 24q_{13}q_9 - 36q_{13} + 32q_{14}q_{15}q_2q_3 - 16q_{14}q_{15}q_2 - 16q_{14}q_{15}q_3 + 32q_{14}q_{15}q_6q_7 - 16q_{14}q_{15}q_6 - 16q_{14}q_{15}q_7 + 24q_{14}q_{15} - 16q_{14}q_2q_3 + 24q_{14}q_2 + 8q_{14}q_3 - 16q_{14}q_4q_6 + 8q_{14}q_4 - 16q_{14}q_5q_6 + 8q_{14}q_5 - 16q_{14}q_6q_7 + 24q_{14}q_6 + 8q_{14}q_7 + 8q_{14}q_8 + 8q_{14}q_9 - 36q_{14} - 16q_{15}q_2q_3 + 8q_{15}q_2 + 24q_{15}q_3 - 16q_{15}q_4q_7 + 8q_{15}q_4 - 16q_{15}q_5q_7 + 8q_{15}q_5 - 16q_{15}q_6q_7 + 8q_{15}q_6 + 24q_{15}q_7 + 8q_{15}q_8 + 8q_{15}q_9 - 36q_{15} + 32q_2q_3q_6q_7 - 16q_2q_3q_6 - 16q_2q_3q_7 + 24q_2q_3 - 16q_2q_4q_6 + 8q_2q_4 - 16q_2q_5q_6 + 8q_2q_5 - 16q_2q_6q_7 + 24q_2q_6 + 8q_2q_7 + 8q_2q_8 + 8q_2q_9 - 36q_2 - 16q_3q_4q_7 + 8q_3q_4 - 16q_3q_5q_7 + 8q_3q_5 - 16q_3q_6q_7 + 8q_3q_6 + 24q_3q_7 + 8q_3q_8 + 8q_3q_9 - 36q_3 + 32q_4q_5q_8q_9 - 16q_4q_5q_8 - 16q_4q_5q_9 + 24q_4q_5 - 16q_4q_6q_8 + 24q_4q_6 - 16q_4q_7q_8 + 24q_4q_7 - 16q_4q_8q_9 + 24q_4q_8 + 8q_4q_9 - 36q_4 - 16q_5q_6q_9 + 24q_5q_6 - 16q_5q_7q_9 + 24q_5q_7 - 16q_5q_8q_9 + 8q_5q_8 + 24q_5q_9 - 36q_5 + 24q_6q_7 + 8q_6q_8 + 8q_6q_9 - 36q_6 + 8q_7q_8 + 8q_7q_9 - 36q_7 + 24q_8q_9 - 36q_8 - 36q_9 + 96$



\subsection*{A complete expression of Eq.(\ref{E2_qi4})}

$ E_2(q_i) = 24q_{0}q_{1} + 8q_{0}q_{10} + 8q_{0}q_{11} + 64q_{0}q_{12} + 8q_{0}q_{13} + 8q_{0}q_{14} + 8q_{0}q_{15} - 128q_{0}q_{16} - 16q_{0}q_{17} - 16q_{0}q_{18} - 16q_{0}q_{19} + 24q_{0}q_{2} - 128q_{0}q_{20} - 16q_{0}q_{21} - 16q_{0}q_{22} - 16q_{0}q_{23} - 128q_{0}q_{24} - 16q_{0}q_{25} - 16q_{0}q_{26} - 16q_{0}q_{27} + 24q_{0}q_{3} + 64q_{0}q_{4} + 8q_{0}q_{5} + 8q_{0}q_{6} + 8q_{0}q_{7} + 64q_{0}q_{8} + 8q_{0}q_{9} - 36q_{0} + 8q_{1}q_{10} + 8q_{1}q_{11} + 8q_{1}q_{12} + 64q_{1}q_{13} + 8q_{1}q_{14} + 8q_{1}q_{15} - 16q_{1}q_{16} - 128q_{1}q_{17} - 16q_{1}q_{18} - 16q_{1}q_{19} + 24q_{1}q_{2} - 16q_{1}q_{20} - 128q_{1}q_{21} - 16q_{1}q_{22} - 16q_{1}q_{23} - 16q_{1}q_{24} - 128q_{1}q_{25} - 16q_{1}q_{26} - 16q_{1}q_{27} + 24q_{1}q_{3} + 8q_{1}q_{4} + 64q_{1}q_{5} + 8q_{1}q_{6} + 8q_{1}q_{7} + 8q_{1}q_{8} + 64q_{1}q_{9} - 36q_{1} + 24q_{10}q_{11} + 8q_{10}q_{12} + 8q_{10}q_{13} + 64q_{10}q_{14} + 8q_{10}q_{15} + 64q_{10}q_{2} - 16q_{10}q_{20} - 16q_{10}q_{21} - 128q_{10}q_{22} - 16q_{10}q_{23} - 16q_{10}q_{28} - 16q_{10}q_{29} + 8q_{10}q_{3} - 128q_{10}q_{30} - 16q_{10}q_{31} - 16q_{10}q_{36} - 16q_{10}q_{37} - 128q_{10}q_{38} - 16q_{10}q_{39} + 8q_{10}q_{4} + 8q_{10}q_{5} + 64q_{10}q_{6} + 8q_{10}q_{7} + 24q_{10}q_{8} + 24q_{10}q_{9} - 36q_{10} + 8q_{11}q_{12} + 8q_{11}q_{13} + 8q_{11}q_{14} + 64q_{11}q_{15} + 8q_{11}q_{2} - 16q_{11}q_{20} - 16q_{11}q_{21} - 16q_{11}q_{22} - 128q_{11}q_{23} - 16q_{11}q_{28} - 16q_{11}q_{29} + 64q_{11}q_{3} - 16q_{11}q_{30} - 128q_{11}q_{31} - 16q_{11}q_{36} - 16q_{11}q_{37} - 16q_{11}q_{38} - 128q_{11}q_{39} + 8q_{11}q_{4} + 8q_{11}q_{5} + 8q_{11}q_{6} + 64q_{11}q_{7} + 24q_{11}q_{8} + 24q_{11}q_{9} - 36q_{11} + 24q_{12}q_{13} + 24q_{12}q_{14} + 24q_{12}q_{15} + 8q_{12}q_{2} - 128q_{12}q_{24} - 16q_{12}q_{25} - 16q_{12}q_{26} - 16q_{12}q_{27} + 8q_{12}q_{3} - 128q_{12}q_{32} - 16q_{12}q_{33} - 16q_{12}q_{34} - 16q_{12}q_{35} - 128q_{12}q_{36} - 16q_{12}q_{37} - 16q_{12}q_{38} - 16q_{12}q_{39} + 64q_{12}q_{4} + 8q_{12}q_{5} + 8q_{12}q_{6} + 8q_{12}q_{7} + 64q_{12}q_{8} + 8q_{12}q_{9} - 36q_{12} + 24q_{13}q_{14} + 24q_{13}q_{15} + 8q_{13}q_{2} - 16q_{13}q_{24} - 128q_{13}q_{25} - 16q_{13}q_{26} - 16q_{13}q_{27} + 8q_{13}q_{3} - 16q_{13}q_{32} - 128q_{13}q_{33} - 16q_{13}q_{34} - 16q_{13}q_{35} - 16q_{13}q_{36} - 128q_{13}q_{37} - 16q_{13}q_{38} - 16q_{13}q_{39} + 8q_{13}q_{4} + 64q_{13}q_{5} + 8q_{13}q_{6} + 8q_{13}q_{7} + 8q_{13}q_{8} + 64q_{13}q_{9} - 36q_{13} + 24q_{14}q_{15} + 64q_{14}q_{2} - 16q_{14}q_{24} - 16q_{14}q_{25} - 128q_{14}q_{26} - 16q_{14}q_{27} + 8q_{14}q_{3} - 16q_{14}q_{32} - 16q_{14}q_{33} - 128q_{14}q_{34} - 16q_{14}q_{35} - 16q_{14}q_{36} - 16q_{14}q_{37} - 128q_{14}q_{38} - 16q_{14}q_{39} + 8q_{14}q_{4} + 8q_{14}q_{5} + 64q_{14}q_{6} + 8q_{14}q_{7} + 8q_{14}q_{8} + 8q_{14}q_{9} - 36q_{14} + 8q_{15}q_{2} - 16q_{15}q_{24} - 16q_{15}q_{25} - 16q_{15}q_{26} - 128q_{15}q_{27} + 64q_{15}q_{3} - 16q_{15}q_{32} - 16q_{15}q_{33} - 16q_{15}q_{34} - 128q_{15}q_{35} - 16q_{15}q_{36} - 16q_{15}q_{37} - 16q_{15}q_{38} - 128q_{15}q_{39} + 8q_{15}q_{4} + 8q_{15}q_{5} + 8q_{15}q_{6} + 64q_{15}q_{7} + 8q_{15}q_{8} + 8q_{15}q_{9} - 36q_{15} + 32q_{16}q_{17} + 32q_{16}q_{18} + 32q_{16}q_{19} - 16q_{16}q_{2} - 16q_{16}q_{3} - 128q_{16}q_{4} - 16q_{16}q_{5} - 16q_{16}q_{6} - 16q_{16}q_{7} + 216q_{16} + 32q_{17}q_{18} + 32q_{17}q_{19} - 16q_{17}q_{2} - 16q_{17}q_{3} - 16q_{17}q_{4} - 128q_{17}q_{5} - 16q_{17}q_{6} - 16q_{17}q_{7} + 216q_{17} + 32q_{18}q_{19} - 128q_{18}q_{2} - 16q_{18}q_{3} - 16q_{18}q_{4} - 16q_{18}q_{5} - 128q_{18}q_{6} - 16q_{18}q_{7} + 216q_{18} - 16q_{19}q_{2} - 128q_{19}q_{3} - 16q_{19}q_{4} - 16q_{19}q_{5} - 16q_{19}q_{6} - 128q_{19}q_{7} + 216q_{19} - 16q_{2}q_{20} - 16q_{2}q_{21} - 128q_{2}q_{22} - 16q_{2}q_{23} - 16q_{2}q_{24} - 16q_{2}q_{25} - 128q_{2}q_{26} - 16q_{2}q_{27} + 24q_{2}q_{3} + 8q_{2}q_{4} + 8q_{2}q_{5} + 64q_{2}q_{6} + 8q_{2}q_{7} + 8q_{2}q_{8} + 8q_{2}q_{9} - 36q_{2} + 32q_{20}q_{21} + 32q_{20}q_{22} + 32q_{20}q_{23} - 16q_{20}q_{3} - 128q_{20}q_{8} - 16q_{20}q_{9} + 216q_{20} + 32q_{21}q_{22} + 32q_{21}q_{23} - 16q_{21}q_{3} - 16q_{21}q_{8} - 128q_{21}q_{9} + 216q_{21} + 32q_{22}q_{23} - 16q_{22}q_{3} - 16q_{22}q_{8} - 16q_{22}q_{9} + 216q_{22} - 128q_{23}q_{3} - 16q_{23}q_{8} - 16q_{23}q_{9} + 216q_{23} + 32q_{24}q_{25} + 32q_{24}q_{26} + 32q_{24}q_{27} - 16q_{24}q_{3} + 216q_{24} + 32q_{25}q_{26} + 32q_{25}q_{27} - 16q_{25}q_{3} + 216q_{25} + 32q_{26}q_{27} - 16q_{26}q_{3} + 216q_{26} - 128q_{27}q_{3} + 216q_{27} + 32q_{28}q_{29} + 32q_{28}q_{30} + 32q_{28}q_{31} - 128q_{28}q_{4} - 16q_{28}q_{5} - 16q_{28}q_{6} - 16q_{28}q_{7} - 128q_{28}q_{8} - 16q_{28}q_{9} + 216q_{28} + 32q_{29}q_{30} + 32q_{29}q_{31} - 16q_{29}q_{4} - 128q_{29}q_{5} - 16q_{29}q_{6} - 16q_{29}q_{7} - 16q_{29}q_{8} - 128q_{29}q_{9} + 216q_{29} + 8q_{3}q_{4} + 8q_{3}q_{5} + 8q_{3}q_{6} + 64q_{3}q_{7} + 8q_{3}q_{8} + 8q_{3}q_{9} - 36q_{3} + 32q_{30}q_{31} - 16q_{30}q_{4} - 16q_{30}q_{5} - 128q_{30}q_{6} - 16q_{30}q_{7} - 16q_{30}q_{8} - 16q_{30}q_{9} + 216q_{30} - 16q_{31}q_{4} - 16q_{31}q_{5} - 16q_{31}q_{6} - 128q_{31}q_{7} - 16q_{31}q_{8} - 16q_{31}q_{9} + 216q_{31} + 32q_{32}q_{33} + 32q_{32}q_{34} + 32q_{32}q_{35} - 128q_{32}q_{4} - 16q_{32}q_{5} - 16q_{32}q_{6} - 16q_{32}q_{7} + 216q_{32} + 32q_{33}q_{34} + 32q_{33}q_{35} - 16q_{33}q_{4} - 128q_{33}q_{5} - 16q_{33}q_{6} - 16q_{33}q_{7} + 216q_{33} + 32q_{34}q_{35} - 16q_{34}q_{4} - 16q_{34}q_{5} - 128q_{34}q_{6} - 16q_{34}q_{7} + 216q_{34} - 16q_{35}q_{4} - 16q_{35}q_{5} - 16q_{35}q_{6} - 128q_{35}q_{7} + 216q_{35} + 32q_{36}q_{37} + 32q_{36}q_{38} + 32q_{36}q_{39} - 128q_{36}q_{8} - 16q_{36}q_{9} + 216q_{36} + 32q_{37}q_{38} + 32q_{37}q_{39} - 16q_{37}q_{8} - 128q_{37}q_{9} + 216q_{37} + 32q_{38}q_{39} - 16q_{38}q_{8} - 16q_{38}q_{9} + 216q_{38} - 16q_{39}q_{8} - 16q_{39}q_{9} + 216q_{39} + 24q_{4}q_{5} + 24q_{4}q_{6} + 24q_{4}q_{7} + 64q_{4}q_{8} + 8q_{4}q_{9} - 36q_{4} + 24q_{5}q_{6} + 24q_{5}q_{7} + 8q_{5}q_{8} + 64q_{5}q_{9} - 36q_{5} + 24q_{6}q_{7} + 8q_{6}q_{8} + 8q_{6}q_{9} - 36q_{6} + 8q_{7}q_{8} + 8q_{7}q_{9} - 36q_{7} + 24q_{8}q_{9} - 36q_{8} - 36q_{9} + 96$



\subsection*{A complete expression of Eq.(\ref{E2_si4})}

$ E_2(s_i)= 6s_{0}s_{1} + 2s_{0}s_{10} + 2s_{0}s_{11} + 16s_{0}s_{12} + 2s_{0}s_{13} + 2s_{0}s_{14} + 2s_{0}s_{15} - 32s_{0}s_{16} - 4s_{0}s_{17} - 4s_{0}s_{18} - 4s_{0}s_{19} + 6s_{0}s_{2} - 32s_{0}s_{20} - 4s_{0}s_{21} - 4s_{0}s_{22} - 4s_{0}s_{23} - 32s_{0}s_{24} - 4s_{0}s_{25} - 4s_{0}s_{26} - 4s_{0}s_{27} + 6s_{0}s_{3} + 16s_{0}s_{4} + 2s_{0}s_{5} + 2s_{0}s_{6} + 2s_{0}s_{7} + 16s_{0}s_{8} + 2s_{0}s_{9} + 66s_{0} + 2s_{1}s_{10} + 2s_{1}s_{11} + 2s_{1}s_{12} + 16s_{1}s_{13} + 2s_{1}s_{14} + 2s_{1}s_{15} - 4s_{1}s_{16} - 32s_{1}s_{17} - 4s_{1}s_{18} - 4s_{1}s_{19} + 6s_{1}s_{2} - 4s_{1}s_{20} - 32s_{1}s_{21} - 4s_{1}s_{22} - 4s_{1}s_{23} - 4s_{1}s_{24} - 32s_{1}s_{25} - 4s_{1}s_{26} - 4s_{1}s_{27} + 6s_{1}s_{3} + 2s_{1}s_{4} + 16s_{1}s_{5} + 2s_{1}s_{6} + 2s_{1}s_{7} + 2s_{1}s_{8} + 16s_{1}s_{9} + 66s_{1} + 6s_{10}s_{11} + 2s_{10}s_{12} + 2s_{10}s_{13} + 16s_{10}s_{14} + 2s_{10}s_{15} + 16s_{10}s_{2} - 4s_{10}s_{20} - 4s_{10}s_{21} - 32s_{10}s_{22} - 4s_{10}s_{23} - 4s_{10}s_{28} - 4s_{10}s_{29} + 2s_{10}s_{3} - 32s_{10}s_{30} - 4s_{10}s_{31} - 4s_{10}s_{36} - 4s_{10}s_{37} - 32s_{10}s_{38} - 4s_{10}s_{39} + 2s_{10}s_{4} + 2s_{10}s_{5} + 16s_{10}s_{6} + 2s_{10}s_{7} + 6s_{10}s_{8} + 6s_{10}s_{9} + 66s_{10} + 2s_{11}s_{12} + 2s_{11}s_{13} + 2s_{11}s_{14} + 16s_{11}s_{15} + 2s_{11}s_{2} - 4s_{11}s_{20} - 4s_{11}s_{21} - 4s_{11}s_{22} - 32s_{11}s_{23} - 4s_{11}s_{28} - 4s_{11}s_{29} + 16s_{11}s_{3} - 4s_{11}s_{30} - 32s_{11}s_{31} - 4s_{11}s_{36} - 4s_{11}s_{37} - 4s_{11}s_{38} - 32s_{11}s_{39} + 2s_{11}s_{4} + 2s_{11}s_{5} + 2s_{11}s_{6} + 16s_{11}s_{7} + 6s_{11}s_{8} + 6s_{11}s_{9} + 66s_{11} + 6s_{12}s_{13} + 6s_{12}s_{14} + 6s_{12}s_{15} + 2s_{12}s_{2} - 32s_{12}s_{24} - 4s_{12}s_{25} - 4s_{12}s_{26} - 4s_{12}s_{27} + 2s_{12}s_{3} - 32s_{12}s_{32} - 4s_{12}s_{33} - 4s_{12}s_{34} - 4s_{12}s_{35} - 32s_{12}s_{36} - 4s_{12}s_{37} - 4s_{12}s_{38} - 4s_{12}s_{39} + 16s_{12}s_{4} + 2s_{12}s_{5} + 2s_{12}s_{6} + 2s_{12}s_{7} + 16s_{12}s_{8} + 2s_{12}s_{9} + 66s_{12} + 6s_{13}s_{14} + 6s_{13}s_{15} + 2s_{13}s_{2} - 4s_{13}s_{24} - 32s_{13}s_{25} - 4s_{13}s_{26} - 4s_{13}s_{27} + 2s_{13}s_{3} - 4s_{13}s_{32} - 32s_{13}s_{33} - 4s_{13}s_{34} - 4s_{13}s_{35} - 4s_{13}s_{36} - 32s_{13}s_{37} - 4s_{13}s_{38} - 4s_{13}s_{39} + 2s_{13}s_{4} + 16s_{13}s_{5} + 2s_{13}s_{6} + 2s_{13}s_{7} + 2s_{13}s_{8} + 16s_{13}s_{9} + 66s_{13} + 6s_{14}s_{15} + 16s_{14}s_{2} - 4s_{14}s_{24} - 4s_{14}s_{25} - 32s_{14}s_{26} - 4s_{14}s_{27} + 2s_{14}s_{3} - 4s_{14}s_{32} - 4s_{14}s_{33} - 32s_{14}s_{34} - 4s_{14}s_{35} - 4s_{14}s_{36} - 4s_{14}s_{37} - 32s_{14}s_{38} - 4s_{14}s_{39} + 2s_{14}s_{4} + 2s_{14}s_{5} + 16s_{14}s_{6} + 2s_{14}s_{7} + 2s_{14}s_{8} + 2s_{14}s_{9} + 66s_{14} + 2s_{15}s_{2} - 4s_{15}s_{24} - 4s_{15}s_{25} - 4s_{15}s_{26} - 32s_{15}s_{27} + 16s_{15}s_{3} - 4s_{15}s_{32} - 4s_{15}s_{33} - 4s_{15}s_{34} - 32s_{15}s_{35} - 4s_{15}s_{36} - 4s_{15}s_{37} - 4s_{15}s_{38} - 32s_{15}s_{39} + 2s_{15}s_{4} + 2s_{15}s_{5} + 2s_{15}s_{6} + 16s_{15}s_{7} + 2s_{15}s_{8} + 2s_{15}s_{9} + 66s_{15} + 8s_{16}s_{17} + 8s_{16}s_{18} + 8s_{16}s_{19} - 4s_{16}s_{2} - 4s_{16}s_{3} - 32s_{16}s_{4} - 4s_{16}s_{5} - 4s_{16}s_{6} - 4s_{16}s_{7} - 44s_{16} + 8s_{17}s_{18} + 8s_{17}s_{19} - 4s_{17}s_{2} - 4s_{17}s_{3} - 4s_{17}s_{4} - 32s_{17}s_{5} - 4s_{17}s_{6} - 4s_{17}s_{7} - 44s_{17} + 8s_{18}s_{19} - 32s_{18}s_{2} - 4s_{18}s_{3} - 4s_{18}s_{4} - 4s_{18}s_{5} - 32s_{18}s_{6} - 4s_{18}s_{7} - 44s_{18} - 4s_{19}s_{2} - 32s_{19}s_{3} - 4s_{19}s_{4} - 4s_{19}s_{5} - 4s_{19}s_{6} - 32s_{19}s_{7} - 44s_{19} - 4s_{2}s_{20} - 4s_{2}s_{21} - 32s_{2}s_{22} - 4s_{2}s_{23} - 4s_{2}s_{24} - 4s_{2}s_{25} - 32s_{2}s_{26} - 4s_{2}s_{27} + 6s_{2}s_{3} + 2s_{2}s_{4} + 2s_{2}s_{5} + 16s_{2}s_{6} + 2s_{2}s_{7} + 2s_{2}s_{8} + 2s_{2}s_{9} + 66s_{2} + 8s_{20}s_{21} + 8s_{20}s_{22} + 8s_{20}s_{23} - 4s_{20}s_{3} - 32s_{20}s_{8} - 4s_{20}s_{9} - 44s_{20} + 8s_{21}s_{22} + 8s_{21}s_{23} - 4s_{21}s_{3} - 4s_{21}s_{8} - 32s_{21}s_{9} - 44s_{21} + 8s_{22}s_{23} - 4s_{22}s_{3} - 4s_{22}s_{8} - 4s_{22}s_{9} - 44s_{22} - 32s_{23}s_{3} - 4s_{23}s_{8} - 4s_{23}s_{9} - 44s_{23} + 8s_{24}s_{25} + 8s_{24}s_{26} + 8s_{24}s_{27} - 4s_{24}s_{3} - 44s_{24} + 8s_{25}s_{26} + 8s_{25}s_{27} - 4s_{25}s_{3} - 44s_{25} + 8s_{26}s_{27} - 4s_{26}s_{3} - 44s_{26} - 32s_{27}s_{3} - 44s_{27} + 8s_{28}s_{29} + 8s_{28}s_{30} + 8s_{28}s_{31} - 32s_{28}s_{4} - 4s_{28}s_{5} - 4s_{28}s_{6} - 4s_{28}s_{7} - 32s_{28}s_{8} - 4s_{28}s_{9} - 44s_{28} + 8s_{29}s_{30} + 8s_{29}s_{31} - 4s_{29}s_{4} - 32s_{29}s_{5} - 4s_{29}s_{6} - 4s_{29}s_{7} - 4s_{29}s_{8} - 32s_{29}s_{9} - 44s_{29} + 2s_{3}s_{4} + 2s_{3}s_{5} + 2s_{3}s_{6} + 16s_{3}s_{7} + 2s_{3}s_{8} + 2s_{3}s_{9} + 66s_{3} + 8s_{30}s_{31} - 4s_{30}s_{4} - 4s_{30}s_{5} - 32s_{30}s_{6} - 4s_{30}s_{7} - 4s_{30}s_{8} - 4s_{30}s_{9} - 44s_{30} - 4s_{31}s_{4} - 4s_{31}s_{5} - 4s_{31}s_{6} - 32s_{31}s_{7} - 4s_{31}s_{8} - 4s_{31}s_{9} - 44s_{31} + 8s_{32}s_{33} + 8s_{32}s_{34} + 8s_{32}s_{35} - 32s_{32}s_{4} - 4s_{32}s_{5} - 4s_{32}s_{6} - 4s_{32}s_{7} - 44s_{32} + 8s_{33}s_{34} + 8s_{33}s_{35} - 4s_{33}s_{4} - 32s_{33}s_{5} - 4s_{33}s_{6} - 4s_{33}s_{7} - 44s_{33} + 8s_{34}s_{35} - 4s_{34}s_{4} - 4s_{34}s_{5} - 32s_{34}s_{6} - 4s_{34}s_{7} - 44s_{34} - 4s_{35}s_{4} - 4s_{35}s_{5} - 4s_{35}s_{6} - 32s_{35}s_{7} - 44s_{35} + 8s_{36}s_{37} + 8s_{36}s_{38} + 8s_{36}s_{39} - 32s_{36}s_{8} - 4s_{36}s_{9} - 44s_{36} + 8s_{37}s_{38} + 8s_{37}s_{39} - 4s_{37}s_{8} - 32s_{37}s_{9} - 44s_{37} + 8s_{38}s_{39} - 4s_{38}s_{8} - 4s_{38}s_{9} - 44s_{38} - 4s_{39}s_{8} - 4s_{39}s_{9} - 44s_{39} + 6s_{4}s_{5} + 6s_{4}s_{6} + 6s_{4}s_{7} + 16s_{4}s_{8} + 2s_{4}s_{9} + 66s_{4} + 6s_{5}s_{6} + 6s_{5}s_{7} + 2s_{5}s_{8} + 16s_{5}s_{9} + 66s_{5} + 6s_{6}s_{7} + 2s_{6}s_{8} + 2s_{6}s_{9} + 66s_{6} + 2s_{7}s_{8} + 2s_{7}s_{9} + 66s_{7} + 6s_{8}s_{9} + 66s_{8} + 66s_{9} + 1,248 $



\subsection*{A complete expression of the Eq.(\ref{H2_SUB_N3_M4})}

$ \hat{H}_2\left(\hat{\sigma}_i^z\right) = 4\hat{\sigma}^z_{0}\hat{\sigma}^z_{1} + 2\hat{\sigma}^z_{0}\hat{\sigma}^z_{10} + 2\hat{\sigma}^z_{0}\hat{\sigma}^z_{11} - 40\hat{\sigma}^z_{0}\hat{\sigma}^z_{12} - 4\hat{\sigma}^z_{0}\hat{\sigma}^z_{13} - 4\hat{\sigma}^z_{0}\hat{\sigma}^z_{14} - 4\hat{\sigma}^z_{0}\hat{\sigma}^z_{15} - 40\hat{\sigma}^z_{0}\hat{\sigma}^z_{16} - 4\hat{\sigma}^z_{0}\hat{\sigma}^z_{17} - 4\hat{\sigma}^z_{0}\hat{\sigma}^z_{18} - 4\hat{\sigma}^z_{0}\hat{\sigma}^z_{19} + 4\hat{\sigma}^z_{0}\hat{\sigma}^z_{2} + 4\hat{\sigma}^z_{0}\hat{\sigma}^z_{3} + 20\hat{\sigma}^z_{0}\hat{\sigma}^z_{4} + 2\hat{\sigma}^z_{0}\hat{\sigma}^z_{5} + 2\hat{\sigma}^z_{0}\hat{\sigma}^z_{6} + 2\hat{\sigma}^z_{0}\hat{\sigma}^z_{7} + 20\hat{\sigma}^z_{0}\hat{\sigma}^z_{8} + 2\hat{\sigma}^z_{0}\hat{\sigma}^z_{9} + 52\hat{\sigma}^z_{0} + 2\hat{\sigma}^z_{1}\hat{\sigma}^z_{10} + 2\hat{\sigma}^z_{1}\hat{\sigma}^z_{11} - 4\hat{\sigma}^z_{1}\hat{\sigma}^z_{12} - 40\hat{\sigma}^z_{1}\hat{\sigma}^z_{13} - 4\hat{\sigma}^z_{1}\hat{\sigma}^z_{14} - 4\hat{\sigma}^z_{1}\hat{\sigma}^z_{15} - 4\hat{\sigma}^z_{1}\hat{\sigma}^z_{16} - 40\hat{\sigma}^z_{1}\hat{\sigma}^z_{17} - 4\hat{\sigma}^z_{1}\hat{\sigma}^z_{18} - 4\hat{\sigma}^z_{1}\hat{\sigma}^z_{19} + 4\hat{\sigma}^z_{1}\hat{\sigma}^z_{2} + 4\hat{\sigma}^z_{1}\hat{\sigma}^z_{3} + 2\hat{\sigma}^z_{1}\hat{\sigma}^z_{4} + 20\hat{\sigma}^z_{1}\hat{\sigma}^z_{5} + 2\hat{\sigma}^z_{1}\hat{\sigma}^z_{6} + 2\hat{\sigma}^z_{1}\hat{\sigma}^z_{7} + 2\hat{\sigma}^z_{1}\hat{\sigma}^z_{8} + 20\hat{\sigma}^z_{1}\hat{\sigma}^z_{9} + 52\hat{\sigma}^z_{1} + 4\hat{\sigma}^z_{10}\hat{\sigma}^z_{11} - 4\hat{\sigma}^z_{10}\hat{\sigma}^z_{16} - 4\hat{\sigma}^z_{10}\hat{\sigma}^z_{17} - 40\hat{\sigma}^z_{10}\hat{\sigma}^z_{18} - 4\hat{\sigma}^z_{10}\hat{\sigma}^z_{19} + 20\hat{\sigma}^z_{10}\hat{\sigma}^z_{2} - 4\hat{\sigma}^z_{10}\hat{\sigma}^z_{20} - 4\hat{\sigma}^z_{10}\hat{\sigma}^z_{21} - 40\hat{\sigma}^z_{10}\hat{\sigma}^z_{22} - 4\hat{\sigma}^z_{10}\hat{\sigma}^z_{23} + 2\hat{\sigma}^z_{10}\hat{\sigma}^z_{3} + 2\hat{\sigma}^z_{10}\hat{\sigma}^z_{4} + 2\hat{\sigma}^z_{10}\hat{\sigma}^z_{5} + 20\hat{\sigma}^z_{10}\hat{\sigma}^z_{6} + 2\hat{\sigma}^z_{10}\hat{\sigma}^z_{7} + 4\hat{\sigma}^z_{10}\hat{\sigma}^z_{8} + 4\hat{\sigma}^z_{10}\hat{\sigma}^z_{9} + 52\hat{\sigma}^z_{10} - 4\hat{\sigma}^z_{11}\hat{\sigma}^z_{16} - 4\hat{\sigma}^z_{11}\hat{\sigma}^z_{17} - 4\hat{\sigma}^z_{11}\hat{\sigma}^z_{18} - 40\hat{\sigma}^z_{11}\hat{\sigma}^z_{19} + 2\hat{\sigma}^z_{11}\hat{\sigma}^z_{2} - 4\hat{\sigma}^z_{11}\hat{\sigma}^z_{20} - 4\hat{\sigma}^z_{11}\hat{\sigma}^z_{21} - 4\hat{\sigma}^z_{11}\hat{\sigma}^z_{22} - 40\hat{\sigma}^z_{11}\hat{\sigma}^z_{23} + 20\hat{\sigma}^z_{11}\hat{\sigma}^z_{3} + 2\hat{\sigma}^z_{11}\hat{\sigma}^z_{4} + 2\hat{\sigma}^z_{11}\hat{\sigma}^z_{5} + 2\hat{\sigma}^z_{11}\hat{\sigma}^z_{6} + 20\hat{\sigma}^z_{11}\hat{\sigma}^z_{7} + 4\hat{\sigma}^z_{11}\hat{\sigma}^z_{8} + 4\hat{\sigma}^z_{11}\hat{\sigma}^z_{9} + 52\hat{\sigma}^z_{11} + 8\hat{\sigma}^z_{12}\hat{\sigma}^z_{13} + 8\hat{\sigma}^z_{12}\hat{\sigma}^z_{14} + 8\hat{\sigma}^z_{12}\hat{\sigma}^z_{15} - 4\hat{\sigma}^z_{12}\hat{\sigma}^z_{2} - 4\hat{\sigma}^z_{12}\hat{\sigma}^z_{3} - 40\hat{\sigma}^z_{12}\hat{\sigma}^z_{4} - 4\hat{\sigma}^z_{12}\hat{\sigma}^z_{5} - 4\hat{\sigma}^z_{12}\hat{\sigma}^z_{6} - 4\hat{\sigma}^z_{12}\hat{\sigma}^z_{7} - 52\hat{\sigma}^z_{12} + 8\hat{\sigma}^z_{13}\hat{\sigma}^z_{14} + 8\hat{\sigma}^z_{13}\hat{\sigma}^z_{15} - 4\hat{\sigma}^z_{13}\hat{\sigma}^z_{2} - 4\hat{\sigma}^z_{13}\hat{\sigma}^z_{3} - 4\hat{\sigma}^z_{13}\hat{\sigma}^z_{4} - 40\hat{\sigma}^z_{13}\hat{\sigma}^z_{5} - 4\hat{\sigma}^z_{13}\hat{\sigma}^z_{6} - 4\hat{\sigma}^z_{13}\hat{\sigma}^z_{7} - 52\hat{\sigma}^z_{13} + 8\hat{\sigma}^z_{14}\hat{\sigma}^z_{15} - 40\hat{\sigma}^z_{14}\hat{\sigma}^z_{2} - 4\hat{\sigma}^z_{14}\hat{\sigma}^z_{3} - 4\hat{\sigma}^z_{14}\hat{\sigma}^z_{4} - 4\hat{\sigma}^z_{14}\hat{\sigma}^z_{5} - 40\hat{\sigma}^z_{14}\hat{\sigma}^z_{6} - 4\hat{\sigma}^z_{14}\hat{\sigma}^z_{7} - 52\hat{\sigma}^z_{14} - 4\hat{\sigma}^z_{15}\hat{\sigma}^z_{2} - 40\hat{\sigma}^z_{15}\hat{\sigma}^z_{3} - 4\hat{\sigma}^z_{15}\hat{\sigma}^z_{4} - 4\hat{\sigma}^z_{15}\hat{\sigma}^z_{5} - 4\hat{\sigma}^z_{15}\hat{\sigma}^z_{6} - 40\hat{\sigma}^z_{15}\hat{\sigma}^z_{7} - 52\hat{\sigma}^z_{15} + 8\hat{\sigma}^z_{16}\hat{\sigma}^z_{17} + 8\hat{\sigma}^z_{16}\hat{\sigma}^z_{18} + 8\hat{\sigma}^z_{16}\hat{\sigma}^z_{19} - 4\hat{\sigma}^z_{16}\hat{\sigma}^z_{2} - 4\hat{\sigma}^z_{16}\hat{\sigma}^z_{3} - 40\hat{\sigma}^z_{16}\hat{\sigma}^z_{8} - 4\hat{\sigma}^z_{16}\hat{\sigma}^z_{9} - 52\hat{\sigma}^z_{16} + 8\hat{\sigma}^z_{17}\hat{\sigma}^z_{18} + 8\hat{\sigma}^z_{17}\hat{\sigma}^z_{19} - 4\hat{\sigma}^z_{17}\hat{\sigma}^z_{2} - 4\hat{\sigma}^z_{17}\hat{\sigma}^z_{3} - 4\hat{\sigma}^z_{17}\hat{\sigma}^z_{8} - 40\hat{\sigma}^z_{17}\hat{\sigma}^z_{9} - 52\hat{\sigma}^z_{17} + 8\hat{\sigma}^z_{18}\hat{\sigma}^z_{19} - 40\hat{\sigma}^z_{18}\hat{\sigma}^z_{2} - 4\hat{\sigma}^z_{18}\hat{\sigma}^z_{3} - 4\hat{\sigma}^z_{18}\hat{\sigma}^z_{8} - 4\hat{\sigma}^z_{18}\hat{\sigma}^z_{9} - 52\hat{\sigma}^z_{18} - 4\hat{\sigma}^z_{19}\hat{\sigma}^z_{2} - 40\hat{\sigma}^z_{19}\hat{\sigma}^z_{3} - 4\hat{\sigma}^z_{19}\hat{\sigma}^z_{8} - 4\hat{\sigma}^z_{19}\hat{\sigma}^z_{9} - 52\hat{\sigma}^z_{19} + 4\hat{\sigma}^z_{2}\hat{\sigma}^z_{3} + 2\hat{\sigma}^z_{2}\hat{\sigma}^z_{4} + 2\hat{\sigma}^z_{2}\hat{\sigma}^z_{5} + 20\hat{\sigma}^z_{2}\hat{\sigma}^z_{6} + 2\hat{\sigma}^z_{2}\hat{\sigma}^z_{7} + 2\hat{\sigma}^z_{2}\hat{\sigma}^z_{8} + 2\hat{\sigma}^z_{2}\hat{\sigma}^z_{9} + 52\hat{\sigma}^z_{2} + 8\hat{\sigma}^z_{20}\hat{\sigma}^z_{21} + 8\hat{\sigma}^z_{20}\hat{\sigma}^z_{22} + 8\hat{\sigma}^z_{20}\hat{\sigma}^z_{23} - 40\hat{\sigma}^z_{20}\hat{\sigma}^z_{4} - 4\hat{\sigma}^z_{20}\hat{\sigma}^z_{5} - 4\hat{\sigma}^z_{20}\hat{\sigma}^z_{6} - 4\hat{\sigma}^z_{20}\hat{\sigma}^z_{7} - 40\hat{\sigma}^z_{20}\hat{\sigma}^z_{8} - 4\hat{\sigma}^z_{20}\hat{\sigma}^z_{9} - 52\hat{\sigma}^z_{20} + 8\hat{\sigma}^z_{21}\hat{\sigma}^z_{22} + 8\hat{\sigma}^z_{21}\hat{\sigma}^z_{23} - 4\hat{\sigma}^z_{21}\hat{\sigma}^z_{4} - 40\hat{\sigma}^z_{21}\hat{\sigma}^z_{5} - 4\hat{\sigma}^z_{21}\hat{\sigma}^z_{6} - 4\hat{\sigma}^z_{21}\hat{\sigma}^z_{7} - 4\hat{\sigma}^z_{21}\hat{\sigma}^z_{8} - 40\hat{\sigma}^z_{21}\hat{\sigma}^z_{9} - 52\hat{\sigma}^z_{21} + 8\hat{\sigma}^z_{22}\hat{\sigma}^z_{23} - 4\hat{\sigma}^z_{22}\hat{\sigma}^z_{4} - 4\hat{\sigma}^z_{22}\hat{\sigma}^z_{5} - 40\hat{\sigma}^z_{22}\hat{\sigma}^z_{6} - 4\hat{\sigma}^z_{22}\hat{\sigma}^z_{7} - 4\hat{\sigma}^z_{22}\hat{\sigma}^z_{8} - 4\hat{\sigma}^z_{22}\hat{\sigma}^z_{9} - 52\hat{\sigma}^z_{22} - 4\hat{\sigma}^z_{23}\hat{\sigma}^z_{4} - 4\hat{\sigma}^z_{23}\hat{\sigma}^z_{5} - 4\hat{\sigma}^z_{23}\hat{\sigma}^z_{6} - 40\hat{\sigma}^z_{23}\hat{\sigma}^z_{7} - 4\hat{\sigma}^z_{23}\hat{\sigma}^z_{8} - 4\hat{\sigma}^z_{23}\hat{\sigma}^z_{9} - 52\hat{\sigma}^z_{23} + 2\hat{\sigma}^z_{3}\hat{\sigma}^z_{4} + 2\hat{\sigma}^z_{3}\hat{\sigma}^z_{5} + 2\hat{\sigma}^z_{3}\hat{\sigma}^z_{6} + 20\hat{\sigma}^z_{3}\hat{\sigma}^z_{7} + 2\hat{\sigma}^z_{3}\hat{\sigma}^z_{8} + 2\hat{\sigma}^z_{3}\hat{\sigma}^z_{9} + 52\hat{\sigma}^z_{3} + 4\hat{\sigma}^z_{4}\hat{\sigma}^z_{5} + 4\hat{\sigma}^z_{4}\hat{\sigma}^z_{6} + 4\hat{\sigma}^z_{4}\hat{\sigma}^z_{7} + 20\hat{\sigma}^z_{4}\hat{\sigma}^z_{8} + 2\hat{\sigma}^z_{4}\hat{\sigma}^z_{9} + 52\hat{\sigma}^z_{4} + 4\hat{\sigma}^z_{5}\hat{\sigma}^z_{6} + 4\hat{\sigma}^z_{5}\hat{\sigma}^z_{7} + 2\hat{\sigma}^z_{5}\hat{\sigma}^z_{8} + 20\hat{\sigma}^z_{5}\hat{\sigma}^z_{9} + 52\hat{\sigma}^z_{5} + 4\hat{\sigma}^z_{6}\hat{\sigma}^z_{7} + 2\hat{\sigma}^z_{6}\hat{\sigma}^z_{8} + 2\hat{\sigma}^z_{6}\hat{\sigma}^z_{9} + 52\hat{\sigma}^z_{6} + 2\hat{\sigma}^z_{7}\hat{\sigma}^z_{8} + 2\hat{\sigma}^z_{7}\hat{\sigma}^z_{9} + 52\hat{\sigma}^z_{7} + 4\hat{\sigma}^z_{8}\hat{\sigma}^z_{9} + 52\hat{\sigma}^z_{8} + 52\hat{\sigma}^z_{9} + 768 $


\subsection*{A complete expression of Eq.(\ref{H2_completion_M4})}

$ \hat{H}_2\left(\hat{\sigma}_i^z\right) = 2\hat{\sigma}^z_{0}\hat{\sigma}^z_{1} - 4\hat{\sigma}^z_{0}\hat{\sigma}^z_{10} - 4\hat{\sigma}^z_{0}\hat{\sigma}^z_{11} + 6\hat{\sigma}^z_{0}\hat{\sigma}^z_{2} + 2\hat{\sigma}^z_{0}\hat{\sigma}^z_{3} + 8\hat{\sigma}^z_{0}\hat{\sigma}^z_{4} + 2\hat{\sigma}^z_{0}\hat{\sigma}^z_{5} + 2\hat{\sigma}^z_{0}\hat{\sigma}^z_{6} + 2\hat{\sigma}^z_{0}\hat{\sigma}^z_{7} - 16\hat{\sigma}^z_{0}\hat{\sigma}^z_{8} - 4\hat{\sigma}^z_{0}\hat{\sigma}^z_{9} + 14\hat{\sigma}^z_{0} - 4\hat{\sigma}^z_{1}\hat{\sigma}^z_{10} - 4\hat{\sigma}^z_{1}\hat{\sigma}^z_{11} + 2\hat{\sigma}^z_{1}\hat{\sigma}^z_{2} + 6\hat{\sigma}^z_{1}\hat{\sigma}^z_{3} + 2\hat{\sigma}^z_{1}\hat{\sigma}^z_{4} + 8\hat{\sigma}^z_{1}\hat{\sigma}^z_{5} + 2\hat{\sigma}^z_{1}\hat{\sigma}^z_{6} + 2\hat{\sigma}^z_{1}\hat{\sigma}^z_{7} - 4\hat{\sigma}^z_{1}\hat{\sigma}^z_{8} - 16\hat{\sigma}^z_{1}\hat{\sigma}^z_{9} + 14\hat{\sigma}^z_{1} + 8\hat{\sigma}^z_{10}\hat{\sigma}^z_{11} - 16\hat{\sigma}^z_{10}\hat{\sigma}^z_{2} - 4\hat{\sigma}^z_{10}\hat{\sigma}^z_{3} - 4\hat{\sigma}^z_{10}\hat{\sigma}^z_{4} - 4\hat{\sigma}^z_{10}\hat{\sigma}^z_{5} - 16\hat{\sigma}^z_{10}\hat{\sigma}^z_{6} - 4\hat{\sigma}^z_{10}\hat{\sigma}^z_{7} + 8\hat{\sigma}^z_{10}\hat{\sigma}^z_{8} + 8\hat{\sigma}^z_{10}\hat{\sigma}^z_{9} - 28\hat{\sigma}^z_{10} - 4\hat{\sigma}^z_{11}\hat{\sigma}^z_{2} - 16\hat{\sigma}^z_{11}\hat{\sigma}^z_{3} - 4\hat{\sigma}^z_{11}\hat{\sigma}^z_{4} - 4\hat{\sigma}^z_{11}\hat{\sigma}^z_{5} - 4\hat{\sigma}^z_{11}\hat{\sigma}^z_{6} - 16\hat{\sigma}^z_{11}\hat{\sigma}^z_{7} + 8\hat{\sigma}^z_{11}\hat{\sigma}^z_{8} + 8\hat{\sigma}^z_{11}\hat{\sigma}^z_{9} - 28\hat{\sigma}^z_{11} + 2\hat{\sigma}^z_{2}\hat{\sigma}^z_{3} + 2\hat{\sigma}^z_{2}\hat{\sigma}^z_{4} + 2\hat{\sigma}^z_{2}\hat{\sigma}^z_{5} + 8\hat{\sigma}^z_{2}\hat{\sigma}^z_{6} + 2\hat{\sigma}^z_{2}\hat{\sigma}^z_{7} - 4\hat{\sigma}^z_{2}\hat{\sigma}^z_{8} - 4\hat{\sigma}^z_{2}\hat{\sigma}^z_{9} + 14\hat{\sigma}^z_{2} + 2\hat{\sigma}^z_{3}\hat{\sigma}^z_{4} + 2\hat{\sigma}^z_{3}\hat{\sigma}^z_{5} + 2\hat{\sigma}^z_{3}\hat{\sigma}^z_{6} + 8\hat{\sigma}^z_{3}\hat{\sigma}^z_{7} - 4\hat{\sigma}^z_{3}\hat{\sigma}^z_{8} - 4\hat{\sigma}^z_{3}\hat{\sigma}^z_{9} + 14\hat{\sigma}^z_{3} + 2\hat{\sigma}^z_{4}\hat{\sigma}^z_{5} + 6\hat{\sigma}^z_{4}\hat{\sigma}^z_{6} + 2\hat{\sigma}^z_{4}\hat{\sigma}^z_{7} - 16\hat{\sigma}^z_{4}\hat{\sigma}^z_{8} - 4\hat{\sigma}^z_{4}\hat{\sigma}^z_{9} + 14\hat{\sigma}^z_{4} + 2\hat{\sigma}^z_{5}\hat{\sigma}^z_{6} + 6\hat{\sigma}^z_{5}\hat{\sigma}^z_{7} - 4\hat{\sigma}^z_{5}\hat{\sigma}^z_{8} - 16\hat{\sigma}^z_{5}\hat{\sigma}^z_{9} + 14\hat{\sigma}^z_{5} + 2\hat{\sigma}^z_{6}\hat{\sigma}^z_{7} - 4\hat{\sigma}^z_{6}\hat{\sigma}^z_{8} - 4\hat{\sigma}^z_{6}\hat{\sigma}^z_{9} + 14\hat{\sigma}^z_{6} - 4\hat{\sigma}^z_{7}\hat{\sigma}^z_{8} - 4\hat{\sigma}^z_{7}\hat{\sigma}^z_{9} + 14\hat{\sigma}^z_{7} + 8\hat{\sigma}^z_{8}\hat{\sigma}^z_{9} - 28\hat{\sigma}^z_{8} - 28\hat{\sigma}^z_{9} + 128 $


\subsection*{A complete expression of Eq.(\ref{H2_SUB_N3_M12})}

$ \hat{H}_2\left(\hat{\sigma}_i^z\right) = 4\hat{\sigma}^z_{0}\hat{\sigma}^z_{1} + 4\hat{\sigma}^z_{0}\hat{\sigma}^z_{10} + 4\hat{\sigma}^z_{0}\hat{\sigma}^z_{11} + 180\hat{\sigma}^z_{0}\hat{\sigma}^z_{12} + 2\hat{\sigma}^z_{0}\hat{\sigma}^z_{13} + 2\hat{\sigma}^z_{0}\hat{\sigma}^z_{14} + 2\hat{\sigma}^z_{0}\hat{\sigma}^z_{15} + 2\hat{\sigma}^z_{0}\hat{\sigma}^z_{16} + 2\hat{\sigma}^z_{0}\hat{\sigma}^z_{17} + 2\hat{\sigma}^z_{0}\hat{\sigma}^z_{18} + 2\hat{\sigma}^z_{0}\hat{\sigma}^z_{19} + 4\hat{\sigma}^z_{0}\hat{\sigma}^z_{2} + 2\hat{\sigma}^z_{0}\hat{\sigma}^z_{20} + 2\hat{\sigma}^z_{0}\hat{\sigma}^z_{21} + 2\hat{\sigma}^z_{0}\hat{\sigma}^z_{22} + 2\hat{\sigma}^z_{0}\hat{\sigma}^z_{23} + 180\hat{\sigma}^z_{0}\hat{\sigma}^z_{24} + 2\hat{\sigma}^z_{0}\hat{\sigma}^z_{25} + 2\hat{\sigma}^z_{0}\hat{\sigma}^z_{26} + 2\hat{\sigma}^z_{0}\hat{\sigma}^z_{27} + 2\hat{\sigma}^z_{0}\hat{\sigma}^z_{28} + 2\hat{\sigma}^z_{0}\hat{\sigma}^z_{29} + 4\hat{\sigma}^z_{0}\hat{\sigma}^z_{3} + 2\hat{\sigma}^z_{0}\hat{\sigma}^z_{30} + 2\hat{\sigma}^z_{0}\hat{\sigma}^z_{31} + 2\hat{\sigma}^z_{0}\hat{\sigma}^z_{32} + 2\hat{\sigma}^z_{0}\hat{\sigma}^z_{33} + 2\hat{\sigma}^z_{0}\hat{\sigma}^z_{34} + 2\hat{\sigma}^z_{0}\hat{\sigma}^z_{35} - 360\hat{\sigma}^z_{0}\hat{\sigma}^z_{36} - 4\hat{\sigma}^z_{0}\hat{\sigma}^z_{37} - 4\hat{\sigma}^z_{0}\hat{\sigma}^z_{38} - 4\hat{\sigma}^z_{0}\hat{\sigma}^z_{39} + 4\hat{\sigma}^z_{0}\hat{\sigma}^z_{4} - 4\hat{\sigma}^z_{0}\hat{\sigma}^z_{40} - 4\hat{\sigma}^z_{0}\hat{\sigma}^z_{41} - 4\hat{\sigma}^z_{0}\hat{\sigma}^z_{42} - 4\hat{\sigma}^z_{0}\hat{\sigma}^z_{43} - 4\hat{\sigma}^z_{0}\hat{\sigma}^z_{44} - 4\hat{\sigma}^z_{0}\hat{\sigma}^z_{45} - 4\hat{\sigma}^z_{0}\hat{\sigma}^z_{46} - 4\hat{\sigma}^z_{0}\hat{\sigma}^z_{47} - 360\hat{\sigma}^z_{0}\hat{\sigma}^z_{48} - 4\hat{\sigma}^z_{0}\hat{\sigma}^z_{49} + 4\hat{\sigma}^z_{0}\hat{\sigma}^z_{5} - 4\hat{\sigma}^z_{0}\hat{\sigma}^z_{50} - 4\hat{\sigma}^z_{0}\hat{\sigma}^z_{51} - 4\hat{\sigma}^z_{0}\hat{\sigma}^z_{52} - 4\hat{\sigma}^z_{0}\hat{\sigma}^z_{53} - 4\hat{\sigma}^z_{0}\hat{\sigma}^z_{54} - 4\hat{\sigma}^z_{0}\hat{\sigma}^z_{55} - 4\hat{\sigma}^z_{0}\hat{\sigma}^z_{56} - 4\hat{\sigma}^z_{0}\hat{\sigma}^z_{57} - 4\hat{\sigma}^z_{0}\hat{\sigma}^z_{58} - 4\hat{\sigma}^z_{0}\hat{\sigma}^z_{59} + 4\hat{\sigma}^z_{0}\hat{\sigma}^z_{6} + 4\hat{\sigma}^z_{0}\hat{\sigma}^z_{7} + 4\hat{\sigma}^z_{0}\hat{\sigma}^z_{8} + 4\hat{\sigma}^z_{0}\hat{\sigma}^z_{9} + 404\hat{\sigma}^z_{0} + 4\hat{\sigma}^z_{1}\hat{\sigma}^z_{10} + 4\hat{\sigma}^z_{1}\hat{\sigma}^z_{11} + 2\hat{\sigma}^z_{1}\hat{\sigma}^z_{12} + 180\hat{\sigma}^z_{1}\hat{\sigma}^z_{13} + 2\hat{\sigma}^z_{1}\hat{\sigma}^z_{14} + 2\hat{\sigma}^z_{1}\hat{\sigma}^z_{15} + 2\hat{\sigma}^z_{1}\hat{\sigma}^z_{16} + 2\hat{\sigma}^z_{1}\hat{\sigma}^z_{17} + 2\hat{\sigma}^z_{1}\hat{\sigma}^z_{18} + 2\hat{\sigma}^z_{1}\hat{\sigma}^z_{19} + 4\hat{\sigma}^z_{1}\hat{\sigma}^z_{2} + 2\hat{\sigma}^z_{1}\hat{\sigma}^z_{20} + 2\hat{\sigma}^z_{1}\hat{\sigma}^z_{21} + 2\hat{\sigma}^z_{1}\hat{\sigma}^z_{22} + 2\hat{\sigma}^z_{1}\hat{\sigma}^z_{23} + 2\hat{\sigma}^z_{1}\hat{\sigma}^z_{24} + 180\hat{\sigma}^z_{1}\hat{\sigma}^z_{25} + 2\hat{\sigma}^z_{1}\hat{\sigma}^z_{26} + 2\hat{\sigma}^z_{1}\hat{\sigma}^z_{27} + 2\hat{\sigma}^z_{1}\hat{\sigma}^z_{28} + 2\hat{\sigma}^z_{1}\hat{\sigma}^z_{29} + 4\hat{\sigma}^z_{1}\hat{\sigma}^z_{3} + 2\hat{\sigma}^z_{1}\hat{\sigma}^z_{30} + 2\hat{\sigma}^z_{1}\hat{\sigma}^z_{31} + 2\hat{\sigma}^z_{1}\hat{\sigma}^z_{32} + 2\hat{\sigma}^z_{1}\hat{\sigma}^z_{33} + 2\hat{\sigma}^z_{1}\hat{\sigma}^z_{34} + 2\hat{\sigma}^z_{1}\hat{\sigma}^z_{35} - 4\hat{\sigma}^z_{1}\hat{\sigma}^z_{36} - 360\hat{\sigma}^z_{1}\hat{\sigma}^z_{37} - 4\hat{\sigma}^z_{1}\hat{\sigma}^z_{38} - 4\hat{\sigma}^z_{1}\hat{\sigma}^z_{39} + 4\hat{\sigma}^z_{1}\hat{\sigma}^z_{4} - 4\hat{\sigma}^z_{1}\hat{\sigma}^z_{40} - 4\hat{\sigma}^z_{1}\hat{\sigma}^z_{41} - 4\hat{\sigma}^z_{1}\hat{\sigma}^z_{42} - 4\hat{\sigma}^z_{1}\hat{\sigma}^z_{43} - 4\hat{\sigma}^z_{1}\hat{\sigma}^z_{44} - 4\hat{\sigma}^z_{1}\hat{\sigma}^z_{45} - 4\hat{\sigma}^z_{1}\hat{\sigma}^z_{46} - 4\hat{\sigma}^z_{1}\hat{\sigma}^z_{47} - 4\hat{\sigma}^z_{1}\hat{\sigma}^z_{48} - 360\hat{\sigma}^z_{1}\hat{\sigma}^z_{49} + 4\hat{\sigma}^z_{1}\hat{\sigma}^z_{5} - 4\hat{\sigma}^z_{1}\hat{\sigma}^z_{50} - 4\hat{\sigma}^z_{1}\hat{\sigma}^z_{51} - 4\hat{\sigma}^z_{1}\hat{\sigma}^z_{52} - 4\hat{\sigma}^z_{1}\hat{\sigma}^z_{53} - 4\hat{\sigma}^z_{1}\hat{\sigma}^z_{54} - 4\hat{\sigma}^z_{1}\hat{\sigma}^z_{55} - 4\hat{\sigma}^z_{1}\hat{\sigma}^z_{56} - 4\hat{\sigma}^z_{1}\hat{\sigma}^z_{57} - 4\hat{\sigma}^z_{1}\hat{\sigma}^z_{58} - 4\hat{\sigma}^z_{1}\hat{\sigma}^z_{59} + 4\hat{\sigma}^z_{1}\hat{\sigma}^z_{6} + 4\hat{\sigma}^z_{1}\hat{\sigma}^z_{7} + 4\hat{\sigma}^z_{1}\hat{\sigma}^z_{8} + 4\hat{\sigma}^z_{1}\hat{\sigma}^z_{9} + 404\hat{\sigma}^z_{1} + 4\hat{\sigma}^z_{10}\hat{\sigma}^z_{11} + 2\hat{\sigma}^z_{10}\hat{\sigma}^z_{12} + 2\hat{\sigma}^z_{10}\hat{\sigma}^z_{13} + 2\hat{\sigma}^z_{10}\hat{\sigma}^z_{14} + 2\hat{\sigma}^z_{10}\hat{\sigma}^z_{15} + 2\hat{\sigma}^z_{10}\hat{\sigma}^z_{16} + 2\hat{\sigma}^z_{10}\hat{\sigma}^z_{17} + 2\hat{\sigma}^z_{10}\hat{\sigma}^z_{18} + 2\hat{\sigma}^z_{10}\hat{\sigma}^z_{19} + 4\hat{\sigma}^z_{10}\hat{\sigma}^z_{2} + 2\hat{\sigma}^z_{10}\hat{\sigma}^z_{20} + 2\hat{\sigma}^z_{10}\hat{\sigma}^z_{21} + 180\hat{\sigma}^z_{10}\hat{\sigma}^z_{22} + 2\hat{\sigma}^z_{10}\hat{\sigma}^z_{23} + 2\hat{\sigma}^z_{10}\hat{\sigma}^z_{24} + 2\hat{\sigma}^z_{10}\hat{\sigma}^z_{25} + 2\hat{\sigma}^z_{10}\hat{\sigma}^z_{26} + 2\hat{\sigma}^z_{10}\hat{\sigma}^z_{27} + 2\hat{\sigma}^z_{10}\hat{\sigma}^z_{28} + 2\hat{\sigma}^z_{10}\hat{\sigma}^z_{29} + 4\hat{\sigma}^z_{10}\hat{\sigma}^z_{3} + 2\hat{\sigma}^z_{10}\hat{\sigma}^z_{30} + 2\hat{\sigma}^z_{10}\hat{\sigma}^z_{31} + 2\hat{\sigma}^z_{10}\hat{\sigma}^z_{32} + 2\hat{\sigma}^z_{10}\hat{\sigma}^z_{33} + 180\hat{\sigma}^z_{10}\hat{\sigma}^z_{34} + 2\hat{\sigma}^z_{10}\hat{\sigma}^z_{35} - 4\hat{\sigma}^z_{10}\hat{\sigma}^z_{36} - 4\hat{\sigma}^z_{10}\hat{\sigma}^z_{37} - 4\hat{\sigma}^z_{10}\hat{\sigma}^z_{38} - 4\hat{\sigma}^z_{10}\hat{\sigma}^z_{39} + 4\hat{\sigma}^z_{10}\hat{\sigma}^z_{4} - 4\hat{\sigma}^z_{10}\hat{\sigma}^z_{40} - 4\hat{\sigma}^z_{10}\hat{\sigma}^z_{41} - 4\hat{\sigma}^z_{10}\hat{\sigma}^z_{42} - 4\hat{\sigma}^z_{10}\hat{\sigma}^z_{43} - 4\hat{\sigma}^z_{10}\hat{\sigma}^z_{44} - 4\hat{\sigma}^z_{10}\hat{\sigma}^z_{45} - 360\hat{\sigma}^z_{10}\hat{\sigma}^z_{46} - 4\hat{\sigma}^z_{10}\hat{\sigma}^z_{47} - 4\hat{\sigma}^z_{10}\hat{\sigma}^z_{48} - 4\hat{\sigma}^z_{10}\hat{\sigma}^z_{49} + 4\hat{\sigma}^z_{10}\hat{\sigma}^z_{5} - 4\hat{\sigma}^z_{10}\hat{\sigma}^z_{50} - 4\hat{\sigma}^z_{10}\hat{\sigma}^z_{51} - 4\hat{\sigma}^z_{10}\hat{\sigma}^z_{52} - 4\hat{\sigma}^z_{10}\hat{\sigma}^z_{53} - 4\hat{\sigma}^z_{10}\hat{\sigma}^z_{54} - 4\hat{\sigma}^z_{10}\hat{\sigma}^z_{55} - 4\hat{\sigma}^z_{10}\hat{\sigma}^z_{56} - 4\hat{\sigma}^z_{10}\hat{\sigma}^z_{57} - 360\hat{\sigma}^z_{10}\hat{\sigma}^z_{58} - 4\hat{\sigma}^z_{10}\hat{\sigma}^z_{59} + 4\hat{\sigma}^z_{10}\hat{\sigma}^z_{6} + 4\hat{\sigma}^z_{10}\hat{\sigma}^z_{7} + 4\hat{\sigma}^z_{10}\hat{\sigma}^z_{8} + 4\hat{\sigma}^z_{10}\hat{\sigma}^z_{9} + 404\hat{\sigma}^z_{10} + 2\hat{\sigma}^z_{11}\hat{\sigma}^z_{12} + 2\hat{\sigma}^z_{11}\hat{\sigma}^z_{13} + 2\hat{\sigma}^z_{11}\hat{\sigma}^z_{14} + 2\hat{\sigma}^z_{11}\hat{\sigma}^z_{15} + 2\hat{\sigma}^z_{11}\hat{\sigma}^z_{16} + 2\hat{\sigma}^z_{11}\hat{\sigma}^z_{17} + 2\hat{\sigma}^z_{11}\hat{\sigma}^z_{18} + 2\hat{\sigma}^z_{11}\hat{\sigma}^z_{19} + 4\hat{\sigma}^z_{11}\hat{\sigma}^z_{2} + 2\hat{\sigma}^z_{11}\hat{\sigma}^z_{20} + 2\hat{\sigma}^z_{11}\hat{\sigma}^z_{21} + 2\hat{\sigma}^z_{11}\hat{\sigma}^z_{22} + 180\hat{\sigma}^z_{11}\hat{\sigma}^z_{23} + 2\hat{\sigma}^z_{11}\hat{\sigma}^z_{24} + 2\hat{\sigma}^z_{11}\hat{\sigma}^z_{25} + 2\hat{\sigma}^z_{11}\hat{\sigma}^z_{26} + 2\hat{\sigma}^z_{11}\hat{\sigma}^z_{27} + 2\hat{\sigma}^z_{11}\hat{\sigma}^z_{28} + 2\hat{\sigma}^z_{11}\hat{\sigma}^z_{29} + 4\hat{\sigma}^z_{11}\hat{\sigma}^z_{3} + 2\hat{\sigma}^z_{11}\hat{\sigma}^z_{30} + 2\hat{\sigma}^z_{11}\hat{\sigma}^z_{31} + 2\hat{\sigma}^z_{11}\hat{\sigma}^z_{32} + 2\hat{\sigma}^z_{11}\hat{\sigma}^z_{33} + 2\hat{\sigma}^z_{11}\hat{\sigma}^z_{34} + 180\hat{\sigma}^z_{11}\hat{\sigma}^z_{35} - 4\hat{\sigma}^z_{11}\hat{\sigma}^z_{36} - 4\hat{\sigma}^z_{11}\hat{\sigma}^z_{37} - 4\hat{\sigma}^z_{11}\hat{\sigma}^z_{38} - 4\hat{\sigma}^z_{11}\hat{\sigma}^z_{39} + 4\hat{\sigma}^z_{11}\hat{\sigma}^z_{4} - 4\hat{\sigma}^z_{11}\hat{\sigma}^z_{40} - 4\hat{\sigma}^z_{11}\hat{\sigma}^z_{41} - 4\hat{\sigma}^z_{11}\hat{\sigma}^z_{42} - 4\hat{\sigma}^z_{11}\hat{\sigma}^z_{43} - 4\hat{\sigma}^z_{11}\hat{\sigma}^z_{44} - 4\hat{\sigma}^z_{11}\hat{\sigma}^z_{45} - 4\hat{\sigma}^z_{11}\hat{\sigma}^z_{46} - 360\hat{\sigma}^z_{11}\hat{\sigma}^z_{47} - 4\hat{\sigma}^z_{11}\hat{\sigma}^z_{48} - 4\hat{\sigma}^z_{11}\hat{\sigma}^z_{49} + 4\hat{\sigma}^z_{11}\hat{\sigma}^z_{5} - 4\hat{\sigma}^z_{11}\hat{\sigma}^z_{50} - 4\hat{\sigma}^z_{11}\hat{\sigma}^z_{51} - 4\hat{\sigma}^z_{11}\hat{\sigma}^z_{52} - 4\hat{\sigma}^z_{11}\hat{\sigma}^z_{53} - 4\hat{\sigma}^z_{11}\hat{\sigma}^z_{54} - 4\hat{\sigma}^z_{11}\hat{\sigma}^z_{55} - 4\hat{\sigma}^z_{11}\hat{\sigma}^z_{56} - 4\hat{\sigma}^z_{11}\hat{\sigma}^z_{57} - 4\hat{\sigma}^z_{11}\hat{\sigma}^z_{58} - 360\hat{\sigma}^z_{11}\hat{\sigma}^z_{59} + 4\hat{\sigma}^z_{11}\hat{\sigma}^z_{6} + 4\hat{\sigma}^z_{11}\hat{\sigma}^z_{7} + 4\hat{\sigma}^z_{11}\hat{\sigma}^z_{8} + 4\hat{\sigma}^z_{11}\hat{\sigma}^z_{9} + 404\hat{\sigma}^z_{11} + 4\hat{\sigma}^z_{12}\hat{\sigma}^z_{13} + 4\hat{\sigma}^z_{12}\hat{\sigma}^z_{14} + 4\hat{\sigma}^z_{12}\hat{\sigma}^z_{15} + 4\hat{\sigma}^z_{12}\hat{\sigma}^z_{16} + 4\hat{\sigma}^z_{12}\hat{\sigma}^z_{17} + 4\hat{\sigma}^z_{12}\hat{\sigma}^z_{18} + 4\hat{\sigma}^z_{12}\hat{\sigma}^z_{19} + 2\hat{\sigma}^z_{12}\hat{\sigma}^z_{2} + 4\hat{\sigma}^z_{12}\hat{\sigma}^z_{20} + 4\hat{\sigma}^z_{12}\hat{\sigma}^z_{21} + 4\hat{\sigma}^z_{12}\hat{\sigma}^z_{22} + 4\hat{\sigma}^z_{12}\hat{\sigma}^z_{23} + 180\hat{\sigma}^z_{12}\hat{\sigma}^z_{24} + 2\hat{\sigma}^z_{12}\hat{\sigma}^z_{25} + 2\hat{\sigma}^z_{12}\hat{\sigma}^z_{26} + 2\hat{\sigma}^z_{12}\hat{\sigma}^z_{27} + 2\hat{\sigma}^z_{12}\hat{\sigma}^z_{28} + 2\hat{\sigma}^z_{12}\hat{\sigma}^z_{29} + 2\hat{\sigma}^z_{12}\hat{\sigma}^z_{3} + 2\hat{\sigma}^z_{12}\hat{\sigma}^z_{30} + 2\hat{\sigma}^z_{12}\hat{\sigma}^z_{31} + 2\hat{\sigma}^z_{12}\hat{\sigma}^z_{32} + 2\hat{\sigma}^z_{12}\hat{\sigma}^z_{33} + 2\hat{\sigma}^z_{12}\hat{\sigma}^z_{34} + 2\hat{\sigma}^z_{12}\hat{\sigma}^z_{35} - 360\hat{\sigma}^z_{12}\hat{\sigma}^z_{36} - 4\hat{\sigma}^z_{12}\hat{\sigma}^z_{37} - 4\hat{\sigma}^z_{12}\hat{\sigma}^z_{38} - 4\hat{\sigma}^z_{12}\hat{\sigma}^z_{39} + 2\hat{\sigma}^z_{12}\hat{\sigma}^z_{4} - 4\hat{\sigma}^z_{12}\hat{\sigma}^z_{40} - 4\hat{\sigma}^z_{12}\hat{\sigma}^z_{41} - 4\hat{\sigma}^z_{12}\hat{\sigma}^z_{42} - 4\hat{\sigma}^z_{12}\hat{\sigma}^z_{43} - 4\hat{\sigma}^z_{12}\hat{\sigma}^z_{44} - 4\hat{\sigma}^z_{12}\hat{\sigma}^z_{45} - 4\hat{\sigma}^z_{12}\hat{\sigma}^z_{46} - 4\hat{\sigma}^z_{12}\hat{\sigma}^z_{47} + 2\hat{\sigma}^z_{12}\hat{\sigma}^z_{5} + 2\hat{\sigma}^z_{12}\hat{\sigma}^z_{6} - 360\hat{\sigma}^z_{12}\hat{\sigma}^z_{60} - 4\hat{\sigma}^z_{12}\hat{\sigma}^z_{61} - 4\hat{\sigma}^z_{12}\hat{\sigma}^z_{62} - 4\hat{\sigma}^z_{12}\hat{\sigma}^z_{63} - 4\hat{\sigma}^z_{12}\hat{\sigma}^z_{64} - 4\hat{\sigma}^z_{12}\hat{\sigma}^z_{65} - 4\hat{\sigma}^z_{12}\hat{\sigma}^z_{66} - 4\hat{\sigma}^z_{12}\hat{\sigma}^z_{67} - 4\hat{\sigma}^z_{12}\hat{\sigma}^z_{68} - 4\hat{\sigma}^z_{12}\hat{\sigma}^z_{69} + 2\hat{\sigma}^z_{12}\hat{\sigma}^z_{7} - 4\hat{\sigma}^z_{12}\hat{\sigma}^z_{70} - 4\hat{\sigma}^z_{12}\hat{\sigma}^z_{71} + 2\hat{\sigma}^z_{12}\hat{\sigma}^z_{8} + 2\hat{\sigma}^z_{12}\hat{\sigma}^z_{9} + 404\hat{\sigma}^z_{12} + 4\hat{\sigma}^z_{13}\hat{\sigma}^z_{14} + 4\hat{\sigma}^z_{13}\hat{\sigma}^z_{15} + 4\hat{\sigma}^z_{13}\hat{\sigma}^z_{16} + 4\hat{\sigma}^z_{13}\hat{\sigma}^z_{17} + 4\hat{\sigma}^z_{13}\hat{\sigma}^z_{18} + 4\hat{\sigma}^z_{13}\hat{\sigma}^z_{19} + 2\hat{\sigma}^z_{13}\hat{\sigma}^z_{2} + 4\hat{\sigma}^z_{13}\hat{\sigma}^z_{20} + 4\hat{\sigma}^z_{13}\hat{\sigma}^z_{21} + 4\hat{\sigma}^z_{13}\hat{\sigma}^z_{22} + 4\hat{\sigma}^z_{13}\hat{\sigma}^z_{23} + 2\hat{\sigma}^z_{13}\hat{\sigma}^z_{24} + 180\hat{\sigma}^z_{13}\hat{\sigma}^z_{25} + 2\hat{\sigma}^z_{13}\hat{\sigma}^z_{26} + 2\hat{\sigma}^z_{13}\hat{\sigma}^z_{27} + 2\hat{\sigma}^z_{13}\hat{\sigma}^z_{28} + 2\hat{\sigma}^z_{13}\hat{\sigma}^z_{29} + 2\hat{\sigma}^z_{13}\hat{\sigma}^z_{3} + 2\hat{\sigma}^z_{13}\hat{\sigma}^z_{30} + 2\hat{\sigma}^z_{13}\hat{\sigma}^z_{31} + 2\hat{\sigma}^z_{13}\hat{\sigma}^z_{32} + 2\hat{\sigma}^z_{13}\hat{\sigma}^z_{33} + 2\hat{\sigma}^z_{13}\hat{\sigma}^z_{34} + 2\hat{\sigma}^z_{13}\hat{\sigma}^z_{35} - 4\hat{\sigma}^z_{13}\hat{\sigma}^z_{36} - 360\hat{\sigma}^z_{13}\hat{\sigma}^z_{37} - 4\hat{\sigma}^z_{13}\hat{\sigma}^z_{38} - 4\hat{\sigma}^z_{13}\hat{\sigma}^z_{39} + 2\hat{\sigma}^z_{13}\hat{\sigma}^z_{4} - 4\hat{\sigma}^z_{13}\hat{\sigma}^z_{40} - 4\hat{\sigma}^z_{13}\hat{\sigma}^z_{41} - 4\hat{\sigma}^z_{13}\hat{\sigma}^z_{42} - 4\hat{\sigma}^z_{13}\hat{\sigma}^z_{43} - 4\hat{\sigma}^z_{13}\hat{\sigma}^z_{44} - 4\hat{\sigma}^z_{13}\hat{\sigma}^z_{45} - 4\hat{\sigma}^z_{13}\hat{\sigma}^z_{46} - 4\hat{\sigma}^z_{13}\hat{\sigma}^z_{47} + 2\hat{\sigma}^z_{13}\hat{\sigma}^z_{5} + 2\hat{\sigma}^z_{13}\hat{\sigma}^z_{6} - 4\hat{\sigma}^z_{13}\hat{\sigma}^z_{60} - 360\hat{\sigma}^z_{13}\hat{\sigma}^z_{61} - 4\hat{\sigma}^z_{13}\hat{\sigma}^z_{62} - 4\hat{\sigma}^z_{13}\hat{\sigma}^z_{63} - 4\hat{\sigma}^z_{13}\hat{\sigma}^z_{64} - 4\hat{\sigma}^z_{13}\hat{\sigma}^z_{65} - 4\hat{\sigma}^z_{13}\hat{\sigma}^z_{66} - 4\hat{\sigma}^z_{13}\hat{\sigma}^z_{67} - 4\hat{\sigma}^z_{13}\hat{\sigma}^z_{68} - 4\hat{\sigma}^z_{13}\hat{\sigma}^z_{69} + 2\hat{\sigma}^z_{13}\hat{\sigma}^z_{7} - 4\hat{\sigma}^z_{13}\hat{\sigma}^z_{70} - 4\hat{\sigma}^z_{13}\hat{\sigma}^z_{71} + 2\hat{\sigma}^z_{13}\hat{\sigma}^z_{8} + 2\hat{\sigma}^z_{13}\hat{\sigma}^z_{9} + 404\hat{\sigma}^z_{13} + 4\hat{\sigma}^z_{14}\hat{\sigma}^z_{15} + 4\hat{\sigma}^z_{14}\hat{\sigma}^z_{16} + 4\hat{\sigma}^z_{14}\hat{\sigma}^z_{17} + 4\hat{\sigma}^z_{14}\hat{\sigma}^z_{18} + 4\hat{\sigma}^z_{14}\hat{\sigma}^z_{19} + 180\hat{\sigma}^z_{14}\hat{\sigma}^z_{2} + 4\hat{\sigma}^z_{14}\hat{\sigma}^z_{20} + 4\hat{\sigma}^z_{14}\hat{\sigma}^z_{21} + 4\hat{\sigma}^z_{14}\hat{\sigma}^z_{22} + 4\hat{\sigma}^z_{14}\hat{\sigma}^z_{23} + 2\hat{\sigma}^z_{14}\hat{\sigma}^z_{24} + 2\hat{\sigma}^z_{14}\hat{\sigma}^z_{25} + 180\hat{\sigma}^z_{14}\hat{\sigma}^z_{26} + 2\hat{\sigma}^z_{14}\hat{\sigma}^z_{27} + 2\hat{\sigma}^z_{14}\hat{\sigma}^z_{28} + 2\hat{\sigma}^z_{14}\hat{\sigma}^z_{29} + 2\hat{\sigma}^z_{14}\hat{\sigma}^z_{3} + 2\hat{\sigma}^z_{14}\hat{\sigma}^z_{30} + 2\hat{\sigma}^z_{14}\hat{\sigma}^z_{31} + 2\hat{\sigma}^z_{14}\hat{\sigma}^z_{32} + 2\hat{\sigma}^z_{14}\hat{\sigma}^z_{33} + 2\hat{\sigma}^z_{14}\hat{\sigma}^z_{34} + 2\hat{\sigma}^z_{14}\hat{\sigma}^z_{35} - 4\hat{\sigma}^z_{14}\hat{\sigma}^z_{36} - 4\hat{\sigma}^z_{14}\hat{\sigma}^z_{37} - 360\hat{\sigma}^z_{14}\hat{\sigma}^z_{38} - 4\hat{\sigma}^z_{14}\hat{\sigma}^z_{39} + 2\hat{\sigma}^z_{14}\hat{\sigma}^z_{4} - 4\hat{\sigma}^z_{14}\hat{\sigma}^z_{40} - 4\hat{\sigma}^z_{14}\hat{\sigma}^z_{41} - 4\hat{\sigma}^z_{14}\hat{\sigma}^z_{42} - 4\hat{\sigma}^z_{14}\hat{\sigma}^z_{43} - 4\hat{\sigma}^z_{14}\hat{\sigma}^z_{44} - 4\hat{\sigma}^z_{14}\hat{\sigma}^z_{45} - 4\hat{\sigma}^z_{14}\hat{\sigma}^z_{46} - 4\hat{\sigma}^z_{14}\hat{\sigma}^z_{47} + 2\hat{\sigma}^z_{14}\hat{\sigma}^z_{5} + 2\hat{\sigma}^z_{14}\hat{\sigma}^z_{6} - 4\hat{\sigma}^z_{14}\hat{\sigma}^z_{60} - 4\hat{\sigma}^z_{14}\hat{\sigma}^z_{61} - 360\hat{\sigma}^z_{14}\hat{\sigma}^z_{62} - 4\hat{\sigma}^z_{14}\hat{\sigma}^z_{63} - 4\hat{\sigma}^z_{14}\hat{\sigma}^z_{64} - 4\hat{\sigma}^z_{14}\hat{\sigma}^z_{65} - 4\hat{\sigma}^z_{14}\hat{\sigma}^z_{66} - 4\hat{\sigma}^z_{14}\hat{\sigma}^z_{67} - 4\hat{\sigma}^z_{14}\hat{\sigma}^z_{68} - 4\hat{\sigma}^z_{14}\hat{\sigma}^z_{69} + 2\hat{\sigma}^z_{14}\hat{\sigma}^z_{7} - 4\hat{\sigma}^z_{14}\hat{\sigma}^z_{70} - 4\hat{\sigma}^z_{14}\hat{\sigma}^z_{71} + 2\hat{\sigma}^z_{14}\hat{\sigma}^z_{8} + 2\hat{\sigma}^z_{14}\hat{\sigma}^z_{9} + 404\hat{\sigma}^z_{14} + 4\hat{\sigma}^z_{15}\hat{\sigma}^z_{16} + 4\hat{\sigma}^z_{15}\hat{\sigma}^z_{17} + 4\hat{\sigma}^z_{15}\hat{\sigma}^z_{18} + 4\hat{\sigma}^z_{15}\hat{\sigma}^z_{19} + 2\hat{\sigma}^z_{15}\hat{\sigma}^z_{2} + 4\hat{\sigma}^z_{15}\hat{\sigma}^z_{20} + 4\hat{\sigma}^z_{15}\hat{\sigma}^z_{21} + 4\hat{\sigma}^z_{15}\hat{\sigma}^z_{22} + 4\hat{\sigma}^z_{15}\hat{\sigma}^z_{23} + 2\hat{\sigma}^z_{15}\hat{\sigma}^z_{24} + 2\hat{\sigma}^z_{15}\hat{\sigma}^z_{25} + 2\hat{\sigma}^z_{15}\hat{\sigma}^z_{26} + 180\hat{\sigma}^z_{15}\hat{\sigma}^z_{27} + 2\hat{\sigma}^z_{15}\hat{\sigma}^z_{28} + 2\hat{\sigma}^z_{15}\hat{\sigma}^z_{29} + 180\hat{\sigma}^z_{15}\hat{\sigma}^z_{3} + 2\hat{\sigma}^z_{15}\hat{\sigma}^z_{30} + 2\hat{\sigma}^z_{15}\hat{\sigma}^z_{31} + 2\hat{\sigma}^z_{15}\hat{\sigma}^z_{32} + 2\hat{\sigma}^z_{15}\hat{\sigma}^z_{33} + 2\hat{\sigma}^z_{15}\hat{\sigma}^z_{34} + 2\hat{\sigma}^z_{15}\hat{\sigma}^z_{35} - 4\hat{\sigma}^z_{15}\hat{\sigma}^z_{36} - 4\hat{\sigma}^z_{15}\hat{\sigma}^z_{37} - 4\hat{\sigma}^z_{15}\hat{\sigma}^z_{38} - 360\hat{\sigma}^z_{15}\hat{\sigma}^z_{39} + 2\hat{\sigma}^z_{15}\hat{\sigma}^z_{4} - 4\hat{\sigma}^z_{15}\hat{\sigma}^z_{40} - 4\hat{\sigma}^z_{15}\hat{\sigma}^z_{41} - 4\hat{\sigma}^z_{15}\hat{\sigma}^z_{42} - 4\hat{\sigma}^z_{15}\hat{\sigma}^z_{43} - 4\hat{\sigma}^z_{15}\hat{\sigma}^z_{44} - 4\hat{\sigma}^z_{15}\hat{\sigma}^z_{45} - 4\hat{\sigma}^z_{15}\hat{\sigma}^z_{46} - 4\hat{\sigma}^z_{15}\hat{\sigma}^z_{47} + 2\hat{\sigma}^z_{15}\hat{\sigma}^z_{5} + 2\hat{\sigma}^z_{15}\hat{\sigma}^z_{6} - 4\hat{\sigma}^z_{15}\hat{\sigma}^z_{60} - 4\hat{\sigma}^z_{15}\hat{\sigma}^z_{61} - 4\hat{\sigma}^z_{15}\hat{\sigma}^z_{62} - 360\hat{\sigma}^z_{15}\hat{\sigma}^z_{63} - 4\hat{\sigma}^z_{15}\hat{\sigma}^z_{64} - 4\hat{\sigma}^z_{15}\hat{\sigma}^z_{65} - 4\hat{\sigma}^z_{15}\hat{\sigma}^z_{66} - 4\hat{\sigma}^z_{15}\hat{\sigma}^z_{67} - 4\hat{\sigma}^z_{15}\hat{\sigma}^z_{68} - 4\hat{\sigma}^z_{15}\hat{\sigma}^z_{69} + 2\hat{\sigma}^z_{15}\hat{\sigma}^z_{7} - 4\hat{\sigma}^z_{15}\hat{\sigma}^z_{70} - 4\hat{\sigma}^z_{15}\hat{\sigma}^z_{71} + 2\hat{\sigma}^z_{15}\hat{\sigma}^z_{8} + 2\hat{\sigma}^z_{15}\hat{\sigma}^z_{9} + 404\hat{\sigma}^z_{15} + 4\hat{\sigma}^z_{16}\hat{\sigma}^z_{17} + 4\hat{\sigma}^z_{16}\hat{\sigma}^z_{18} + 4\hat{\sigma}^z_{16}\hat{\sigma}^z_{19} + 2\hat{\sigma}^z_{16}\hat{\sigma}^z_{2} + 4\hat{\sigma}^z_{16}\hat{\sigma}^z_{20} + 4\hat{\sigma}^z_{16}\hat{\sigma}^z_{21} + 4\hat{\sigma}^z_{16}\hat{\sigma}^z_{22} + 4\hat{\sigma}^z_{16}\hat{\sigma}^z_{23} + 2\hat{\sigma}^z_{16}\hat{\sigma}^z_{24} + 2\hat{\sigma}^z_{16}\hat{\sigma}^z_{25} + 2\hat{\sigma}^z_{16}\hat{\sigma}^z_{26} + 2\hat{\sigma}^z_{16}\hat{\sigma}^z_{27} + 180\hat{\sigma}^z_{16}\hat{\sigma}^z_{28} + 2\hat{\sigma}^z_{16}\hat{\sigma}^z_{29} + 2\hat{\sigma}^z_{16}\hat{\sigma}^z_{3} + 2\hat{\sigma}^z_{16}\hat{\sigma}^z_{30} + 2\hat{\sigma}^z_{16}\hat{\sigma}^z_{31} + 2\hat{\sigma}^z_{16}\hat{\sigma}^z_{32} + 2\hat{\sigma}^z_{16}\hat{\sigma}^z_{33} + 2\hat{\sigma}^z_{16}\hat{\sigma}^z_{34} + 2\hat{\sigma}^z_{16}\hat{\sigma}^z_{35} - 4\hat{\sigma}^z_{16}\hat{\sigma}^z_{36} - 4\hat{\sigma}^z_{16}\hat{\sigma}^z_{37} - 4\hat{\sigma}^z_{16}\hat{\sigma}^z_{38} - 4\hat{\sigma}^z_{16}\hat{\sigma}^z_{39} + 180\hat{\sigma}^z_{16}\hat{\sigma}^z_{4} - 360\hat{\sigma}^z_{16}\hat{\sigma}^z_{40} - 4\hat{\sigma}^z_{16}\hat{\sigma}^z_{41} - 4\hat{\sigma}^z_{16}\hat{\sigma}^z_{42} - 4\hat{\sigma}^z_{16}\hat{\sigma}^z_{43} - 4\hat{\sigma}^z_{16}\hat{\sigma}^z_{44} - 4\hat{\sigma}^z_{16}\hat{\sigma}^z_{45} - 4\hat{\sigma}^z_{16}\hat{\sigma}^z_{46} - 4\hat{\sigma}^z_{16}\hat{\sigma}^z_{47} + 2\hat{\sigma}^z_{16}\hat{\sigma}^z_{5} + 2\hat{\sigma}^z_{16}\hat{\sigma}^z_{6} - 4\hat{\sigma}^z_{16}\hat{\sigma}^z_{60} - 4\hat{\sigma}^z_{16}\hat{\sigma}^z_{61} - 4\hat{\sigma}^z_{16}\hat{\sigma}^z_{62} - 4\hat{\sigma}^z_{16}\hat{\sigma}^z_{63} - 360\hat{\sigma}^z_{16}\hat{\sigma}^z_{64} - 4\hat{\sigma}^z_{16}\hat{\sigma}^z_{65} - 4\hat{\sigma}^z_{16}\hat{\sigma}^z_{66} - 4\hat{\sigma}^z_{16}\hat{\sigma}^z_{67} - 4\hat{\sigma}^z_{16}\hat{\sigma}^z_{68} - 4\hat{\sigma}^z_{16}\hat{\sigma}^z_{69} + 2\hat{\sigma}^z_{16}\hat{\sigma}^z_{7} - 4\hat{\sigma}^z_{16}\hat{\sigma}^z_{70} - 4\hat{\sigma}^z_{16}\hat{\sigma}^z_{71} + 2\hat{\sigma}^z_{16}\hat{\sigma}^z_{8} + 2\hat{\sigma}^z_{16}\hat{\sigma}^z_{9} + 404\hat{\sigma}^z_{16} + 4\hat{\sigma}^z_{17}\hat{\sigma}^z_{18} + 4\hat{\sigma}^z_{17}\hat{\sigma}^z_{19} + 2\hat{\sigma}^z_{17}\hat{\sigma}^z_{2} + 4\hat{\sigma}^z_{17}\hat{\sigma}^z_{20} + 4\hat{\sigma}^z_{17}\hat{\sigma}^z_{21} + 4\hat{\sigma}^z_{17}\hat{\sigma}^z_{22} + 4\hat{\sigma}^z_{17}\hat{\sigma}^z_{23} + 2\hat{\sigma}^z_{17}\hat{\sigma}^z_{24} + 2\hat{\sigma}^z_{17}\hat{\sigma}^z_{25} + 2\hat{\sigma}^z_{17}\hat{\sigma}^z_{26} + 2\hat{\sigma}^z_{17}\hat{\sigma}^z_{27} + 2\hat{\sigma}^z_{17}\hat{\sigma}^z_{28} + 180\hat{\sigma}^z_{17}\hat{\sigma}^z_{29} + 2\hat{\sigma}^z_{17}\hat{\sigma}^z_{3} + 2\hat{\sigma}^z_{17}\hat{\sigma}^z_{30} + 2\hat{\sigma}^z_{17}\hat{\sigma}^z_{31} + 2\hat{\sigma}^z_{17}\hat{\sigma}^z_{32} + 2\hat{\sigma}^z_{17}\hat{\sigma}^z_{33} + 2\hat{\sigma}^z_{17}\hat{\sigma}^z_{34} + 2\hat{\sigma}^z_{17}\hat{\sigma}^z_{35} - 4\hat{\sigma}^z_{17}\hat{\sigma}^z_{36} - 4\hat{\sigma}^z_{17}\hat{\sigma}^z_{37} - 4\hat{\sigma}^z_{17}\hat{\sigma}^z_{38} - 4\hat{\sigma}^z_{17}\hat{\sigma}^z_{39} + 2\hat{\sigma}^z_{17}\hat{\sigma}^z_{4} - 4\hat{\sigma}^z_{17}\hat{\sigma}^z_{40} - 360\hat{\sigma}^z_{17}\hat{\sigma}^z_{41} - 4\hat{\sigma}^z_{17}\hat{\sigma}^z_{42} - 4\hat{\sigma}^z_{17}\hat{\sigma}^z_{43} - 4\hat{\sigma}^z_{17}\hat{\sigma}^z_{44} - 4\hat{\sigma}^z_{17}\hat{\sigma}^z_{45} - 4\hat{\sigma}^z_{17}\hat{\sigma}^z_{46} - 4\hat{\sigma}^z_{17}\hat{\sigma}^z_{47} + 180\hat{\sigma}^z_{17}\hat{\sigma}^z_{5} + 2\hat{\sigma}^z_{17}\hat{\sigma}^z_{6} - 4\hat{\sigma}^z_{17}\hat{\sigma}^z_{60} - 4\hat{\sigma}^z_{17}\hat{\sigma}^z_{61} - 4\hat{\sigma}^z_{17}\hat{\sigma}^z_{62} - 4\hat{\sigma}^z_{17}\hat{\sigma}^z_{63} - 4\hat{\sigma}^z_{17}\hat{\sigma}^z_{64} - 360\hat{\sigma}^z_{17}\hat{\sigma}^z_{65} - 4\hat{\sigma}^z_{17}\hat{\sigma}^z_{66} - 4\hat{\sigma}^z_{17}\hat{\sigma}^z_{67} - 4\hat{\sigma}^z_{17}\hat{\sigma}^z_{68} - 4\hat{\sigma}^z_{17}\hat{\sigma}^z_{69} + 2\hat{\sigma}^z_{17}\hat{\sigma}^z_{7} - 4\hat{\sigma}^z_{17}\hat{\sigma}^z_{70} - 4\hat{\sigma}^z_{17}\hat{\sigma}^z_{71} + 2\hat{\sigma}^z_{17}\hat{\sigma}^z_{8} + 2\hat{\sigma}^z_{17}\hat{\sigma}^z_{9} + 404\hat{\sigma}^z_{17} + 4\hat{\sigma}^z_{18}\hat{\sigma}^z_{19} + 2\hat{\sigma}^z_{18}\hat{\sigma}^z_{2} + 4\hat{\sigma}^z_{18}\hat{\sigma}^z_{20} + 4\hat{\sigma}^z_{18}\hat{\sigma}^z_{21} + 4\hat{\sigma}^z_{18}\hat{\sigma}^z_{22} + 4\hat{\sigma}^z_{18}\hat{\sigma}^z_{23} + 2\hat{\sigma}^z_{18}\hat{\sigma}^z_{24} + 2\hat{\sigma}^z_{18}\hat{\sigma}^z_{25} + 2\hat{\sigma}^z_{18}\hat{\sigma}^z_{26} + 2\hat{\sigma}^z_{18}\hat{\sigma}^z_{27} + 2\hat{\sigma}^z_{18}\hat{\sigma}^z_{28} + 2\hat{\sigma}^z_{18}\hat{\sigma}^z_{29} + 2\hat{\sigma}^z_{18}\hat{\sigma}^z_{3} + 180\hat{\sigma}^z_{18}\hat{\sigma}^z_{30} + 2\hat{\sigma}^z_{18}\hat{\sigma}^z_{31} + 2\hat{\sigma}^z_{18}\hat{\sigma}^z_{32} + 2\hat{\sigma}^z_{18}\hat{\sigma}^z_{33} + 2\hat{\sigma}^z_{18}\hat{\sigma}^z_{34} + 2\hat{\sigma}^z_{18}\hat{\sigma}^z_{35} - 4\hat{\sigma}^z_{18}\hat{\sigma}^z_{36} - 4\hat{\sigma}^z_{18}\hat{\sigma}^z_{37} - 4\hat{\sigma}^z_{18}\hat{\sigma}^z_{38} - 4\hat{\sigma}^z_{18}\hat{\sigma}^z_{39} + 2\hat{\sigma}^z_{18}\hat{\sigma}^z_{4} - 4\hat{\sigma}^z_{18}\hat{\sigma}^z_{40} - 4\hat{\sigma}^z_{18}\hat{\sigma}^z_{41} - 360\hat{\sigma}^z_{18}\hat{\sigma}^z_{42} - 4\hat{\sigma}^z_{18}\hat{\sigma}^z_{43} - 4\hat{\sigma}^z_{18}\hat{\sigma}^z_{44} - 4\hat{\sigma}^z_{18}\hat{\sigma}^z_{45} - 4\hat{\sigma}^z_{18}\hat{\sigma}^z_{46} - 4\hat{\sigma}^z_{18}\hat{\sigma}^z_{47} + 2\hat{\sigma}^z_{18}\hat{\sigma}^z_{5} + 180\hat{\sigma}^z_{18}\hat{\sigma}^z_{6} - 4\hat{\sigma}^z_{18}\hat{\sigma}^z_{60} - 4\hat{\sigma}^z_{18}\hat{\sigma}^z_{61} - 4\hat{\sigma}^z_{18}\hat{\sigma}^z_{62} - 4\hat{\sigma}^z_{18}\hat{\sigma}^z_{63} - 4\hat{\sigma}^z_{18}\hat{\sigma}^z_{64} - 4\hat{\sigma}^z_{18}\hat{\sigma}^z_{65} - 360\hat{\sigma}^z_{18}\hat{\sigma}^z_{66} - 4\hat{\sigma}^z_{18}\hat{\sigma}^z_{67} - 4\hat{\sigma}^z_{18}\hat{\sigma}^z_{68} - 4\hat{\sigma}^z_{18}\hat{\sigma}^z_{69} + 2\hat{\sigma}^z_{18}\hat{\sigma}^z_{7} - 4\hat{\sigma}^z_{18}\hat{\sigma}^z_{70} - 4\hat{\sigma}^z_{18}\hat{\sigma}^z_{71} + 2\hat{\sigma}^z_{18}\hat{\sigma}^z_{8} + 2\hat{\sigma}^z_{18}\hat{\sigma}^z_{9} + 404\hat{\sigma}^z_{18} + 2\hat{\sigma}^z_{19}\hat{\sigma}^z_{2} + 4\hat{\sigma}^z_{19}\hat{\sigma}^z_{20} + 4\hat{\sigma}^z_{19}\hat{\sigma}^z_{21} + 4\hat{\sigma}^z_{19}\hat{\sigma}^z_{22} + 4\hat{\sigma}^z_{19}\hat{\sigma}^z_{23} + 2\hat{\sigma}^z_{19}\hat{\sigma}^z_{24} + 2\hat{\sigma}^z_{19}\hat{\sigma}^z_{25} + 2\hat{\sigma}^z_{19}\hat{\sigma}^z_{26} + 2\hat{\sigma}^z_{19}\hat{\sigma}^z_{27} + 2\hat{\sigma}^z_{19}\hat{\sigma}^z_{28} + 2\hat{\sigma}^z_{19}\hat{\sigma}^z_{29} + 2\hat{\sigma}^z_{19}\hat{\sigma}^z_{3} + 2\hat{\sigma}^z_{19}\hat{\sigma}^z_{30} + 180\hat{\sigma}^z_{19}\hat{\sigma}^z_{31} + 2\hat{\sigma}^z_{19}\hat{\sigma}^z_{32} + 2\hat{\sigma}^z_{19}\hat{\sigma}^z_{33} + 2\hat{\sigma}^z_{19}\hat{\sigma}^z_{34} + 2\hat{\sigma}^z_{19}\hat{\sigma}^z_{35} - 4\hat{\sigma}^z_{19}\hat{\sigma}^z_{36} - 4\hat{\sigma}^z_{19}\hat{\sigma}^z_{37} - 4\hat{\sigma}^z_{19}\hat{\sigma}^z_{38} - 4\hat{\sigma}^z_{19}\hat{\sigma}^z_{39} + 2\hat{\sigma}^z_{19}\hat{\sigma}^z_{4} - 4\hat{\sigma}^z_{19}\hat{\sigma}^z_{40} - 4\hat{\sigma}^z_{19}\hat{\sigma}^z_{41} - 4\hat{\sigma}^z_{19}\hat{\sigma}^z_{42} - 360\hat{\sigma}^z_{19}\hat{\sigma}^z_{43} - 4\hat{\sigma}^z_{19}\hat{\sigma}^z_{44} - 4\hat{\sigma}^z_{19}\hat{\sigma}^z_{45} - 4\hat{\sigma}^z_{19}\hat{\sigma}^z_{46} - 4\hat{\sigma}^z_{19}\hat{\sigma}^z_{47} + 2\hat{\sigma}^z_{19}\hat{\sigma}^z_{5} + 2\hat{\sigma}^z_{19}\hat{\sigma}^z_{6} - 4\hat{\sigma}^z_{19}\hat{\sigma}^z_{60} - 4\hat{\sigma}^z_{19}\hat{\sigma}^z_{61} - 4\hat{\sigma}^z_{19}\hat{\sigma}^z_{62} - 4\hat{\sigma}^z_{19}\hat{\sigma}^z_{63} - 4\hat{\sigma}^z_{19}\hat{\sigma}^z_{64} - 4\hat{\sigma}^z_{19}\hat{\sigma}^z_{65} - 4\hat{\sigma}^z_{19}\hat{\sigma}^z_{66} - 360\hat{\sigma}^z_{19}\hat{\sigma}^z_{67} - 4\hat{\sigma}^z_{19}\hat{\sigma}^z_{68} - 4\hat{\sigma}^z_{19}\hat{\sigma}^z_{69} + 180\hat{\sigma}^z_{19}\hat{\sigma}^z_{7} - 4\hat{\sigma}^z_{19}\hat{\sigma}^z_{70} - 4\hat{\sigma}^z_{19}\hat{\sigma}^z_{71} + 2\hat{\sigma}^z_{19}\hat{\sigma}^z_{8} + 2\hat{\sigma}^z_{19}\hat{\sigma}^z_{9} + 404\hat{\sigma}^z_{19} + 2\hat{\sigma}^z_{2}\hat{\sigma}^z_{20} + 2\hat{\sigma}^z_{2}\hat{\sigma}^z_{21} + 2\hat{\sigma}^z_{2}\hat{\sigma}^z_{22} + 2\hat{\sigma}^z_{2}\hat{\sigma}^z_{23} + 2\hat{\sigma}^z_{2}\hat{\sigma}^z_{24} + 2\hat{\sigma}^z_{2}\hat{\sigma}^z_{25} + 180\hat{\sigma}^z_{2}\hat{\sigma}^z_{26} + 2\hat{\sigma}^z_{2}\hat{\sigma}^z_{27} + 2\hat{\sigma}^z_{2}\hat{\sigma}^z_{28} + 2\hat{\sigma}^z_{2}\hat{\sigma}^z_{29} + 4\hat{\sigma}^z_{2}\hat{\sigma}^z_{3} + 2\hat{\sigma}^z_{2}\hat{\sigma}^z_{30} + 2\hat{\sigma}^z_{2}\hat{\sigma}^z_{31} + 2\hat{\sigma}^z_{2}\hat{\sigma}^z_{32} + 2\hat{\sigma}^z_{2}\hat{\sigma}^z_{33} + 2\hat{\sigma}^z_{2}\hat{\sigma}^z_{34} + 2\hat{\sigma}^z_{2}\hat{\sigma}^z_{35} - 4\hat{\sigma}^z_{2}\hat{\sigma}^z_{36} - 4\hat{\sigma}^z_{2}\hat{\sigma}^z_{37} - 360\hat{\sigma}^z_{2}\hat{\sigma}^z_{38} - 4\hat{\sigma}^z_{2}\hat{\sigma}^z_{39} + 4\hat{\sigma}^z_{2}\hat{\sigma}^z_{4} - 4\hat{\sigma}^z_{2}\hat{\sigma}^z_{40} - 4\hat{\sigma}^z_{2}\hat{\sigma}^z_{41} - 4\hat{\sigma}^z_{2}\hat{\sigma}^z_{42} - 4\hat{\sigma}^z_{2}\hat{\sigma}^z_{43} - 4\hat{\sigma}^z_{2}\hat{\sigma}^z_{44} - 4\hat{\sigma}^z_{2}\hat{\sigma}^z_{45} - 4\hat{\sigma}^z_{2}\hat{\sigma}^z_{46} - 4\hat{\sigma}^z_{2}\hat{\sigma}^z_{47} - 4\hat{\sigma}^z_{2}\hat{\sigma}^z_{48} - 4\hat{\sigma}^z_{2}\hat{\sigma}^z_{49} + 4\hat{\sigma}^z_{2}\hat{\sigma}^z_{5} - 360\hat{\sigma}^z_{2}\hat{\sigma}^z_{50} - 4\hat{\sigma}^z_{2}\hat{\sigma}^z_{51} - 4\hat{\sigma}^z_{2}\hat{\sigma}^z_{52} - 4\hat{\sigma}^z_{2}\hat{\sigma}^z_{53} - 4\hat{\sigma}^z_{2}\hat{\sigma}^z_{54} - 4\hat{\sigma}^z_{2}\hat{\sigma}^z_{55} - 4\hat{\sigma}^z_{2}\hat{\sigma}^z_{56} - 4\hat{\sigma}^z_{2}\hat{\sigma}^z_{57} - 4\hat{\sigma}^z_{2}\hat{\sigma}^z_{58} - 4\hat{\sigma}^z_{2}\hat{\sigma}^z_{59} + 4\hat{\sigma}^z_{2}\hat{\sigma}^z_{6} + 4\hat{\sigma}^z_{2}\hat{\sigma}^z_{7} + 4\hat{\sigma}^z_{2}\hat{\sigma}^z_{8} + 4\hat{\sigma}^z_{2}\hat{\sigma}^z_{9} + 404\hat{\sigma}^z_{2} + 4\hat{\sigma}^z_{20}\hat{\sigma}^z_{21} + 4\hat{\sigma}^z_{20}\hat{\sigma}^z_{22} + 4\hat{\sigma}^z_{20}\hat{\sigma}^z_{23} + 2\hat{\sigma}^z_{20}\hat{\sigma}^z_{24} + 2\hat{\sigma}^z_{20}\hat{\sigma}^z_{25} + 2\hat{\sigma}^z_{20}\hat{\sigma}^z_{26} + 2\hat{\sigma}^z_{20}\hat{\sigma}^z_{27} + 2\hat{\sigma}^z_{20}\hat{\sigma}^z_{28} + 2\hat{\sigma}^z_{20}\hat{\sigma}^z_{29} + 2\hat{\sigma}^z_{20}\hat{\sigma}^z_{3} + 2\hat{\sigma}^z_{20}\hat{\sigma}^z_{30} + 2\hat{\sigma}^z_{20}\hat{\sigma}^z_{31} + 180\hat{\sigma}^z_{20}\hat{\sigma}^z_{32} + 2\hat{\sigma}^z_{20}\hat{\sigma}^z_{33} + 2\hat{\sigma}^z_{20}\hat{\sigma}^z_{34} + 2\hat{\sigma}^z_{20}\hat{\sigma}^z_{35} - 4\hat{\sigma}^z_{20}\hat{\sigma}^z_{36} - 4\hat{\sigma}^z_{20}\hat{\sigma}^z_{37} - 4\hat{\sigma}^z_{20}\hat{\sigma}^z_{38} - 4\hat{\sigma}^z_{20}\hat{\sigma}^z_{39} + 2\hat{\sigma}^z_{20}\hat{\sigma}^z_{4} - 4\hat{\sigma}^z_{20}\hat{\sigma}^z_{40} - 4\hat{\sigma}^z_{20}\hat{\sigma}^z_{41} - 4\hat{\sigma}^z_{20}\hat{\sigma}^z_{42} - 4\hat{\sigma}^z_{20}\hat{\sigma}^z_{43} - 360\hat{\sigma}^z_{20}\hat{\sigma}^z_{44} - 4\hat{\sigma}^z_{20}\hat{\sigma}^z_{45} - 4\hat{\sigma}^z_{20}\hat{\sigma}^z_{46} - 4\hat{\sigma}^z_{20}\hat{\sigma}^z_{47} + 2\hat{\sigma}^z_{20}\hat{\sigma}^z_{5} + 2\hat{\sigma}^z_{20}\hat{\sigma}^z_{6} - 4\hat{\sigma}^z_{20}\hat{\sigma}^z_{60} - 4\hat{\sigma}^z_{20}\hat{\sigma}^z_{61} - 4\hat{\sigma}^z_{20}\hat{\sigma}^z_{62} - 4\hat{\sigma}^z_{20}\hat{\sigma}^z_{63} - 4\hat{\sigma}^z_{20}\hat{\sigma}^z_{64} - 4\hat{\sigma}^z_{20}\hat{\sigma}^z_{65} - 4\hat{\sigma}^z_{20}\hat{\sigma}^z_{66} - 4\hat{\sigma}^z_{20}\hat{\sigma}^z_{67} - 360\hat{\sigma}^z_{20}\hat{\sigma}^z_{68} - 4\hat{\sigma}^z_{20}\hat{\sigma}^z_{69} + 2\hat{\sigma}^z_{20}\hat{\sigma}^z_{7} - 4\hat{\sigma}^z_{20}\hat{\sigma}^z_{70} - 4\hat{\sigma}^z_{20}\hat{\sigma}^z_{71} + 180\hat{\sigma}^z_{20}\hat{\sigma}^z_{8} + 2\hat{\sigma}^z_{20}\hat{\sigma}^z_{9} + 404\hat{\sigma}^z_{20} + 4\hat{\sigma}^z_{21}\hat{\sigma}^z_{22} + 4\hat{\sigma}^z_{21}\hat{\sigma}^z_{23} + 2\hat{\sigma}^z_{21}\hat{\sigma}^z_{24} + 2\hat{\sigma}^z_{21}\hat{\sigma}^z_{25} + 2\hat{\sigma}^z_{21}\hat{\sigma}^z_{26} + 2\hat{\sigma}^z_{21}\hat{\sigma}^z_{27} + 2\hat{\sigma}^z_{21}\hat{\sigma}^z_{28} + 2\hat{\sigma}^z_{21}\hat{\sigma}^z_{29} + 2\hat{\sigma}^z_{21}\hat{\sigma}^z_{3} + 2\hat{\sigma}^z_{21}\hat{\sigma}^z_{30} + 2\hat{\sigma}^z_{21}\hat{\sigma}^z_{31} + 2\hat{\sigma}^z_{21}\hat{\sigma}^z_{32} + 180\hat{\sigma}^z_{21}\hat{\sigma}^z_{33} + 2\hat{\sigma}^z_{21}\hat{\sigma}^z_{34} + 2\hat{\sigma}^z_{21}\hat{\sigma}^z_{35} - 4\hat{\sigma}^z_{21}\hat{\sigma}^z_{36} - 4\hat{\sigma}^z_{21}\hat{\sigma}^z_{37} - 4\hat{\sigma}^z_{21}\hat{\sigma}^z_{38} - 4\hat{\sigma}^z_{21}\hat{\sigma}^z_{39} + 2\hat{\sigma}^z_{21}\hat{\sigma}^z_{4} - 4\hat{\sigma}^z_{21}\hat{\sigma}^z_{40} - 4\hat{\sigma}^z_{21}\hat{\sigma}^z_{41} - 4\hat{\sigma}^z_{21}\hat{\sigma}^z_{42} - 4\hat{\sigma}^z_{21}\hat{\sigma}^z_{43} - 4\hat{\sigma}^z_{21}\hat{\sigma}^z_{44} - 360\hat{\sigma}^z_{21}\hat{\sigma}^z_{45} - 4\hat{\sigma}^z_{21}\hat{\sigma}^z_{46} - 4\hat{\sigma}^z_{21}\hat{\sigma}^z_{47} + 2\hat{\sigma}^z_{21}\hat{\sigma}^z_{5} + 2\hat{\sigma}^z_{21}\hat{\sigma}^z_{6} - 4\hat{\sigma}^z_{21}\hat{\sigma}^z_{60} - 4\hat{\sigma}^z_{21}\hat{\sigma}^z_{61} - 4\hat{\sigma}^z_{21}\hat{\sigma}^z_{62} - 4\hat{\sigma}^z_{21}\hat{\sigma}^z_{63} - 4\hat{\sigma}^z_{21}\hat{\sigma}^z_{64} - 4\hat{\sigma}^z_{21}\hat{\sigma}^z_{65} - 4\hat{\sigma}^z_{21}\hat{\sigma}^z_{66} - 4\hat{\sigma}^z_{21}\hat{\sigma}^z_{67} - 4\hat{\sigma}^z_{21}\hat{\sigma}^z_{68} - 360\hat{\sigma}^z_{21}\hat{\sigma}^z_{69} + 2\hat{\sigma}^z_{21}\hat{\sigma}^z_{7} - 4\hat{\sigma}^z_{21}\hat{\sigma}^z_{70} - 4\hat{\sigma}^z_{21}\hat{\sigma}^z_{71} + 2\hat{\sigma}^z_{21}\hat{\sigma}^z_{8} + 180\hat{\sigma}^z_{21}\hat{\sigma}^z_{9} + 404\hat{\sigma}^z_{21} + 4\hat{\sigma}^z_{22}\hat{\sigma}^z_{23} + 2\hat{\sigma}^z_{22}\hat{\sigma}^z_{24} + 2\hat{\sigma}^z_{22}\hat{\sigma}^z_{25} + 2\hat{\sigma}^z_{22}\hat{\sigma}^z_{26} + 2\hat{\sigma}^z_{22}\hat{\sigma}^z_{27} + 2\hat{\sigma}^z_{22}\hat{\sigma}^z_{28} + 2\hat{\sigma}^z_{22}\hat{\sigma}^z_{29} + 2\hat{\sigma}^z_{22}\hat{\sigma}^z_{3} + 2\hat{\sigma}^z_{22}\hat{\sigma}^z_{30} + 2\hat{\sigma}^z_{22}\hat{\sigma}^z_{31} + 2\hat{\sigma}^z_{22}\hat{\sigma}^z_{32} + 2\hat{\sigma}^z_{22}\hat{\sigma}^z_{33} + 180\hat{\sigma}^z_{22}\hat{\sigma}^z_{34} + 2\hat{\sigma}^z_{22}\hat{\sigma}^z_{35} - 4\hat{\sigma}^z_{22}\hat{\sigma}^z_{36} - 4\hat{\sigma}^z_{22}\hat{\sigma}^z_{37} - 4\hat{\sigma}^z_{22}\hat{\sigma}^z_{38} - 4\hat{\sigma}^z_{22}\hat{\sigma}^z_{39} + 2\hat{\sigma}^z_{22}\hat{\sigma}^z_{4} - 4\hat{\sigma}^z_{22}\hat{\sigma}^z_{40} - 4\hat{\sigma}^z_{22}\hat{\sigma}^z_{41} - 4\hat{\sigma}^z_{22}\hat{\sigma}^z_{42} - 4\hat{\sigma}^z_{22}\hat{\sigma}^z_{43} - 4\hat{\sigma}^z_{22}\hat{\sigma}^z_{44} - 4\hat{\sigma}^z_{22}\hat{\sigma}^z_{45} - 360\hat{\sigma}^z_{22}\hat{\sigma}^z_{46} - 4\hat{\sigma}^z_{22}\hat{\sigma}^z_{47} + 2\hat{\sigma}^z_{22}\hat{\sigma}^z_{5} + 2\hat{\sigma}^z_{22}\hat{\sigma}^z_{6} - 4\hat{\sigma}^z_{22}\hat{\sigma}^z_{60} - 4\hat{\sigma}^z_{22}\hat{\sigma}^z_{61} - 4\hat{\sigma}^z_{22}\hat{\sigma}^z_{62} - 4\hat{\sigma}^z_{22}\hat{\sigma}^z_{63} - 4\hat{\sigma}^z_{22}\hat{\sigma}^z_{64} - 4\hat{\sigma}^z_{22}\hat{\sigma}^z_{65} - 4\hat{\sigma}^z_{22}\hat{\sigma}^z_{66} - 4\hat{\sigma}^z_{22}\hat{\sigma}^z_{67} - 4\hat{\sigma}^z_{22}\hat{\sigma}^z_{68} - 4\hat{\sigma}^z_{22}\hat{\sigma}^z_{69} + 2\hat{\sigma}^z_{22}\hat{\sigma}^z_{7} - 360\hat{\sigma}^z_{22}\hat{\sigma}^z_{70} - 4\hat{\sigma}^z_{22}\hat{\sigma}^z_{71} + 2\hat{\sigma}^z_{22}\hat{\sigma}^z_{8} + 2\hat{\sigma}^z_{22}\hat{\sigma}^z_{9} + 404\hat{\sigma}^z_{22} + 2\hat{\sigma}^z_{23}\hat{\sigma}^z_{24} + 2\hat{\sigma}^z_{23}\hat{\sigma}^z_{25} + 2\hat{\sigma}^z_{23}\hat{\sigma}^z_{26} + 2\hat{\sigma}^z_{23}\hat{\sigma}^z_{27} + 2\hat{\sigma}^z_{23}\hat{\sigma}^z_{28} + 2\hat{\sigma}^z_{23}\hat{\sigma}^z_{29} + 2\hat{\sigma}^z_{23}\hat{\sigma}^z_{3} + 2\hat{\sigma}^z_{23}\hat{\sigma}^z_{30} + 2\hat{\sigma}^z_{23}\hat{\sigma}^z_{31} + 2\hat{\sigma}^z_{23}\hat{\sigma}^z_{32} + 2\hat{\sigma}^z_{23}\hat{\sigma}^z_{33} + 2\hat{\sigma}^z_{23}\hat{\sigma}^z_{34} + 180\hat{\sigma}^z_{23}\hat{\sigma}^z_{35} - 4\hat{\sigma}^z_{23}\hat{\sigma}^z_{36} - 4\hat{\sigma}^z_{23}\hat{\sigma}^z_{37} - 4\hat{\sigma}^z_{23}\hat{\sigma}^z_{38} - 4\hat{\sigma}^z_{23}\hat{\sigma}^z_{39} + 2\hat{\sigma}^z_{23}\hat{\sigma}^z_{4} - 4\hat{\sigma}^z_{23}\hat{\sigma}^z_{40} - 4\hat{\sigma}^z_{23}\hat{\sigma}^z_{41} - 4\hat{\sigma}^z_{23}\hat{\sigma}^z_{42} - 4\hat{\sigma}^z_{23}\hat{\sigma}^z_{43} - 4\hat{\sigma}^z_{23}\hat{\sigma}^z_{44} - 4\hat{\sigma}^z_{23}\hat{\sigma}^z_{45} - 4\hat{\sigma}^z_{23}\hat{\sigma}^z_{46} - 360\hat{\sigma}^z_{23}\hat{\sigma}^z_{47} + 2\hat{\sigma}^z_{23}\hat{\sigma}^z_{5} + 2\hat{\sigma}^z_{23}\hat{\sigma}^z_{6} - 4\hat{\sigma}^z_{23}\hat{\sigma}^z_{60} - 4\hat{\sigma}^z_{23}\hat{\sigma}^z_{61} - 4\hat{\sigma}^z_{23}\hat{\sigma}^z_{62} - 4\hat{\sigma}^z_{23}\hat{\sigma}^z_{63} - 4\hat{\sigma}^z_{23}\hat{\sigma}^z_{64} - 4\hat{\sigma}^z_{23}\hat{\sigma}^z_{65} - 4\hat{\sigma}^z_{23}\hat{\sigma}^z_{66} - 4\hat{\sigma}^z_{23}\hat{\sigma}^z_{67} - 4\hat{\sigma}^z_{23}\hat{\sigma}^z_{68} - 4\hat{\sigma}^z_{23}\hat{\sigma}^z_{69} + 2\hat{\sigma}^z_{23}\hat{\sigma}^z_{7} - 4\hat{\sigma}^z_{23}\hat{\sigma}^z_{70} - 360\hat{\sigma}^z_{23}\hat{\sigma}^z_{71} + 2\hat{\sigma}^z_{23}\hat{\sigma}^z_{8} + 2\hat{\sigma}^z_{23}\hat{\sigma}^z_{9} + 404\hat{\sigma}^z_{23} + 4\hat{\sigma}^z_{24}\hat{\sigma}^z_{25} + 4\hat{\sigma}^z_{24}\hat{\sigma}^z_{26} + 4\hat{\sigma}^z_{24}\hat{\sigma}^z_{27} + 4\hat{\sigma}^z_{24}\hat{\sigma}^z_{28} + 4\hat{\sigma}^z_{24}\hat{\sigma}^z_{29} + 2\hat{\sigma}^z_{24}\hat{\sigma}^z_{3} + 4\hat{\sigma}^z_{24}\hat{\sigma}^z_{30} + 4\hat{\sigma}^z_{24}\hat{\sigma}^z_{31} + 4\hat{\sigma}^z_{24}\hat{\sigma}^z_{32} + 4\hat{\sigma}^z_{24}\hat{\sigma}^z_{33} + 4\hat{\sigma}^z_{24}\hat{\sigma}^z_{34} + 4\hat{\sigma}^z_{24}\hat{\sigma}^z_{35} + 2\hat{\sigma}^z_{24}\hat{\sigma}^z_{4} - 360\hat{\sigma}^z_{24}\hat{\sigma}^z_{48} - 4\hat{\sigma}^z_{24}\hat{\sigma}^z_{49} + 2\hat{\sigma}^z_{24}\hat{\sigma}^z_{5} - 4\hat{\sigma}^z_{24}\hat{\sigma}^z_{50} - 4\hat{\sigma}^z_{24}\hat{\sigma}^z_{51} - 4\hat{\sigma}^z_{24}\hat{\sigma}^z_{52} - 4\hat{\sigma}^z_{24}\hat{\sigma}^z_{53} - 4\hat{\sigma}^z_{24}\hat{\sigma}^z_{54} - 4\hat{\sigma}^z_{24}\hat{\sigma}^z_{55} - 4\hat{\sigma}^z_{24}\hat{\sigma}^z_{56} - 4\hat{\sigma}^z_{24}\hat{\sigma}^z_{57} - 4\hat{\sigma}^z_{24}\hat{\sigma}^z_{58} - 4\hat{\sigma}^z_{24}\hat{\sigma}^z_{59} + 2\hat{\sigma}^z_{24}\hat{\sigma}^z_{6} - 360\hat{\sigma}^z_{24}\hat{\sigma}^z_{60} - 4\hat{\sigma}^z_{24}\hat{\sigma}^z_{61} - 4\hat{\sigma}^z_{24}\hat{\sigma}^z_{62} - 4\hat{\sigma}^z_{24}\hat{\sigma}^z_{63} - 4\hat{\sigma}^z_{24}\hat{\sigma}^z_{64} - 4\hat{\sigma}^z_{24}\hat{\sigma}^z_{65} - 4\hat{\sigma}^z_{24}\hat{\sigma}^z_{66} - 4\hat{\sigma}^z_{24}\hat{\sigma}^z_{67} - 4\hat{\sigma}^z_{24}\hat{\sigma}^z_{68} - 4\hat{\sigma}^z_{24}\hat{\sigma}^z_{69} + 2\hat{\sigma}^z_{24}\hat{\sigma}^z_{7} - 4\hat{\sigma}^z_{24}\hat{\sigma}^z_{70} - 4\hat{\sigma}^z_{24}\hat{\sigma}^z_{71} + 2\hat{\sigma}^z_{24}\hat{\sigma}^z_{8} + 2\hat{\sigma}^z_{24}\hat{\sigma}^z_{9} + 404\hat{\sigma}^z_{24} + 4\hat{\sigma}^z_{25}\hat{\sigma}^z_{26} + 4\hat{\sigma}^z_{25}\hat{\sigma}^z_{27} + 4\hat{\sigma}^z_{25}\hat{\sigma}^z_{28} + 4\hat{\sigma}^z_{25}\hat{\sigma}^z_{29} + 2\hat{\sigma}^z_{25}\hat{\sigma}^z_{3} + 4\hat{\sigma}^z_{25}\hat{\sigma}^z_{30} + 4\hat{\sigma}^z_{25}\hat{\sigma}^z_{31} + 4\hat{\sigma}^z_{25}\hat{\sigma}^z_{32} + 4\hat{\sigma}^z_{25}\hat{\sigma}^z_{33} + 4\hat{\sigma}^z_{25}\hat{\sigma}^z_{34} + 4\hat{\sigma}^z_{25}\hat{\sigma}^z_{35} + 2\hat{\sigma}^z_{25}\hat{\sigma}^z_{4} - 4\hat{\sigma}^z_{25}\hat{\sigma}^z_{48} - 360\hat{\sigma}^z_{25}\hat{\sigma}^z_{49} + 2\hat{\sigma}^z_{25}\hat{\sigma}^z_{5} - 4\hat{\sigma}^z_{25}\hat{\sigma}^z_{50} - 4\hat{\sigma}^z_{25}\hat{\sigma}^z_{51} - 4\hat{\sigma}^z_{25}\hat{\sigma}^z_{52} - 4\hat{\sigma}^z_{25}\hat{\sigma}^z_{53} - 4\hat{\sigma}^z_{25}\hat{\sigma}^z_{54} - 4\hat{\sigma}^z_{25}\hat{\sigma}^z_{55} - 4\hat{\sigma}^z_{25}\hat{\sigma}^z_{56} - 4\hat{\sigma}^z_{25}\hat{\sigma}^z_{57} - 4\hat{\sigma}^z_{25}\hat{\sigma}^z_{58} - 4\hat{\sigma}^z_{25}\hat{\sigma}^z_{59} + 2\hat{\sigma}^z_{25}\hat{\sigma}^z_{6} - 4\hat{\sigma}^z_{25}\hat{\sigma}^z_{60} - 360\hat{\sigma}^z_{25}\hat{\sigma}^z_{61} - 4\hat{\sigma}^z_{25}\hat{\sigma}^z_{62} - 4\hat{\sigma}^z_{25}\hat{\sigma}^z_{63} - 4\hat{\sigma}^z_{25}\hat{\sigma}^z_{64} - 4\hat{\sigma}^z_{25}\hat{\sigma}^z_{65} - 4\hat{\sigma}^z_{25}\hat{\sigma}^z_{66} - 4\hat{\sigma}^z_{25}\hat{\sigma}^z_{67} - 4\hat{\sigma}^z_{25}\hat{\sigma}^z_{68} - 4\hat{\sigma}^z_{25}\hat{\sigma}^z_{69} + 2\hat{\sigma}^z_{25}\hat{\sigma}^z_{7} - 4\hat{\sigma}^z_{25}\hat{\sigma}^z_{70} - 4\hat{\sigma}^z_{25}\hat{\sigma}^z_{71} + 2\hat{\sigma}^z_{25}\hat{\sigma}^z_{8} + 2\hat{\sigma}^z_{25}\hat{\sigma}^z_{9} + 404\hat{\sigma}^z_{25} + 4\hat{\sigma}^z_{26}\hat{\sigma}^z_{27} + 4\hat{\sigma}^z_{26}\hat{\sigma}^z_{28} + 4\hat{\sigma}^z_{26}\hat{\sigma}^z_{29} + 2\hat{\sigma}^z_{26}\hat{\sigma}^z_{3} + 4\hat{\sigma}^z_{26}\hat{\sigma}^z_{30} + 4\hat{\sigma}^z_{26}\hat{\sigma}^z_{31} + 4\hat{\sigma}^z_{26}\hat{\sigma}^z_{32} + 4\hat{\sigma}^z_{26}\hat{\sigma}^z_{33} + 4\hat{\sigma}^z_{26}\hat{\sigma}^z_{34} + 4\hat{\sigma}^z_{26}\hat{\sigma}^z_{35} + 2\hat{\sigma}^z_{26}\hat{\sigma}^z_{4} - 4\hat{\sigma}^z_{26}\hat{\sigma}^z_{48} - 4\hat{\sigma}^z_{26}\hat{\sigma}^z_{49} + 2\hat{\sigma}^z_{26}\hat{\sigma}^z_{5} - 360\hat{\sigma}^z_{26}\hat{\sigma}^z_{50} - 4\hat{\sigma}^z_{26}\hat{\sigma}^z_{51} - 4\hat{\sigma}^z_{26}\hat{\sigma}^z_{52} - 4\hat{\sigma}^z_{26}\hat{\sigma}^z_{53} - 4\hat{\sigma}^z_{26}\hat{\sigma}^z_{54} - 4\hat{\sigma}^z_{26}\hat{\sigma}^z_{55} - 4\hat{\sigma}^z_{26}\hat{\sigma}^z_{56} - 4\hat{\sigma}^z_{26}\hat{\sigma}^z_{57} - 4\hat{\sigma}^z_{26}\hat{\sigma}^z_{58} - 4\hat{\sigma}^z_{26}\hat{\sigma}^z_{59} + 2\hat{\sigma}^z_{26}\hat{\sigma}^z_{6} - 4\hat{\sigma}^z_{26}\hat{\sigma}^z_{60} - 4\hat{\sigma}^z_{26}\hat{\sigma}^z_{61} - 360\hat{\sigma}^z_{26}\hat{\sigma}^z_{62} - 4\hat{\sigma}^z_{26}\hat{\sigma}^z_{63} - 4\hat{\sigma}^z_{26}\hat{\sigma}^z_{64} - 4\hat{\sigma}^z_{26}\hat{\sigma}^z_{65} - 4\hat{\sigma}^z_{26}\hat{\sigma}^z_{66} - 4\hat{\sigma}^z_{26}\hat{\sigma}^z_{67} - 4\hat{\sigma}^z_{26}\hat{\sigma}^z_{68} - 4\hat{\sigma}^z_{26}\hat{\sigma}^z_{69} + 2\hat{\sigma}^z_{26}\hat{\sigma}^z_{7} - 4\hat{\sigma}^z_{26}\hat{\sigma}^z_{70} - 4\hat{\sigma}^z_{26}\hat{\sigma}^z_{71} + 2\hat{\sigma}^z_{26}\hat{\sigma}^z_{8} + 2\hat{\sigma}^z_{26}\hat{\sigma}^z_{9} + 404\hat{\sigma}^z_{26} + 4\hat{\sigma}^z_{27}\hat{\sigma}^z_{28} + 4\hat{\sigma}^z_{27}\hat{\sigma}^z_{29} + 180\hat{\sigma}^z_{27}\hat{\sigma}^z_{3} + 4\hat{\sigma}^z_{27}\hat{\sigma}^z_{30} + 4\hat{\sigma}^z_{27}\hat{\sigma}^z_{31} + 4\hat{\sigma}^z_{27}\hat{\sigma}^z_{32} + 4\hat{\sigma}^z_{27}\hat{\sigma}^z_{33} + 4\hat{\sigma}^z_{27}\hat{\sigma}^z_{34} + 4\hat{\sigma}^z_{27}\hat{\sigma}^z_{35} + 2\hat{\sigma}^z_{27}\hat{\sigma}^z_{4} - 4\hat{\sigma}^z_{27}\hat{\sigma}^z_{48} - 4\hat{\sigma}^z_{27}\hat{\sigma}^z_{49} + 2\hat{\sigma}^z_{27}\hat{\sigma}^z_{5} - 4\hat{\sigma}^z_{27}\hat{\sigma}^z_{50} - 360\hat{\sigma}^z_{27}\hat{\sigma}^z_{51} - 4\hat{\sigma}^z_{27}\hat{\sigma}^z_{52} - 4\hat{\sigma}^z_{27}\hat{\sigma}^z_{53} - 4\hat{\sigma}^z_{27}\hat{\sigma}^z_{54} - 4\hat{\sigma}^z_{27}\hat{\sigma}^z_{55} - 4\hat{\sigma}^z_{27}\hat{\sigma}^z_{56} - 4\hat{\sigma}^z_{27}\hat{\sigma}^z_{57} - 4\hat{\sigma}^z_{27}\hat{\sigma}^z_{58} - 4\hat{\sigma}^z_{27}\hat{\sigma}^z_{59} + 2\hat{\sigma}^z_{27}\hat{\sigma}^z_{6} - 4\hat{\sigma}^z_{27}\hat{\sigma}^z_{60} - 4\hat{\sigma}^z_{27}\hat{\sigma}^z_{61} - 4\hat{\sigma}^z_{27}\hat{\sigma}^z_{62} - 360\hat{\sigma}^z_{27}\hat{\sigma}^z_{63} - 4\hat{\sigma}^z_{27}\hat{\sigma}^z_{64} - 4\hat{\sigma}^z_{27}\hat{\sigma}^z_{65} - 4\hat{\sigma}^z_{27}\hat{\sigma}^z_{66} - 4\hat{\sigma}^z_{27}\hat{\sigma}^z_{67} - 4\hat{\sigma}^z_{27}\hat{\sigma}^z_{68} - 4\hat{\sigma}^z_{27}\hat{\sigma}^z_{69} + 2\hat{\sigma}^z_{27}\hat{\sigma}^z_{7} - 4\hat{\sigma}^z_{27}\hat{\sigma}^z_{70} - 4\hat{\sigma}^z_{27}\hat{\sigma}^z_{71} + 2\hat{\sigma}^z_{27}\hat{\sigma}^z_{8} + 2\hat{\sigma}^z_{27}\hat{\sigma}^z_{9} + 404\hat{\sigma}^z_{27} + 4\hat{\sigma}^z_{28}\hat{\sigma}^z_{29} + 2\hat{\sigma}^z_{28}\hat{\sigma}^z_{3} + 4\hat{\sigma}^z_{28}\hat{\sigma}^z_{30} + 4\hat{\sigma}^z_{28}\hat{\sigma}^z_{31} + 4\hat{\sigma}^z_{28}\hat{\sigma}^z_{32} + 4\hat{\sigma}^z_{28}\hat{\sigma}^z_{33} + 4\hat{\sigma}^z_{28}\hat{\sigma}^z_{34} + 4\hat{\sigma}^z_{28}\hat{\sigma}^z_{35} + 180\hat{\sigma}^z_{28}\hat{\sigma}^z_{4} - 4\hat{\sigma}^z_{28}\hat{\sigma}^z_{48} - 4\hat{\sigma}^z_{28}\hat{\sigma}^z_{49} + 2\hat{\sigma}^z_{28}\hat{\sigma}^z_{5} - 4\hat{\sigma}^z_{28}\hat{\sigma}^z_{50} - 4\hat{\sigma}^z_{28}\hat{\sigma}^z_{51} - 360\hat{\sigma}^z_{28}\hat{\sigma}^z_{52} - 4\hat{\sigma}^z_{28}\hat{\sigma}^z_{53} - 4\hat{\sigma}^z_{28}\hat{\sigma}^z_{54} - 4\hat{\sigma}^z_{28}\hat{\sigma}^z_{55} - 4\hat{\sigma}^z_{28}\hat{\sigma}^z_{56} - 4\hat{\sigma}^z_{28}\hat{\sigma}^z_{57} - 4\hat{\sigma}^z_{28}\hat{\sigma}^z_{58} - 4\hat{\sigma}^z_{28}\hat{\sigma}^z_{59} + 2\hat{\sigma}^z_{28}\hat{\sigma}^z_{6} - 4\hat{\sigma}^z_{28}\hat{\sigma}^z_{60} - 4\hat{\sigma}^z_{28}\hat{\sigma}^z_{61} - 4\hat{\sigma}^z_{28}\hat{\sigma}^z_{62} - 4\hat{\sigma}^z_{28}\hat{\sigma}^z_{63} - 360\hat{\sigma}^z_{28}\hat{\sigma}^z_{64} - 4\hat{\sigma}^z_{28}\hat{\sigma}^z_{65} - 4\hat{\sigma}^z_{28}\hat{\sigma}^z_{66} - 4\hat{\sigma}^z_{28}\hat{\sigma}^z_{67} - 4\hat{\sigma}^z_{28}\hat{\sigma}^z_{68} - 4\hat{\sigma}^z_{28}\hat{\sigma}^z_{69} + 2\hat{\sigma}^z_{28}\hat{\sigma}^z_{7} - 4\hat{\sigma}^z_{28}\hat{\sigma}^z_{70} - 4\hat{\sigma}^z_{28}\hat{\sigma}^z_{71} + 2\hat{\sigma}^z_{28}\hat{\sigma}^z_{8} + 2\hat{\sigma}^z_{28}\hat{\sigma}^z_{9} + 404\hat{\sigma}^z_{28} + 2\hat{\sigma}^z_{29}\hat{\sigma}^z_{3} + 4\hat{\sigma}^z_{29}\hat{\sigma}^z_{30} + 4\hat{\sigma}^z_{29}\hat{\sigma}^z_{31} + 4\hat{\sigma}^z_{29}\hat{\sigma}^z_{32} + 4\hat{\sigma}^z_{29}\hat{\sigma}^z_{33} + 4\hat{\sigma}^z_{29}\hat{\sigma}^z_{34} + 4\hat{\sigma}^z_{29}\hat{\sigma}^z_{35} + 2\hat{\sigma}^z_{29}\hat{\sigma}^z_{4} - 4\hat{\sigma}^z_{29}\hat{\sigma}^z_{48} - 4\hat{\sigma}^z_{29}\hat{\sigma}^z_{49} + 180\hat{\sigma}^z_{29}\hat{\sigma}^z_{5} - 4\hat{\sigma}^z_{29}\hat{\sigma}^z_{50} - 4\hat{\sigma}^z_{29}\hat{\sigma}^z_{51} - 4\hat{\sigma}^z_{29}\hat{\sigma}^z_{52} - 360\hat{\sigma}^z_{29}\hat{\sigma}^z_{53} - 4\hat{\sigma}^z_{29}\hat{\sigma}^z_{54} - 4\hat{\sigma}^z_{29}\hat{\sigma}^z_{55} - 4\hat{\sigma}^z_{29}\hat{\sigma}^z_{56} - 4\hat{\sigma}^z_{29}\hat{\sigma}^z_{57} - 4\hat{\sigma}^z_{29}\hat{\sigma}^z_{58} - 4\hat{\sigma}^z_{29}\hat{\sigma}^z_{59} + 2\hat{\sigma}^z_{29}\hat{\sigma}^z_{6} - 4\hat{\sigma}^z_{29}\hat{\sigma}^z_{60} - 4\hat{\sigma}^z_{29}\hat{\sigma}^z_{61} - 4\hat{\sigma}^z_{29}\hat{\sigma}^z_{62} - 4\hat{\sigma}^z_{29}\hat{\sigma}^z_{63} - 4\hat{\sigma}^z_{29}\hat{\sigma}^z_{64} - 360\hat{\sigma}^z_{29}\hat{\sigma}^z_{65} - 4\hat{\sigma}^z_{29}\hat{\sigma}^z_{66} - 4\hat{\sigma}^z_{29}\hat{\sigma}^z_{67} - 4\hat{\sigma}^z_{29}\hat{\sigma}^z_{68} - 4\hat{\sigma}^z_{29}\hat{\sigma}^z_{69} + 2\hat{\sigma}^z_{29}\hat{\sigma}^z_{7} - 4\hat{\sigma}^z_{29}\hat{\sigma}^z_{70} - 4\hat{\sigma}^z_{29}\hat{\sigma}^z_{71} + 2\hat{\sigma}^z_{29}\hat{\sigma}^z_{8} + 2\hat{\sigma}^z_{29}\hat{\sigma}^z_{9} + 404\hat{\sigma}^z_{29} + 2\hat{\sigma}^z_{3}\hat{\sigma}^z_{30} + 2\hat{\sigma}^z_{3}\hat{\sigma}^z_{31} + 2\hat{\sigma}^z_{3}\hat{\sigma}^z_{32} + 2\hat{\sigma}^z_{3}\hat{\sigma}^z_{33} + 2\hat{\sigma}^z_{3}\hat{\sigma}^z_{34} + 2\hat{\sigma}^z_{3}\hat{\sigma}^z_{35} - 4\hat{\sigma}^z_{3}\hat{\sigma}^z_{36} - 4\hat{\sigma}^z_{3}\hat{\sigma}^z_{37} - 4\hat{\sigma}^z_{3}\hat{\sigma}^z_{38} - 360\hat{\sigma}^z_{3}\hat{\sigma}^z_{39} + 4\hat{\sigma}^z_{3}\hat{\sigma}^z_{4} - 4\hat{\sigma}^z_{3}\hat{\sigma}^z_{40} - 4\hat{\sigma}^z_{3}\hat{\sigma}^z_{41} - 4\hat{\sigma}^z_{3}\hat{\sigma}^z_{42} - 4\hat{\sigma}^z_{3}\hat{\sigma}^z_{43} - 4\hat{\sigma}^z_{3}\hat{\sigma}^z_{44} - 4\hat{\sigma}^z_{3}\hat{\sigma}^z_{45} - 4\hat{\sigma}^z_{3}\hat{\sigma}^z_{46} - 4\hat{\sigma}^z_{3}\hat{\sigma}^z_{47} - 4\hat{\sigma}^z_{3}\hat{\sigma}^z_{48} - 4\hat{\sigma}^z_{3}\hat{\sigma}^z_{49} + 4\hat{\sigma}^z_{3}\hat{\sigma}^z_{5} - 4\hat{\sigma}^z_{3}\hat{\sigma}^z_{50} - 360\hat{\sigma}^z_{3}\hat{\sigma}^z_{51} - 4\hat{\sigma}^z_{3}\hat{\sigma}^z_{52} - 4\hat{\sigma}^z_{3}\hat{\sigma}^z_{53} - 4\hat{\sigma}^z_{3}\hat{\sigma}^z_{54} - 4\hat{\sigma}^z_{3}\hat{\sigma}^z_{55} - 4\hat{\sigma}^z_{3}\hat{\sigma}^z_{56} - 4\hat{\sigma}^z_{3}\hat{\sigma}^z_{57} - 4\hat{\sigma}^z_{3}\hat{\sigma}^z_{58} - 4\hat{\sigma}^z_{3}\hat{\sigma}^z_{59} + 4\hat{\sigma}^z_{3}\hat{\sigma}^z_{6} + 4\hat{\sigma}^z_{3}\hat{\sigma}^z_{7} + 4\hat{\sigma}^z_{3}\hat{\sigma}^z_{8} + 4\hat{\sigma}^z_{3}\hat{\sigma}^z_{9} + 404\hat{\sigma}^z_{3} + 4\hat{\sigma}^z_{30}\hat{\sigma}^z_{31} + 4\hat{\sigma}^z_{30}\hat{\sigma}^z_{32} + 4\hat{\sigma}^z_{30}\hat{\sigma}^z_{33} + 4\hat{\sigma}^z_{30}\hat{\sigma}^z_{34} + 4\hat{\sigma}^z_{30}\hat{\sigma}^z_{35} + 2\hat{\sigma}^z_{30}\hat{\sigma}^z_{4} - 4\hat{\sigma}^z_{30}\hat{\sigma}^z_{48} - 4\hat{\sigma}^z_{30}\hat{\sigma}^z_{49} + 2\hat{\sigma}^z_{30}\hat{\sigma}^z_{5} - 4\hat{\sigma}^z_{30}\hat{\sigma}^z_{50} - 4\hat{\sigma}^z_{30}\hat{\sigma}^z_{51} - 4\hat{\sigma}^z_{30}\hat{\sigma}^z_{52} - 4\hat{\sigma}^z_{30}\hat{\sigma}^z_{53} - 360\hat{\sigma}^z_{30}\hat{\sigma}^z_{54} - 4\hat{\sigma}^z_{30}\hat{\sigma}^z_{55} - 4\hat{\sigma}^z_{30}\hat{\sigma}^z_{56} - 4\hat{\sigma}^z_{30}\hat{\sigma}^z_{57} - 4\hat{\sigma}^z_{30}\hat{\sigma}^z_{58} - 4\hat{\sigma}^z_{30}\hat{\sigma}^z_{59} + 180\hat{\sigma}^z_{30}\hat{\sigma}^z_{6} - 4\hat{\sigma}^z_{30}\hat{\sigma}^z_{60} - 4\hat{\sigma}^z_{30}\hat{\sigma}^z_{61} - 4\hat{\sigma}^z_{30}\hat{\sigma}^z_{62} - 4\hat{\sigma}^z_{30}\hat{\sigma}^z_{63} - 4\hat{\sigma}^z_{30}\hat{\sigma}^z_{64} - 4\hat{\sigma}^z_{30}\hat{\sigma}^z_{65} - 360\hat{\sigma}^z_{30}\hat{\sigma}^z_{66} - 4\hat{\sigma}^z_{30}\hat{\sigma}^z_{67} - 4\hat{\sigma}^z_{30}\hat{\sigma}^z_{68} - 4\hat{\sigma}^z_{30}\hat{\sigma}^z_{69} + 2\hat{\sigma}^z_{30}\hat{\sigma}^z_{7} - 4\hat{\sigma}^z_{30}\hat{\sigma}^z_{70} - 4\hat{\sigma}^z_{30}\hat{\sigma}^z_{71} + 2\hat{\sigma}^z_{30}\hat{\sigma}^z_{8} + 2\hat{\sigma}^z_{30}\hat{\sigma}^z_{9} + 404\hat{\sigma}^z_{30} + 4\hat{\sigma}^z_{31}\hat{\sigma}^z_{32} + 4\hat{\sigma}^z_{31}\hat{\sigma}^z_{33} + 4\hat{\sigma}^z_{31}\hat{\sigma}^z_{34} + 4\hat{\sigma}^z_{31}\hat{\sigma}^z_{35} + 2\hat{\sigma}^z_{31}\hat{\sigma}^z_{4} - 4\hat{\sigma}^z_{31}\hat{\sigma}^z_{48} - 4\hat{\sigma}^z_{31}\hat{\sigma}^z_{49} + 2\hat{\sigma}^z_{31}\hat{\sigma}^z_{5} - 4\hat{\sigma}^z_{31}\hat{\sigma}^z_{50} - 4\hat{\sigma}^z_{31}\hat{\sigma}^z_{51} - 4\hat{\sigma}^z_{31}\hat{\sigma}^z_{52} - 4\hat{\sigma}^z_{31}\hat{\sigma}^z_{53} - 4\hat{\sigma}^z_{31}\hat{\sigma}^z_{54} - 360\hat{\sigma}^z_{31}\hat{\sigma}^z_{55} - 4\hat{\sigma}^z_{31}\hat{\sigma}^z_{56} - 4\hat{\sigma}^z_{31}\hat{\sigma}^z_{57} - 4\hat{\sigma}^z_{31}\hat{\sigma}^z_{58} - 4\hat{\sigma}^z_{31}\hat{\sigma}^z_{59} + 2\hat{\sigma}^z_{31}\hat{\sigma}^z_{6} - 4\hat{\sigma}^z_{31}\hat{\sigma}^z_{60} - 4\hat{\sigma}^z_{31}\hat{\sigma}^z_{61} - 4\hat{\sigma}^z_{31}\hat{\sigma}^z_{62} - 4\hat{\sigma}^z_{31}\hat{\sigma}^z_{63} - 4\hat{\sigma}^z_{31}\hat{\sigma}^z_{64} - 4\hat{\sigma}^z_{31}\hat{\sigma}^z_{65} - 4\hat{\sigma}^z_{31}\hat{\sigma}^z_{66} - 360\hat{\sigma}^z_{31}\hat{\sigma}^z_{67} - 4\hat{\sigma}^z_{31}\hat{\sigma}^z_{68} - 4\hat{\sigma}^z_{31}\hat{\sigma}^z_{69} + 180\hat{\sigma}^z_{31}\hat{\sigma}^z_{7} - 4\hat{\sigma}^z_{31}\hat{\sigma}^z_{70} - 4\hat{\sigma}^z_{31}\hat{\sigma}^z_{71} + 2\hat{\sigma}^z_{31}\hat{\sigma}^z_{8} + 2\hat{\sigma}^z_{31}\hat{\sigma}^z_{9} + 404\hat{\sigma}^z_{31} + 4\hat{\sigma}^z_{32}\hat{\sigma}^z_{33} + 4\hat{\sigma}^z_{32}\hat{\sigma}^z_{34} + 4\hat{\sigma}^z_{32}\hat{\sigma}^z_{35} + 2\hat{\sigma}^z_{32}\hat{\sigma}^z_{4} - 4\hat{\sigma}^z_{32}\hat{\sigma}^z_{48} - 4\hat{\sigma}^z_{32}\hat{\sigma}^z_{49} + 2\hat{\sigma}^z_{32}\hat{\sigma}^z_{5} - 4\hat{\sigma}^z_{32}\hat{\sigma}^z_{50} - 4\hat{\sigma}^z_{32}\hat{\sigma}^z_{51} - 4\hat{\sigma}^z_{32}\hat{\sigma}^z_{52} - 4\hat{\sigma}^z_{32}\hat{\sigma}^z_{53} - 4\hat{\sigma}^z_{32}\hat{\sigma}^z_{54} - 4\hat{\sigma}^z_{32}\hat{\sigma}^z_{55} - 360\hat{\sigma}^z_{32}\hat{\sigma}^z_{56} - 4\hat{\sigma}^z_{32}\hat{\sigma}^z_{57} - 4\hat{\sigma}^z_{32}\hat{\sigma}^z_{58} - 4\hat{\sigma}^z_{32}\hat{\sigma}^z_{59} + 2\hat{\sigma}^z_{32}\hat{\sigma}^z_{6} - 4\hat{\sigma}^z_{32}\hat{\sigma}^z_{60} - 4\hat{\sigma}^z_{32}\hat{\sigma}^z_{61} - 4\hat{\sigma}^z_{32}\hat{\sigma}^z_{62} - 4\hat{\sigma}^z_{32}\hat{\sigma}^z_{63} - 4\hat{\sigma}^z_{32}\hat{\sigma}^z_{64} - 4\hat{\sigma}^z_{32}\hat{\sigma}^z_{65} - 4\hat{\sigma}^z_{32}\hat{\sigma}^z_{66} - 4\hat{\sigma}^z_{32}\hat{\sigma}^z_{67} - 360\hat{\sigma}^z_{32}\hat{\sigma}^z_{68} - 4\hat{\sigma}^z_{32}\hat{\sigma}^z_{69} + 2\hat{\sigma}^z_{32}\hat{\sigma}^z_{7} - 4\hat{\sigma}^z_{32}\hat{\sigma}^z_{70} - 4\hat{\sigma}^z_{32}\hat{\sigma}^z_{71} + 180\hat{\sigma}^z_{32}\hat{\sigma}^z_{8} + 2\hat{\sigma}^z_{32}\hat{\sigma}^z_{9} + 404\hat{\sigma}^z_{32} + 4\hat{\sigma}^z_{33}\hat{\sigma}^z_{34} + 4\hat{\sigma}^z_{33}\hat{\sigma}^z_{35} + 2\hat{\sigma}^z_{33}\hat{\sigma}^z_{4} - 4\hat{\sigma}^z_{33}\hat{\sigma}^z_{48} - 4\hat{\sigma}^z_{33}\hat{\sigma}^z_{49} + 2\hat{\sigma}^z_{33}\hat{\sigma}^z_{5} - 4\hat{\sigma}^z_{33}\hat{\sigma}^z_{50} - 4\hat{\sigma}^z_{33}\hat{\sigma}^z_{51} - 4\hat{\sigma}^z_{33}\hat{\sigma}^z_{52} - 4\hat{\sigma}^z_{33}\hat{\sigma}^z_{53} - 4\hat{\sigma}^z_{33}\hat{\sigma}^z_{54} - 4\hat{\sigma}^z_{33}\hat{\sigma}^z_{55} - 4\hat{\sigma}^z_{33}\hat{\sigma}^z_{56} - 360\hat{\sigma}^z_{33}\hat{\sigma}^z_{57} - 4\hat{\sigma}^z_{33}\hat{\sigma}^z_{58} - 4\hat{\sigma}^z_{33}\hat{\sigma}^z_{59} + 2\hat{\sigma}^z_{33}\hat{\sigma}^z_{6} - 4\hat{\sigma}^z_{33}\hat{\sigma}^z_{60} - 4\hat{\sigma}^z_{33}\hat{\sigma}^z_{61} - 4\hat{\sigma}^z_{33}\hat{\sigma}^z_{62} - 4\hat{\sigma}^z_{33}\hat{\sigma}^z_{63} - 4\hat{\sigma}^z_{33}\hat{\sigma}^z_{64} - 4\hat{\sigma}^z_{33}\hat{\sigma}^z_{65} - 4\hat{\sigma}^z_{33}\hat{\sigma}^z_{66} - 4\hat{\sigma}^z_{33}\hat{\sigma}^z_{67} - 4\hat{\sigma}^z_{33}\hat{\sigma}^z_{68} - 360\hat{\sigma}^z_{33}\hat{\sigma}^z_{69} + 2\hat{\sigma}^z_{33}\hat{\sigma}^z_{7} - 4\hat{\sigma}^z_{33}\hat{\sigma}^z_{70} - 4\hat{\sigma}^z_{33}\hat{\sigma}^z_{71} + 2\hat{\sigma}^z_{33}\hat{\sigma}^z_{8} + 180\hat{\sigma}^z_{33}\hat{\sigma}^z_{9} + 404\hat{\sigma}^z_{33} + 4\hat{\sigma}^z_{34}\hat{\sigma}^z_{35} + 2\hat{\sigma}^z_{34}\hat{\sigma}^z_{4} - 4\hat{\sigma}^z_{34}\hat{\sigma}^z_{48} - 4\hat{\sigma}^z_{34}\hat{\sigma}^z_{49} + 2\hat{\sigma}^z_{34}\hat{\sigma}^z_{5} - 4\hat{\sigma}^z_{34}\hat{\sigma}^z_{50} - 4\hat{\sigma}^z_{34}\hat{\sigma}^z_{51} - 4\hat{\sigma}^z_{34}\hat{\sigma}^z_{52} - 4\hat{\sigma}^z_{34}\hat{\sigma}^z_{53} - 4\hat{\sigma}^z_{34}\hat{\sigma}^z_{54} - 4\hat{\sigma}^z_{34}\hat{\sigma}^z_{55} - 4\hat{\sigma}^z_{34}\hat{\sigma}^z_{56} - 4\hat{\sigma}^z_{34}\hat{\sigma}^z_{57} - 360\hat{\sigma}^z_{34}\hat{\sigma}^z_{58} - 4\hat{\sigma}^z_{34}\hat{\sigma}^z_{59} + 2\hat{\sigma}^z_{34}\hat{\sigma}^z_{6} - 4\hat{\sigma}^z_{34}\hat{\sigma}^z_{60} - 4\hat{\sigma}^z_{34}\hat{\sigma}^z_{61} - 4\hat{\sigma}^z_{34}\hat{\sigma}^z_{62} - 4\hat{\sigma}^z_{34}\hat{\sigma}^z_{63} - 4\hat{\sigma}^z_{34}\hat{\sigma}^z_{64} - 4\hat{\sigma}^z_{34}\hat{\sigma}^z_{65} - 4\hat{\sigma}^z_{34}\hat{\sigma}^z_{66} - 4\hat{\sigma}^z_{34}\hat{\sigma}^z_{67} - 4\hat{\sigma}^z_{34}\hat{\sigma}^z_{68} - 4\hat{\sigma}^z_{34}\hat{\sigma}^z_{69} + 2\hat{\sigma}^z_{34}\hat{\sigma}^z_{7} - 360\hat{\sigma}^z_{34}\hat{\sigma}^z_{70} - 4\hat{\sigma}^z_{34}\hat{\sigma}^z_{71} + 2\hat{\sigma}^z_{34}\hat{\sigma}^z_{8} + 2\hat{\sigma}^z_{34}\hat{\sigma}^z_{9} + 404\hat{\sigma}^z_{34} + 2\hat{\sigma}^z_{35}\hat{\sigma}^z_{4} - 4\hat{\sigma}^z_{35}\hat{\sigma}^z_{48} - 4\hat{\sigma}^z_{35}\hat{\sigma}^z_{49} + 2\hat{\sigma}^z_{35}\hat{\sigma}^z_{5} - 4\hat{\sigma}^z_{35}\hat{\sigma}^z_{50} - 4\hat{\sigma}^z_{35}\hat{\sigma}^z_{51} - 4\hat{\sigma}^z_{35}\hat{\sigma}^z_{52} - 4\hat{\sigma}^z_{35}\hat{\sigma}^z_{53} - 4\hat{\sigma}^z_{35}\hat{\sigma}^z_{54} - 4\hat{\sigma}^z_{35}\hat{\sigma}^z_{55} - 4\hat{\sigma}^z_{35}\hat{\sigma}^z_{56} - 4\hat{\sigma}^z_{35}\hat{\sigma}^z_{57} - 4\hat{\sigma}^z_{35}\hat{\sigma}^z_{58} - 360\hat{\sigma}^z_{35}\hat{\sigma}^z_{59} + 2\hat{\sigma}^z_{35}\hat{\sigma}^z_{6} - 4\hat{\sigma}^z_{35}\hat{\sigma}^z_{60} - 4\hat{\sigma}^z_{35}\hat{\sigma}^z_{61} - 4\hat{\sigma}^z_{35}\hat{\sigma}^z_{62} - 4\hat{\sigma}^z_{35}\hat{\sigma}^z_{63} - 4\hat{\sigma}^z_{35}\hat{\sigma}^z_{64} - 4\hat{\sigma}^z_{35}\hat{\sigma}^z_{65} - 4\hat{\sigma}^z_{35}\hat{\sigma}^z_{66} - 4\hat{\sigma}^z_{35}\hat{\sigma}^z_{67} - 4\hat{\sigma}^z_{35}\hat{\sigma}^z_{68} - 4\hat{\sigma}^z_{35}\hat{\sigma}^z_{69} + 2\hat{\sigma}^z_{35}\hat{\sigma}^z_{7} - 4\hat{\sigma}^z_{35}\hat{\sigma}^z_{70} - 360\hat{\sigma}^z_{35}\hat{\sigma}^z_{71} + 2\hat{\sigma}^z_{35}\hat{\sigma}^z_{8} + 2\hat{\sigma}^z_{35}\hat{\sigma}^z_{9} + 404\hat{\sigma}^z_{35} + 8\hat{\sigma}^z_{36}\hat{\sigma}^z_{37} + 8\hat{\sigma}^z_{36}\hat{\sigma}^z_{38} + 8\hat{\sigma}^z_{36}\hat{\sigma}^z_{39} - 4\hat{\sigma}^z_{36}\hat{\sigma}^z_{4} + 8\hat{\sigma}^z_{36}\hat{\sigma}^z_{40} + 8\hat{\sigma}^z_{36}\hat{\sigma}^z_{41} + 8\hat{\sigma}^z_{36}\hat{\sigma}^z_{42} + 8\hat{\sigma}^z_{36}\hat{\sigma}^z_{43} + 8\hat{\sigma}^z_{36}\hat{\sigma}^z_{44} + 8\hat{\sigma}^z_{36}\hat{\sigma}^z_{45} + 8\hat{\sigma}^z_{36}\hat{\sigma}^z_{46} + 8\hat{\sigma}^z_{36}\hat{\sigma}^z_{47} - 4\hat{\sigma}^z_{36}\hat{\sigma}^z_{5} - 4\hat{\sigma}^z_{36}\hat{\sigma}^z_{6} - 4\hat{\sigma}^z_{36}\hat{\sigma}^z_{7} - 4\hat{\sigma}^z_{36}\hat{\sigma}^z_{8} - 4\hat{\sigma}^z_{36}\hat{\sigma}^z_{9} - 404\hat{\sigma}^z_{36} + 8\hat{\sigma}^z_{37}\hat{\sigma}^z_{38} + 8\hat{\sigma}^z_{37}\hat{\sigma}^z_{39} - 4\hat{\sigma}^z_{37}\hat{\sigma}^z_{4} + 8\hat{\sigma}^z_{37}\hat{\sigma}^z_{40} + 8\hat{\sigma}^z_{37}\hat{\sigma}^z_{41} + 8\hat{\sigma}^z_{37}\hat{\sigma}^z_{42} + 8\hat{\sigma}^z_{37}\hat{\sigma}^z_{43} + 8\hat{\sigma}^z_{37}\hat{\sigma}^z_{44} + 8\hat{\sigma}^z_{37}\hat{\sigma}^z_{45} + 8\hat{\sigma}^z_{37}\hat{\sigma}^z_{46} + 8\hat{\sigma}^z_{37}\hat{\sigma}^z_{47} - 4\hat{\sigma}^z_{37}\hat{\sigma}^z_{5} - 4\hat{\sigma}^z_{37}\hat{\sigma}^z_{6} - 4\hat{\sigma}^z_{37}\hat{\sigma}^z_{7} - 4\hat{\sigma}^z_{37}\hat{\sigma}^z_{8} - 4\hat{\sigma}^z_{37}\hat{\sigma}^z_{9} - 404\hat{\sigma}^z_{37} + 8\hat{\sigma}^z_{38}\hat{\sigma}^z_{39} - 4\hat{\sigma}^z_{38}\hat{\sigma}^z_{4} + 8\hat{\sigma}^z_{38}\hat{\sigma}^z_{40} + 8\hat{\sigma}^z_{38}\hat{\sigma}^z_{41} + 8\hat{\sigma}^z_{38}\hat{\sigma}^z_{42} + 8\hat{\sigma}^z_{38}\hat{\sigma}^z_{43} + 8\hat{\sigma}^z_{38}\hat{\sigma}^z_{44} + 8\hat{\sigma}^z_{38}\hat{\sigma}^z_{45} + 8\hat{\sigma}^z_{38}\hat{\sigma}^z_{46} + 8\hat{\sigma}^z_{38}\hat{\sigma}^z_{47} - 4\hat{\sigma}^z_{38}\hat{\sigma}^z_{5} - 4\hat{\sigma}^z_{38}\hat{\sigma}^z_{6} - 4\hat{\sigma}^z_{38}\hat{\sigma}^z_{7} - 4\hat{\sigma}^z_{38}\hat{\sigma}^z_{8} - 4\hat{\sigma}^z_{38}\hat{\sigma}^z_{9} - 404\hat{\sigma}^z_{38} - 4\hat{\sigma}^z_{39}\hat{\sigma}^z_{4} + 8\hat{\sigma}^z_{39}\hat{\sigma}^z_{40} + 8\hat{\sigma}^z_{39}\hat{\sigma}^z_{41} + 8\hat{\sigma}^z_{39}\hat{\sigma}^z_{42} + 8\hat{\sigma}^z_{39}\hat{\sigma}^z_{43} + 8\hat{\sigma}^z_{39}\hat{\sigma}^z_{44} + 8\hat{\sigma}^z_{39}\hat{\sigma}^z_{45} + 8\hat{\sigma}^z_{39}\hat{\sigma}^z_{46} + 8\hat{\sigma}^z_{39}\hat{\sigma}^z_{47} - 4\hat{\sigma}^z_{39}\hat{\sigma}^z_{5} - 4\hat{\sigma}^z_{39}\hat{\sigma}^z_{6} - 4\hat{\sigma}^z_{39}\hat{\sigma}^z_{7} - 4\hat{\sigma}^z_{39}\hat{\sigma}^z_{8} - 4\hat{\sigma}^z_{39}\hat{\sigma}^z_{9} - 404\hat{\sigma}^z_{39} - 360\hat{\sigma}^z_{4}\hat{\sigma}^z_{40} - 4\hat{\sigma}^z_{4}\hat{\sigma}^z_{41} - 4\hat{\sigma}^z_{4}\hat{\sigma}^z_{42} - 4\hat{\sigma}^z_{4}\hat{\sigma}^z_{43} - 4\hat{\sigma}^z_{4}\hat{\sigma}^z_{44} - 4\hat{\sigma}^z_{4}\hat{\sigma}^z_{45} - 4\hat{\sigma}^z_{4}\hat{\sigma}^z_{46} - 4\hat{\sigma}^z_{4}\hat{\sigma}^z_{47} - 4\hat{\sigma}^z_{4}\hat{\sigma}^z_{48} - 4\hat{\sigma}^z_{4}\hat{\sigma}^z_{49} + 4\hat{\sigma}^z_{4}\hat{\sigma}^z_{5} - 4\hat{\sigma}^z_{4}\hat{\sigma}^z_{50} - 4\hat{\sigma}^z_{4}\hat{\sigma}^z_{51} - 360\hat{\sigma}^z_{4}\hat{\sigma}^z_{52} - 4\hat{\sigma}^z_{4}\hat{\sigma}^z_{53} - 4\hat{\sigma}^z_{4}\hat{\sigma}^z_{54} - 4\hat{\sigma}^z_{4}\hat{\sigma}^z_{55} - 4\hat{\sigma}^z_{4}\hat{\sigma}^z_{56} - 4\hat{\sigma}^z_{4}\hat{\sigma}^z_{57} - 4\hat{\sigma}^z_{4}\hat{\sigma}^z_{58} - 4\hat{\sigma}^z_{4}\hat{\sigma}^z_{59} + 4\hat{\sigma}^z_{4}\hat{\sigma}^z_{6} + 4\hat{\sigma}^z_{4}\hat{\sigma}^z_{7} + 4\hat{\sigma}^z_{4}\hat{\sigma}^z_{8} + 4\hat{\sigma}^z_{4}\hat{\sigma}^z_{9} + 404\hat{\sigma}^z_{4} + 8\hat{\sigma}^z_{40}\hat{\sigma}^z_{41} + 8\hat{\sigma}^z_{40}\hat{\sigma}^z_{42} + 8\hat{\sigma}^z_{40}\hat{\sigma}^z_{43} + 8\hat{\sigma}^z_{40}\hat{\sigma}^z_{44} + 8\hat{\sigma}^z_{40}\hat{\sigma}^z_{45} + 8\hat{\sigma}^z_{40}\hat{\sigma}^z_{46} + 8\hat{\sigma}^z_{40}\hat{\sigma}^z_{47} - 4\hat{\sigma}^z_{40}\hat{\sigma}^z_{5} - 4\hat{\sigma}^z_{40}\hat{\sigma}^z_{6} - 4\hat{\sigma}^z_{40}\hat{\sigma}^z_{7} - 4\hat{\sigma}^z_{40}\hat{\sigma}^z_{8} - 4\hat{\sigma}^z_{40}\hat{\sigma}^z_{9} - 404\hat{\sigma}^z_{40} + 8\hat{\sigma}^z_{41}\hat{\sigma}^z_{42} + 8\hat{\sigma}^z_{41}\hat{\sigma}^z_{43} + 8\hat{\sigma}^z_{41}\hat{\sigma}^z_{44} + 8\hat{\sigma}^z_{41}\hat{\sigma}^z_{45} + 8\hat{\sigma}^z_{41}\hat{\sigma}^z_{46} + 8\hat{\sigma}^z_{41}\hat{\sigma}^z_{47} - 360\hat{\sigma}^z_{41}\hat{\sigma}^z_{5} - 4\hat{\sigma}^z_{41}\hat{\sigma}^z_{6} - 4\hat{\sigma}^z_{41}\hat{\sigma}^z_{7} - 4\hat{\sigma}^z_{41}\hat{\sigma}^z_{8} - 4\hat{\sigma}^z_{41}\hat{\sigma}^z_{9} - 404\hat{\sigma}^z_{41} + 8\hat{\sigma}^z_{42}\hat{\sigma}^z_{43} + 8\hat{\sigma}^z_{42}\hat{\sigma}^z_{44} + 8\hat{\sigma}^z_{42}\hat{\sigma}^z_{45} + 8\hat{\sigma}^z_{42}\hat{\sigma}^z_{46} + 8\hat{\sigma}^z_{42}\hat{\sigma}^z_{47} - 4\hat{\sigma}^z_{42}\hat{\sigma}^z_{5} - 360\hat{\sigma}^z_{42}\hat{\sigma}^z_{6} - 4\hat{\sigma}^z_{42}\hat{\sigma}^z_{7} - 4\hat{\sigma}^z_{42}\hat{\sigma}^z_{8} - 4\hat{\sigma}^z_{42}\hat{\sigma}^z_{9} - 404\hat{\sigma}^z_{42} + 8\hat{\sigma}^z_{43}\hat{\sigma}^z_{44} + 8\hat{\sigma}^z_{43}\hat{\sigma}^z_{45} + 8\hat{\sigma}^z_{43}\hat{\sigma}^z_{46} + 8\hat{\sigma}^z_{43}\hat{\sigma}^z_{47} - 4\hat{\sigma}^z_{43}\hat{\sigma}^z_{5} - 4\hat{\sigma}^z_{43}\hat{\sigma}^z_{6} - 360\hat{\sigma}^z_{43}\hat{\sigma}^z_{7} - 4\hat{\sigma}^z_{43}\hat{\sigma}^z_{8} - 4\hat{\sigma}^z_{43}\hat{\sigma}^z_{9} - 404\hat{\sigma}^z_{43} + 8\hat{\sigma}^z_{44}\hat{\sigma}^z_{45} + 8\hat{\sigma}^z_{44}\hat{\sigma}^z_{46} + 8\hat{\sigma}^z_{44}\hat{\sigma}^z_{47} - 4\hat{\sigma}^z_{44}\hat{\sigma}^z_{5} - 4\hat{\sigma}^z_{44}\hat{\sigma}^z_{6} - 4\hat{\sigma}^z_{44}\hat{\sigma}^z_{7} - 360\hat{\sigma}^z_{44}\hat{\sigma}^z_{8} - 4\hat{\sigma}^z_{44}\hat{\sigma}^z_{9} - 404\hat{\sigma}^z_{44} + 8\hat{\sigma}^z_{45}\hat{\sigma}^z_{46} + 8\hat{\sigma}^z_{45}\hat{\sigma}^z_{47} - 4\hat{\sigma}^z_{45}\hat{\sigma}^z_{5} - 4\hat{\sigma}^z_{45}\hat{\sigma}^z_{6} - 4\hat{\sigma}^z_{45}\hat{\sigma}^z_{7} - 4\hat{\sigma}^z_{45}\hat{\sigma}^z_{8} - 360\hat{\sigma}^z_{45}\hat{\sigma}^z_{9} - 404\hat{\sigma}^z_{45} + 8\hat{\sigma}^z_{46}\hat{\sigma}^z_{47} - 4\hat{\sigma}^z_{46}\hat{\sigma}^z_{5} - 4\hat{\sigma}^z_{46}\hat{\sigma}^z_{6} - 4\hat{\sigma}^z_{46}\hat{\sigma}^z_{7} - 4\hat{\sigma}^z_{46}\hat{\sigma}^z_{8} - 4\hat{\sigma}^z_{46}\hat{\sigma}^z_{9} - 404\hat{\sigma}^z_{46} - 4\hat{\sigma}^z_{47}\hat{\sigma}^z_{5} - 4\hat{\sigma}^z_{47}\hat{\sigma}^z_{6} - 4\hat{\sigma}^z_{47}\hat{\sigma}^z_{7} - 4\hat{\sigma}^z_{47}\hat{\sigma}^z_{8} - 4\hat{\sigma}^z_{47}\hat{\sigma}^z_{9} - 404\hat{\sigma}^z_{47} + 8\hat{\sigma}^z_{48}\hat{\sigma}^z_{49} - 4\hat{\sigma}^z_{48}\hat{\sigma}^z_{5} + 8\hat{\sigma}^z_{48}\hat{\sigma}^z_{50} + 8\hat{\sigma}^z_{48}\hat{\sigma}^z_{51} + 8\hat{\sigma}^z_{48}\hat{\sigma}^z_{52} + 8\hat{\sigma}^z_{48}\hat{\sigma}^z_{53} + 8\hat{\sigma}^z_{48}\hat{\sigma}^z_{54} + 8\hat{\sigma}^z_{48}\hat{\sigma}^z_{55} + 8\hat{\sigma}^z_{48}\hat{\sigma}^z_{56} + 8\hat{\sigma}^z_{48}\hat{\sigma}^z_{57} + 8\hat{\sigma}^z_{48}\hat{\sigma}^z_{58} + 8\hat{\sigma}^z_{48}\hat{\sigma}^z_{59} - 4\hat{\sigma}^z_{48}\hat{\sigma}^z_{6} - 4\hat{\sigma}^z_{48}\hat{\sigma}^z_{7} - 4\hat{\sigma}^z_{48}\hat{\sigma}^z_{8} - 4\hat{\sigma}^z_{48}\hat{\sigma}^z_{9} - 404\hat{\sigma}^z_{48} - 4\hat{\sigma}^z_{49}\hat{\sigma}^z_{5} + 8\hat{\sigma}^z_{49}\hat{\sigma}^z_{50} + 8\hat{\sigma}^z_{49}\hat{\sigma}^z_{51} + 8\hat{\sigma}^z_{49}\hat{\sigma}^z_{52} + 8\hat{\sigma}^z_{49}\hat{\sigma}^z_{53} + 8\hat{\sigma}^z_{49}\hat{\sigma}^z_{54} + 8\hat{\sigma}^z_{49}\hat{\sigma}^z_{55} + 8\hat{\sigma}^z_{49}\hat{\sigma}^z_{56} + 8\hat{\sigma}^z_{49}\hat{\sigma}^z_{57} + 8\hat{\sigma}^z_{49}\hat{\sigma}^z_{58} + 8\hat{\sigma}^z_{49}\hat{\sigma}^z_{59} - 4\hat{\sigma}^z_{49}\hat{\sigma}^z_{6} - 4\hat{\sigma}^z_{49}\hat{\sigma}^z_{7} - 4\hat{\sigma}^z_{49}\hat{\sigma}^z_{8} - 4\hat{\sigma}^z_{49}\hat{\sigma}^z_{9} - 404\hat{\sigma}^z_{49} - 4\hat{\sigma}^z_{5}\hat{\sigma}^z_{50} - 4\hat{\sigma}^z_{5}\hat{\sigma}^z_{51} - 4\hat{\sigma}^z_{5}\hat{\sigma}^z_{52} - 360\hat{\sigma}^z_{5}\hat{\sigma}^z_{53} - 4\hat{\sigma}^z_{5}\hat{\sigma}^z_{54} - 4\hat{\sigma}^z_{5}\hat{\sigma}^z_{55} - 4\hat{\sigma}^z_{5}\hat{\sigma}^z_{56} - 4\hat{\sigma}^z_{5}\hat{\sigma}^z_{57} - 4\hat{\sigma}^z_{5}\hat{\sigma}^z_{58} - 4\hat{\sigma}^z_{5}\hat{\sigma}^z_{59} + 4\hat{\sigma}^z_{5}\hat{\sigma}^z_{6} + 4\hat{\sigma}^z_{5}\hat{\sigma}^z_{7} + 4\hat{\sigma}^z_{5}\hat{\sigma}^z_{8} + 4\hat{\sigma}^z_{5}\hat{\sigma}^z_{9} + 404\hat{\sigma}^z_{5} + 8\hat{\sigma}^z_{50}\hat{\sigma}^z_{51} + 8\hat{\sigma}^z_{50}\hat{\sigma}^z_{52} + 8\hat{\sigma}^z_{50}\hat{\sigma}^z_{53} + 8\hat{\sigma}^z_{50}\hat{\sigma}^z_{54} + 8\hat{\sigma}^z_{50}\hat{\sigma}^z_{55} + 8\hat{\sigma}^z_{50}\hat{\sigma}^z_{56} + 8\hat{\sigma}^z_{50}\hat{\sigma}^z_{57} + 8\hat{\sigma}^z_{50}\hat{\sigma}^z_{58} + 8\hat{\sigma}^z_{50}\hat{\sigma}^z_{59} - 4\hat{\sigma}^z_{50}\hat{\sigma}^z_{6} - 4\hat{\sigma}^z_{50}\hat{\sigma}^z_{7} - 4\hat{\sigma}^z_{50}\hat{\sigma}^z_{8} - 4\hat{\sigma}^z_{50}\hat{\sigma}^z_{9} - 404\hat{\sigma}^z_{50} + 8\hat{\sigma}^z_{51}\hat{\sigma}^z_{52} + 8\hat{\sigma}^z_{51}\hat{\sigma}^z_{53} + 8\hat{\sigma}^z_{51}\hat{\sigma}^z_{54} + 8\hat{\sigma}^z_{51}\hat{\sigma}^z_{55} + 8\hat{\sigma}^z_{51}\hat{\sigma}^z_{56} + 8\hat{\sigma}^z_{51}\hat{\sigma}^z_{57} + 8\hat{\sigma}^z_{51}\hat{\sigma}^z_{58} + 8\hat{\sigma}^z_{51}\hat{\sigma}^z_{59} - 4\hat{\sigma}^z_{51}\hat{\sigma}^z_{6} - 4\hat{\sigma}^z_{51}\hat{\sigma}^z_{7} - 4\hat{\sigma}^z_{51}\hat{\sigma}^z_{8} - 4\hat{\sigma}^z_{51}\hat{\sigma}^z_{9} - 404\hat{\sigma}^z_{51} + 8\hat{\sigma}^z_{52}\hat{\sigma}^z_{53} + 8\hat{\sigma}^z_{52}\hat{\sigma}^z_{54} + 8\hat{\sigma}^z_{52}\hat{\sigma}^z_{55} + 8\hat{\sigma}^z_{52}\hat{\sigma}^z_{56} + 8\hat{\sigma}^z_{52}\hat{\sigma}^z_{57} + 8\hat{\sigma}^z_{52}\hat{\sigma}^z_{58} + 8\hat{\sigma}^z_{52}\hat{\sigma}^z_{59} - 4\hat{\sigma}^z_{52}\hat{\sigma}^z_{6} - 4\hat{\sigma}^z_{52}\hat{\sigma}^z_{7} - 4\hat{\sigma}^z_{52}\hat{\sigma}^z_{8} - 4\hat{\sigma}^z_{52}\hat{\sigma}^z_{9} - 404\hat{\sigma}^z_{52} + 8\hat{\sigma}^z_{53}\hat{\sigma}^z_{54} + 8\hat{\sigma}^z_{53}\hat{\sigma}^z_{55} + 8\hat{\sigma}^z_{53}\hat{\sigma}^z_{56} + 8\hat{\sigma}^z_{53}\hat{\sigma}^z_{57} + 8\hat{\sigma}^z_{53}\hat{\sigma}^z_{58} + 8\hat{\sigma}^z_{53}\hat{\sigma}^z_{59} - 4\hat{\sigma}^z_{53}\hat{\sigma}^z_{6} - 4\hat{\sigma}^z_{53}\hat{\sigma}^z_{7} - 4\hat{\sigma}^z_{53}\hat{\sigma}^z_{8} - 4\hat{\sigma}^z_{53}\hat{\sigma}^z_{9} - 404\hat{\sigma}^z_{53} + 8\hat{\sigma}^z_{54}\hat{\sigma}^z_{55} + 8\hat{\sigma}^z_{54}\hat{\sigma}^z_{56} + 8\hat{\sigma}^z_{54}\hat{\sigma}^z_{57} + 8\hat{\sigma}^z_{54}\hat{\sigma}^z_{58} + 8\hat{\sigma}^z_{54}\hat{\sigma}^z_{59} - 360\hat{\sigma}^z_{54}\hat{\sigma}^z_{6} - 4\hat{\sigma}^z_{54}\hat{\sigma}^z_{7} - 4\hat{\sigma}^z_{54}\hat{\sigma}^z_{8} - 4\hat{\sigma}^z_{54}\hat{\sigma}^z_{9} - 404\hat{\sigma}^z_{54} + 8\hat{\sigma}^z_{55}\hat{\sigma}^z_{56} + 8\hat{\sigma}^z_{55}\hat{\sigma}^z_{57} + 8\hat{\sigma}^z_{55}\hat{\sigma}^z_{58} + 8\hat{\sigma}^z_{55}\hat{\sigma}^z_{59} - 4\hat{\sigma}^z_{55}\hat{\sigma}^z_{6} - 360\hat{\sigma}^z_{55}\hat{\sigma}^z_{7} - 4\hat{\sigma}^z_{55}\hat{\sigma}^z_{8} - 4\hat{\sigma}^z_{55}\hat{\sigma}^z_{9} - 404\hat{\sigma}^z_{55} + 8\hat{\sigma}^z_{56}\hat{\sigma}^z_{57} + 8\hat{\sigma}^z_{56}\hat{\sigma}^z_{58} + 8\hat{\sigma}^z_{56}\hat{\sigma}^z_{59} - 4\hat{\sigma}^z_{56}\hat{\sigma}^z_{6} - 4\hat{\sigma}^z_{56}\hat{\sigma}^z_{7} - 360\hat{\sigma}^z_{56}\hat{\sigma}^z_{8} - 4\hat{\sigma}^z_{56}\hat{\sigma}^z_{9} - 404\hat{\sigma}^z_{56} + 8\hat{\sigma}^z_{57}\hat{\sigma}^z_{58} + 8\hat{\sigma}^z_{57}\hat{\sigma}^z_{59} - 4\hat{\sigma}^z_{57}\hat{\sigma}^z_{6} - 4\hat{\sigma}^z_{57}\hat{\sigma}^z_{7} - 4\hat{\sigma}^z_{57}\hat{\sigma}^z_{8} - 360\hat{\sigma}^z_{57}\hat{\sigma}^z_{9} - 404\hat{\sigma}^z_{57} + 8\hat{\sigma}^z_{58}\hat{\sigma}^z_{59} - 4\hat{\sigma}^z_{58}\hat{\sigma}^z_{6} - 4\hat{\sigma}^z_{58}\hat{\sigma}^z_{7} - 4\hat{\sigma}^z_{58}\hat{\sigma}^z_{8} - 4\hat{\sigma}^z_{58}\hat{\sigma}^z_{9} - 404\hat{\sigma}^z_{58} - 4\hat{\sigma}^z_{59}\hat{\sigma}^z_{6} - 4\hat{\sigma}^z_{59}\hat{\sigma}^z_{7} - 4\hat{\sigma}^z_{59}\hat{\sigma}^z_{8} - 4\hat{\sigma}^z_{59}\hat{\sigma}^z_{9} - 404\hat{\sigma}^z_{59} + 4\hat{\sigma}^z_{6}\hat{\sigma}^z_{7} + 4\hat{\sigma}^z_{6}\hat{\sigma}^z_{8} + 4\hat{\sigma}^z_{6}\hat{\sigma}^z_{9} + 404\hat{\sigma}^z_{6} + 8\hat{\sigma}^z_{60}\hat{\sigma}^z_{61} + 8\hat{\sigma}^z_{60}\hat{\sigma}^z_{62} + 8\hat{\sigma}^z_{60}\hat{\sigma}^z_{63} + 8\hat{\sigma}^z_{60}\hat{\sigma}^z_{64} + 8\hat{\sigma}^z_{60}\hat{\sigma}^z_{65} + 8\hat{\sigma}^z_{60}\hat{\sigma}^z_{66} + 8\hat{\sigma}^z_{60}\hat{\sigma}^z_{67} + 8\hat{\sigma}^z_{60}\hat{\sigma}^z_{68} + 8\hat{\sigma}^z_{60}\hat{\sigma}^z_{69} + 8\hat{\sigma}^z_{60}\hat{\sigma}^z_{70} + 8\hat{\sigma}^z_{60}\hat{\sigma}^z_{71} - 404\hat{\sigma}^z_{60} + 8\hat{\sigma}^z_{61}\hat{\sigma}^z_{62} + 8\hat{\sigma}^z_{61}\hat{\sigma}^z_{63} + 8\hat{\sigma}^z_{61}\hat{\sigma}^z_{64} + 8\hat{\sigma}^z_{61}\hat{\sigma}^z_{65} + 8\hat{\sigma}^z_{61}\hat{\sigma}^z_{66} + 8\hat{\sigma}^z_{61}\hat{\sigma}^z_{67} + 8\hat{\sigma}^z_{61}\hat{\sigma}^z_{68} + 8\hat{\sigma}^z_{61}\hat{\sigma}^z_{69} + 8\hat{\sigma}^z_{61}\hat{\sigma}^z_{70} + 8\hat{\sigma}^z_{61}\hat{\sigma}^z_{71} - 404\hat{\sigma}^z_{61} + 8\hat{\sigma}^z_{62}\hat{\sigma}^z_{63} + 8\hat{\sigma}^z_{62}\hat{\sigma}^z_{64} + 8\hat{\sigma}^z_{62}\hat{\sigma}^z_{65} + 8\hat{\sigma}^z_{62}\hat{\sigma}^z_{66} + 8\hat{\sigma}^z_{62}\hat{\sigma}^z_{67} + 8\hat{\sigma}^z_{62}\hat{\sigma}^z_{68} + 8\hat{\sigma}^z_{62}\hat{\sigma}^z_{69} + 8\hat{\sigma}^z_{62}\hat{\sigma}^z_{70} + 8\hat{\sigma}^z_{62}\hat{\sigma}^z_{71} - 404\hat{\sigma}^z_{62} + 8\hat{\sigma}^z_{63}\hat{\sigma}^z_{64} + 8\hat{\sigma}^z_{63}\hat{\sigma}^z_{65} + 8\hat{\sigma}^z_{63}\hat{\sigma}^z_{66} + 8\hat{\sigma}^z_{63}\hat{\sigma}^z_{67} + 8\hat{\sigma}^z_{63}\hat{\sigma}^z_{68} + 8\hat{\sigma}^z_{63}\hat{\sigma}^z_{69} + 8\hat{\sigma}^z_{63}\hat{\sigma}^z_{70} + 8\hat{\sigma}^z_{63}\hat{\sigma}^z_{71} - 404\hat{\sigma}^z_{63} + 8\hat{\sigma}^z_{64}\hat{\sigma}^z_{65} + 8\hat{\sigma}^z_{64}\hat{\sigma}^z_{66} + 8\hat{\sigma}^z_{64}\hat{\sigma}^z_{67} + 8\hat{\sigma}^z_{64}\hat{\sigma}^z_{68} + 8\hat{\sigma}^z_{64}\hat{\sigma}^z_{69} + 8\hat{\sigma}^z_{64}\hat{\sigma}^z_{70} + 8\hat{\sigma}^z_{64}\hat{\sigma}^z_{71} - 404\hat{\sigma}^z_{64} + 8\hat{\sigma}^z_{65}\hat{\sigma}^z_{66} + 8\hat{\sigma}^z_{65}\hat{\sigma}^z_{67} + 8\hat{\sigma}^z_{65}\hat{\sigma}^z_{68} + 8\hat{\sigma}^z_{65}\hat{\sigma}^z_{69} + 8\hat{\sigma}^z_{65}\hat{\sigma}^z_{70} + 8\hat{\sigma}^z_{65}\hat{\sigma}^z_{71} - 404\hat{\sigma}^z_{65} + 8\hat{\sigma}^z_{66}\hat{\sigma}^z_{67} + 8\hat{\sigma}^z_{66}\hat{\sigma}^z_{68} + 8\hat{\sigma}^z_{66}\hat{\sigma}^z_{69} + 8\hat{\sigma}^z_{66}\hat{\sigma}^z_{70} + 8\hat{\sigma}^z_{66}\hat{\sigma}^z_{71} - 404\hat{\sigma}^z_{66} + 8\hat{\sigma}^z_{67}\hat{\sigma}^z_{68} + 8\hat{\sigma}^z_{67}\hat{\sigma}^z_{69} + 8\hat{\sigma}^z_{67}\hat{\sigma}^z_{70} + 8\hat{\sigma}^z_{67}\hat{\sigma}^z_{71} - 404\hat{\sigma}^z_{67} + 8\hat{\sigma}^z_{68}\hat{\sigma}^z_{69} + 8\hat{\sigma}^z_{68}\hat{\sigma}^z_{70} + 8\hat{\sigma}^z_{68}\hat{\sigma}^z_{71} - 404\hat{\sigma}^z_{68} + 8\hat{\sigma}^z_{69}\hat{\sigma}^z_{70} + 8\hat{\sigma}^z_{69}\hat{\sigma}^z_{71} - 404\hat{\sigma}^z_{69} + 4\hat{\sigma}^z_{7}\hat{\sigma}^z_{8} + 4\hat{\sigma}^z_{7}\hat{\sigma}^z_{9} + 404\hat{\sigma}^z_{7} + 8\hat{\sigma}^z_{70}\hat{\sigma}^z_{71} - 404\hat{\sigma}^z_{70} - 404\hat{\sigma}^z_{71} + 4\hat{\sigma}^z_{8}\hat{\sigma}^z_{9} + 404\hat{\sigma}^z_{8} + 404\hat{\sigma}^z_{9} + 19,872$


\subsection*{A complete expression of Eq.(\ref{H2_COMPLETION_M12})}

$ \hat{H}_2\left(\hat{\sigma}_i^z\right) = 2\hat{\sigma}^z_{0}\hat{\sigma}^z_{1} + 2\hat{\sigma}^z_{0}\hat{\sigma}^z_{10} - 2\hat{\sigma}^z_{0}\hat{\sigma}^z_{11} + 2\hat{\sigma}^z_{0}\hat{\sigma}^z_{12} - 2\hat{\sigma}^z_{0}\hat{\sigma}^z_{13} + 2\hat{\sigma}^z_{0}\hat{\sigma}^z_{14} - 2\hat{\sigma}^z_{0}\hat{\sigma}^z_{15} + 2\hat{\sigma}^z_{0}\hat{\sigma}^z_{16} + 2\hat{\sigma}^z_{0}\hat{\sigma}^z_{17} - 2\hat{\sigma}^z_{0}\hat{\sigma}^z_{18} - 2\hat{\sigma}^z_{0}\hat{\sigma}^z_{19} - 2\hat{\sigma}^z_{0}\hat{\sigma}^z_{2} - 2\hat{\sigma}^z_{0}\hat{\sigma}^z_{20} - 2\hat{\sigma}^z_{0}\hat{\sigma}^z_{21} + 2\hat{\sigma}^z_{0}\hat{\sigma}^z_{22} + 2\hat{\sigma}^z_{0}\hat{\sigma}^z_{23} - 2\hat{\sigma}^z_{0}\hat{\sigma}^z_{24} + 2\hat{\sigma}^z_{0}\hat{\sigma}^z_{25} + 2\hat{\sigma}^z_{0}\hat{\sigma}^z_{26} + 2\hat{\sigma}^z_{0}\hat{\sigma}^z_{27} - 2\hat{\sigma}^z_{0}\hat{\sigma}^z_{3} + 2\hat{\sigma}^z_{0}\hat{\sigma}^z_{4} + 2\hat{\sigma}^z_{0}\hat{\sigma}^z_{5} + 2\hat{\sigma}^z_{0}\hat{\sigma}^z_{6} + 2\hat{\sigma}^z_{0}\hat{\sigma}^z_{7} - 2\hat{\sigma}^z_{0}\hat{\sigma}^z_{8} - 2\hat{\sigma}^z_{0}\hat{\sigma}^z_{9} - 2\hat{\sigma}^z_{1}\hat{\sigma}^z_{10} + 2\hat{\sigma}^z_{1}\hat{\sigma}^z_{11} - 2\hat{\sigma}^z_{1}\hat{\sigma}^z_{12} + 2\hat{\sigma}^z_{1}\hat{\sigma}^z_{13} - 2\hat{\sigma}^z_{1}\hat{\sigma}^z_{14} + 2\hat{\sigma}^z_{1}\hat{\sigma}^z_{15} - 2\hat{\sigma}^z_{1}\hat{\sigma}^z_{16} - 2\hat{\sigma}^z_{1}\hat{\sigma}^z_{17} + 2\hat{\sigma}^z_{1}\hat{\sigma}^z_{18} + 2\hat{\sigma}^z_{1}\hat{\sigma}^z_{19} + 2\hat{\sigma}^z_{1}\hat{\sigma}^z_{2} + 2\hat{\sigma}^z_{1}\hat{\sigma}^z_{20} + 2\hat{\sigma}^z_{1}\hat{\sigma}^z_{21} - 2\hat{\sigma}^z_{1}\hat{\sigma}^z_{22} - 2\hat{\sigma}^z_{1}\hat{\sigma}^z_{23} + 2\hat{\sigma}^z_{1}\hat{\sigma}^z_{24} - 2\hat{\sigma}^z_{1}\hat{\sigma}^z_{25} - 2\hat{\sigma}^z_{1}\hat{\sigma}^z_{26} - 2\hat{\sigma}^z_{1}\hat{\sigma}^z_{27} + 2\hat{\sigma}^z_{1}\hat{\sigma}^z_{3} - 2\hat{\sigma}^z_{1}\hat{\sigma}^z_{4} - 2\hat{\sigma}^z_{1}\hat{\sigma}^z_{5} - 2\hat{\sigma}^z_{1}\hat{\sigma}^z_{6} - 2\hat{\sigma}^z_{1}\hat{\sigma}^z_{7} + 2\hat{\sigma}^z_{1}\hat{\sigma}^z_{8} + 2\hat{\sigma}^z_{1}\hat{\sigma}^z_{9} + 2\hat{\sigma}^z_{10}\hat{\sigma}^z_{11} - 2\hat{\sigma}^z_{10}\hat{\sigma}^z_{12} + 2\hat{\sigma}^z_{10}\hat{\sigma}^z_{13} - 2\hat{\sigma}^z_{10}\hat{\sigma}^z_{14} + 2\hat{\sigma}^z_{10}\hat{\sigma}^z_{15} - 2\hat{\sigma}^z_{10}\hat{\sigma}^z_{16} - 2\hat{\sigma}^z_{10}\hat{\sigma}^z_{17} + 2\hat{\sigma}^z_{10}\hat{\sigma}^z_{18} + 2\hat{\sigma}^z_{10}\hat{\sigma}^z_{19} + 2\hat{\sigma}^z_{10}\hat{\sigma}^z_{2} + 2\hat{\sigma}^z_{10}\hat{\sigma}^z_{20} + 2\hat{\sigma}^z_{10}\hat{\sigma}^z_{21} - 2\hat{\sigma}^z_{10}\hat{\sigma}^z_{22} - 2\hat{\sigma}^z_{10}\hat{\sigma}^z_{23} + 2\hat{\sigma}^z_{10}\hat{\sigma}^z_{24} - 2\hat{\sigma}^z_{10}\hat{\sigma}^z_{25} - 2\hat{\sigma}^z_{10}\hat{\sigma}^z_{26} - 2\hat{\sigma}^z_{10}\hat{\sigma}^z_{27} + 2\hat{\sigma}^z_{10}\hat{\sigma}^z_{3} - 2\hat{\sigma}^z_{10}\hat{\sigma}^z_{4} - 2\hat{\sigma}^z_{10}\hat{\sigma}^z_{5} - 2\hat{\sigma}^z_{10}\hat{\sigma}^z_{6} - 2\hat{\sigma}^z_{10}\hat{\sigma}^z_{7} + 2\hat{\sigma}^z_{10}\hat{\sigma}^z_{8} + 2\hat{\sigma}^z_{10}\hat{\sigma}^z_{9} + 2\hat{\sigma}^z_{11}\hat{\sigma}^z_{12} - 2\hat{\sigma}^z_{11}\hat{\sigma}^z_{13} + 2\hat{\sigma}^z_{11}\hat{\sigma}^z_{14} - 2\hat{\sigma}^z_{11}\hat{\sigma}^z_{15} + 2\hat{\sigma}^z_{11}\hat{\sigma}^z_{16} + 2\hat{\sigma}^z_{11}\hat{\sigma}^z_{17} - 2\hat{\sigma}^z_{11}\hat{\sigma}^z_{18} - 2\hat{\sigma}^z_{11}\hat{\sigma}^z_{19} - 2\hat{\sigma}^z_{11}\hat{\sigma}^z_{2} - 2\hat{\sigma}^z_{11}\hat{\sigma}^z_{20} - 2\hat{\sigma}^z_{11}\hat{\sigma}^z_{21} + 2\hat{\sigma}^z_{11}\hat{\sigma}^z_{22} + 2\hat{\sigma}^z_{11}\hat{\sigma}^z_{23} - 2\hat{\sigma}^z_{11}\hat{\sigma}^z_{24} + 2\hat{\sigma}^z_{11}\hat{\sigma}^z_{25} + 2\hat{\sigma}^z_{11}\hat{\sigma}^z_{26} + 2\hat{\sigma}^z_{11}\hat{\sigma}^z_{27} - 2\hat{\sigma}^z_{11}\hat{\sigma}^z_{3} + 2\hat{\sigma}^z_{11}\hat{\sigma}^z_{4} + 2\hat{\sigma}^z_{11}\hat{\sigma}^z_{5} + 2\hat{\sigma}^z_{11}\hat{\sigma}^z_{6} + 2\hat{\sigma}^z_{11}\hat{\sigma}^z_{7} - 2\hat{\sigma}^z_{11}\hat{\sigma}^z_{8} - 2\hat{\sigma}^z_{11}\hat{\sigma}^z_{9} + 2\hat{\sigma}^z_{12}\hat{\sigma}^z_{13} - 2\hat{\sigma}^z_{12}\hat{\sigma}^z_{14} + 2\hat{\sigma}^z_{12}\hat{\sigma}^z_{15} - 2\hat{\sigma}^z_{12}\hat{\sigma}^z_{16} - 2\hat{\sigma}^z_{12}\hat{\sigma}^z_{17} + 2\hat{\sigma}^z_{12}\hat{\sigma}^z_{18} + 2\hat{\sigma}^z_{12}\hat{\sigma}^z_{19} + 2\hat{\sigma}^z_{12}\hat{\sigma}^z_{2} + 2\hat{\sigma}^z_{12}\hat{\sigma}^z_{20} + 2\hat{\sigma}^z_{12}\hat{\sigma}^z_{21} - 2\hat{\sigma}^z_{12}\hat{\sigma}^z_{22} - 2\hat{\sigma}^z_{12}\hat{\sigma}^z_{23} + 2\hat{\sigma}^z_{12}\hat{\sigma}^z_{24} - 2\hat{\sigma}^z_{12}\hat{\sigma}^z_{25} - 2\hat{\sigma}^z_{12}\hat{\sigma}^z_{26} - 2\hat{\sigma}^z_{12}\hat{\sigma}^z_{27} + 2\hat{\sigma}^z_{12}\hat{\sigma}^z_{3} - 2\hat{\sigma}^z_{12}\hat{\sigma}^z_{4} - 2\hat{\sigma}^z_{12}\hat{\sigma}^z_{5} - 2\hat{\sigma}^z_{12}\hat{\sigma}^z_{6} - 2\hat{\sigma}^z_{12}\hat{\sigma}^z_{7} + 2\hat{\sigma}^z_{12}\hat{\sigma}^z_{8} + 2\hat{\sigma}^z_{12}\hat{\sigma}^z_{9} + 2\hat{\sigma}^z_{13}\hat{\sigma}^z_{14} - 2\hat{\sigma}^z_{13}\hat{\sigma}^z_{15} + 2\hat{\sigma}^z_{13}\hat{\sigma}^z_{16} + 2\hat{\sigma}^z_{13}\hat{\sigma}^z_{17} - 2\hat{\sigma}^z_{13}\hat{\sigma}^z_{18} - 2\hat{\sigma}^z_{13}\hat{\sigma}^z_{19} - 2\hat{\sigma}^z_{13}\hat{\sigma}^z_{2} - 2\hat{\sigma}^z_{13}\hat{\sigma}^z_{20} - 2\hat{\sigma}^z_{13}\hat{\sigma}^z_{21} + 2\hat{\sigma}^z_{13}\hat{\sigma}^z_{22} + 2\hat{\sigma}^z_{13}\hat{\sigma}^z_{23} - 2\hat{\sigma}^z_{13}\hat{\sigma}^z_{24} + 2\hat{\sigma}^z_{13}\hat{\sigma}^z_{25} + 2\hat{\sigma}^z_{13}\hat{\sigma}^z_{26} + 2\hat{\sigma}^z_{13}\hat{\sigma}^z_{27} - 2\hat{\sigma}^z_{13}\hat{\sigma}^z_{3} + 2\hat{\sigma}^z_{13}\hat{\sigma}^z_{4} + 2\hat{\sigma}^z_{13}\hat{\sigma}^z_{5} + 2\hat{\sigma}^z_{13}\hat{\sigma}^z_{6} + 2\hat{\sigma}^z_{13}\hat{\sigma}^z_{7} - 2\hat{\sigma}^z_{13}\hat{\sigma}^z_{8} - 2\hat{\sigma}^z_{13}\hat{\sigma}^z_{9} + 2\hat{\sigma}^z_{14}\hat{\sigma}^z_{15} - 2\hat{\sigma}^z_{14}\hat{\sigma}^z_{16} - 2\hat{\sigma}^z_{14}\hat{\sigma}^z_{17} + 2\hat{\sigma}^z_{14}\hat{\sigma}^z_{18} + 2\hat{\sigma}^z_{14}\hat{\sigma}^z_{19} + 2\hat{\sigma}^z_{14}\hat{\sigma}^z_{2} + 2\hat{\sigma}^z_{14}\hat{\sigma}^z_{20} + 2\hat{\sigma}^z_{14}\hat{\sigma}^z_{21} - 2\hat{\sigma}^z_{14}\hat{\sigma}^z_{22} - 2\hat{\sigma}^z_{14}\hat{\sigma}^z_{23} + 2\hat{\sigma}^z_{14}\hat{\sigma}^z_{24} - 2\hat{\sigma}^z_{14}\hat{\sigma}^z_{25} - 2\hat{\sigma}^z_{14}\hat{\sigma}^z_{26} - 2\hat{\sigma}^z_{14}\hat{\sigma}^z_{27} + 2\hat{\sigma}^z_{14}\hat{\sigma}^z_{3} - 2\hat{\sigma}^z_{14}\hat{\sigma}^z_{4} - 2\hat{\sigma}^z_{14}\hat{\sigma}^z_{5} - 2\hat{\sigma}^z_{14}\hat{\sigma}^z_{6} - 2\hat{\sigma}^z_{14}\hat{\sigma}^z_{7} + 2\hat{\sigma}^z_{14}\hat{\sigma}^z_{8} + 2\hat{\sigma}^z_{14}\hat{\sigma}^z_{9} + 2\hat{\sigma}^z_{15}\hat{\sigma}^z_{16} + 2\hat{\sigma}^z_{15}\hat{\sigma}^z_{17} - 2\hat{\sigma}^z_{15}\hat{\sigma}^z_{18} - 2\hat{\sigma}^z_{15}\hat{\sigma}^z_{19} - 2\hat{\sigma}^z_{15}\hat{\sigma}^z_{2} - 2\hat{\sigma}^z_{15}\hat{\sigma}^z_{20} - 2\hat{\sigma}^z_{15}\hat{\sigma}^z_{21} + 2\hat{\sigma}^z_{15}\hat{\sigma}^z_{22} + 2\hat{\sigma}^z_{15}\hat{\sigma}^z_{23} - 2\hat{\sigma}^z_{15}\hat{\sigma}^z_{24} + 2\hat{\sigma}^z_{15}\hat{\sigma}^z_{25} + 2\hat{\sigma}^z_{15}\hat{\sigma}^z_{26} + 2\hat{\sigma}^z_{15}\hat{\sigma}^z_{27} - 2\hat{\sigma}^z_{15}\hat{\sigma}^z_{3} + 2\hat{\sigma}^z_{15}\hat{\sigma}^z_{4} + 2\hat{\sigma}^z_{15}\hat{\sigma}^z_{5} + 2\hat{\sigma}^z_{15}\hat{\sigma}^z_{6} + 2\hat{\sigma}^z_{15}\hat{\sigma}^z_{7} - 2\hat{\sigma}^z_{15}\hat{\sigma}^z_{8} - 2\hat{\sigma}^z_{15}\hat{\sigma}^z_{9} - 2\hat{\sigma}^z_{16}\hat{\sigma}^z_{17} + 2\hat{\sigma}^z_{16}\hat{\sigma}^z_{18} + 2\hat{\sigma}^z_{16}\hat{\sigma}^z_{19} + 2\hat{\sigma}^z_{16}\hat{\sigma}^z_{2} + 2\hat{\sigma}^z_{16}\hat{\sigma}^z_{20} + 2\hat{\sigma}^z_{16}\hat{\sigma}^z_{21} - 2\hat{\sigma}^z_{16}\hat{\sigma}^z_{22} - 2\hat{\sigma}^z_{16}\hat{\sigma}^z_{23} + 2\hat{\sigma}^z_{16}\hat{\sigma}^z_{24} - 2\hat{\sigma}^z_{16}\hat{\sigma}^z_{25} - 2\hat{\sigma}^z_{16}\hat{\sigma}^z_{26} - 2\hat{\sigma}^z_{16}\hat{\sigma}^z_{27} + 2\hat{\sigma}^z_{16}\hat{\sigma}^z_{3} - 2\hat{\sigma}^z_{16}\hat{\sigma}^z_{4} - 2\hat{\sigma}^z_{16}\hat{\sigma}^z_{5} - 2\hat{\sigma}^z_{16}\hat{\sigma}^z_{6} - 2\hat{\sigma}^z_{16}\hat{\sigma}^z_{7} + 2\hat{\sigma}^z_{16}\hat{\sigma}^z_{8} + 2\hat{\sigma}^z_{16}\hat{\sigma}^z_{9} + 2\hat{\sigma}^z_{17}\hat{\sigma}^z_{18} + 2\hat{\sigma}^z_{17}\hat{\sigma}^z_{19} + 2\hat{\sigma}^z_{17}\hat{\sigma}^z_{2} + 2\hat{\sigma}^z_{17}\hat{\sigma}^z_{20} + 2\hat{\sigma}^z_{17}\hat{\sigma}^z_{21} - 2\hat{\sigma}^z_{17}\hat{\sigma}^z_{22} - 2\hat{\sigma}^z_{17}\hat{\sigma}^z_{23} + 2\hat{\sigma}^z_{17}\hat{\sigma}^z_{24} - 2\hat{\sigma}^z_{17}\hat{\sigma}^z_{25} - 2\hat{\sigma}^z_{17}\hat{\sigma}^z_{26} - 2\hat{\sigma}^z_{17}\hat{\sigma}^z_{27} + 2\hat{\sigma}^z_{17}\hat{\sigma}^z_{3} - 2\hat{\sigma}^z_{17}\hat{\sigma}^z_{4} - 2\hat{\sigma}^z_{17}\hat{\sigma}^z_{5} - 2\hat{\sigma}^z_{17}\hat{\sigma}^z_{6} - 2\hat{\sigma}^z_{17}\hat{\sigma}^z_{7} + 2\hat{\sigma}^z_{17}\hat{\sigma}^z_{8} + 2\hat{\sigma}^z_{17}\hat{\sigma}^z_{9} - 2\hat{\sigma}^z_{18}\hat{\sigma}^z_{19} - 2\hat{\sigma}^z_{18}\hat{\sigma}^z_{2} - 2\hat{\sigma}^z_{18}\hat{\sigma}^z_{20} - 2\hat{\sigma}^z_{18}\hat{\sigma}^z_{21} + 2\hat{\sigma}^z_{18}\hat{\sigma}^z_{22} + 2\hat{\sigma}^z_{18}\hat{\sigma}^z_{23} - 2\hat{\sigma}^z_{18}\hat{\sigma}^z_{24} + 2\hat{\sigma}^z_{18}\hat{\sigma}^z_{25} + 2\hat{\sigma}^z_{18}\hat{\sigma}^z_{26} + 2\hat{\sigma}^z_{18}\hat{\sigma}^z_{27} - 2\hat{\sigma}^z_{18}\hat{\sigma}^z_{3} + 2\hat{\sigma}^z_{18}\hat{\sigma}^z_{4} + 2\hat{\sigma}^z_{18}\hat{\sigma}^z_{5} + 2\hat{\sigma}^z_{18}\hat{\sigma}^z_{6} + 2\hat{\sigma}^z_{18}\hat{\sigma}^z_{7} - 2\hat{\sigma}^z_{18}\hat{\sigma}^z_{8} - 2\hat{\sigma}^z_{18}\hat{\sigma}^z_{9} - 2\hat{\sigma}^z_{19}\hat{\sigma}^z_{2} - 2\hat{\sigma}^z_{19}\hat{\sigma}^z_{20} - 2\hat{\sigma}^z_{19}\hat{\sigma}^z_{21} + 2\hat{\sigma}^z_{19}\hat{\sigma}^z_{22} + 2\hat{\sigma}^z_{19}\hat{\sigma}^z_{23} - 2\hat{\sigma}^z_{19}\hat{\sigma}^z_{24} + 2\hat{\sigma}^z_{19}\hat{\sigma}^z_{25} + 2\hat{\sigma}^z_{19}\hat{\sigma}^z_{26} + 2\hat{\sigma}^z_{19}\hat{\sigma}^z_{27} - 2\hat{\sigma}^z_{19}\hat{\sigma}^z_{3} + 2\hat{\sigma}^z_{19}\hat{\sigma}^z_{4} + 2\hat{\sigma}^z_{19}\hat{\sigma}^z_{5} + 2\hat{\sigma}^z_{19}\hat{\sigma}^z_{6} + 2\hat{\sigma}^z_{19}\hat{\sigma}^z_{7} - 2\hat{\sigma}^z_{19}\hat{\sigma}^z_{8} - 2\hat{\sigma}^z_{19}\hat{\sigma}^z_{9} - 2\hat{\sigma}^z_{2}\hat{\sigma}^z_{20} - 2\hat{\sigma}^z_{2}\hat{\sigma}^z_{21} + 2\hat{\sigma}^z_{2}\hat{\sigma}^z_{22} + 2\hat{\sigma}^z_{2}\hat{\sigma}^z_{23} - 2\hat{\sigma}^z_{2}\hat{\sigma}^z_{24} + 2\hat{\sigma}^z_{2}\hat{\sigma}^z_{25} + 2\hat{\sigma}^z_{2}\hat{\sigma}^z_{26} + 2\hat{\sigma}^z_{2}\hat{\sigma}^z_{27} - 2\hat{\sigma}^z_{2}\hat{\sigma}^z_{3} + 2\hat{\sigma}^z_{2}\hat{\sigma}^z_{4} + 2\hat{\sigma}^z_{2}\hat{\sigma}^z_{5} + 2\hat{\sigma}^z_{2}\hat{\sigma}^z_{6} + 2\hat{\sigma}^z_{2}\hat{\sigma}^z_{7} - 2\hat{\sigma}^z_{2}\hat{\sigma}^z_{8} - 2\hat{\sigma}^z_{2}\hat{\sigma}^z_{9} - 2\hat{\sigma}^z_{20}\hat{\sigma}^z_{21} + 2\hat{\sigma}^z_{20}\hat{\sigma}^z_{22} + 2\hat{\sigma}^z_{20}\hat{\sigma}^z_{23} - 2\hat{\sigma}^z_{20}\hat{\sigma}^z_{24} + 2\hat{\sigma}^z_{20}\hat{\sigma}^z_{25} + 2\hat{\sigma}^z_{20}\hat{\sigma}^z_{26} + 2\hat{\sigma}^z_{20}\hat{\sigma}^z_{27} - 2\hat{\sigma}^z_{20}\hat{\sigma}^z_{3} + 2\hat{\sigma}^z_{20}\hat{\sigma}^z_{4} + 2\hat{\sigma}^z_{20}\hat{\sigma}^z_{5} + 2\hat{\sigma}^z_{20}\hat{\sigma}^z_{6} + 2\hat{\sigma}^z_{20}\hat{\sigma}^z_{7} - 2\hat{\sigma}^z_{20}\hat{\sigma}^z_{8} - 2\hat{\sigma}^z_{20}\hat{\sigma}^z_{9} + 2\hat{\sigma}^z_{21}\hat{\sigma}^z_{22} + 2\hat{\sigma}^z_{21}\hat{\sigma}^z_{23} - 2\hat{\sigma}^z_{21}\hat{\sigma}^z_{24} + 2\hat{\sigma}^z_{21}\hat{\sigma}^z_{25} + 2\hat{\sigma}^z_{21}\hat{\sigma}^z_{26} + 2\hat{\sigma}^z_{21}\hat{\sigma}^z_{27} - 2\hat{\sigma}^z_{21}\hat{\sigma}^z_{3} + 2\hat{\sigma}^z_{21}\hat{\sigma}^z_{4} + 2\hat{\sigma}^z_{21}\hat{\sigma}^z_{5} + 2\hat{\sigma}^z_{21}\hat{\sigma}^z_{6} + 2\hat{\sigma}^z_{21}\hat{\sigma}^z_{7} - 2\hat{\sigma}^z_{21}\hat{\sigma}^z_{8} - 2\hat{\sigma}^z_{21}\hat{\sigma}^z_{9} - 2\hat{\sigma}^z_{22}\hat{\sigma}^z_{23} + 2\hat{\sigma}^z_{22}\hat{\sigma}^z_{24} - 2\hat{\sigma}^z_{22}\hat{\sigma}^z_{25} - 2\hat{\sigma}^z_{22}\hat{\sigma}^z_{26} - 2\hat{\sigma}^z_{22}\hat{\sigma}^z_{27} + 2\hat{\sigma}^z_{22}\hat{\sigma}^z_{3} - 2\hat{\sigma}^z_{22}\hat{\sigma}^z_{4} - 2\hat{\sigma}^z_{22}\hat{\sigma}^z_{5} - 2\hat{\sigma}^z_{22}\hat{\sigma}^z_{6} - 2\hat{\sigma}^z_{22}\hat{\sigma}^z_{7} + 2\hat{\sigma}^z_{22}\hat{\sigma}^z_{8} + 2\hat{\sigma}^z_{22}\hat{\sigma}^z_{9} + 2\hat{\sigma}^z_{23}\hat{\sigma}^z_{24} - 2\hat{\sigma}^z_{23}\hat{\sigma}^z_{25} - 2\hat{\sigma}^z_{23}\hat{\sigma}^z_{26} - 2\hat{\sigma}^z_{23}\hat{\sigma}^z_{27} + 2\hat{\sigma}^z_{23}\hat{\sigma}^z_{3} - 2\hat{\sigma}^z_{23}\hat{\sigma}^z_{4} - 2\hat{\sigma}^z_{23}\hat{\sigma}^z_{5} - 2\hat{\sigma}^z_{23}\hat{\sigma}^z_{6} - 2\hat{\sigma}^z_{23}\hat{\sigma}^z_{7} + 2\hat{\sigma}^z_{23}\hat{\sigma}^z_{8} + 2\hat{\sigma}^z_{23}\hat{\sigma}^z_{9} + 2\hat{\sigma}^z_{24}\hat{\sigma}^z_{25} + 2\hat{\sigma}^z_{24}\hat{\sigma}^z_{26} + 2\hat{\sigma}^z_{24}\hat{\sigma}^z_{27} - 2\hat{\sigma}^z_{24}\hat{\sigma}^z_{3} + 2\hat{\sigma}^z_{24}\hat{\sigma}^z_{4} + 2\hat{\sigma}^z_{24}\hat{\sigma}^z_{5} + 2\hat{\sigma}^z_{24}\hat{\sigma}^z_{6} + 2\hat{\sigma}^z_{24}\hat{\sigma}^z_{7} - 2\hat{\sigma}^z_{24}\hat{\sigma}^z_{8} - 2\hat{\sigma}^z_{24}\hat{\sigma}^z_{9} - 2\hat{\sigma}^z_{25}\hat{\sigma}^z_{26} - 2\hat{\sigma}^z_{25}\hat{\sigma}^z_{27} + 2\hat{\sigma}^z_{25}\hat{\sigma}^z_{3} - 2\hat{\sigma}^z_{25}\hat{\sigma}^z_{4} - 2\hat{\sigma}^z_{25}\hat{\sigma}^z_{5} - 2\hat{\sigma}^z_{25}\hat{\sigma}^z_{6} - 2\hat{\sigma}^z_{25}\hat{\sigma}^z_{7} + 2\hat{\sigma}^z_{25}\hat{\sigma}^z_{8} + 2\hat{\sigma}^z_{25}\hat{\sigma}^z_{9} - 2\hat{\sigma}^z_{26}\hat{\sigma}^z_{27} + 2\hat{\sigma}^z_{26}\hat{\sigma}^z_{3} - 2\hat{\sigma}^z_{26}\hat{\sigma}^z_{4} - 2\hat{\sigma}^z_{26}\hat{\sigma}^z_{5} - 2\hat{\sigma}^z_{26}\hat{\sigma}^z_{6} - 2\hat{\sigma}^z_{26}\hat{\sigma}^z_{7} + 2\hat{\sigma}^z_{26}\hat{\sigma}^z_{8} + 2\hat{\sigma}^z_{26}\hat{\sigma}^z_{9} + 2\hat{\sigma}^z_{27}\hat{\sigma}^z_{3} - 2\hat{\sigma}^z_{27}\hat{\sigma}^z_{4} - 2\hat{\sigma}^z_{27}\hat{\sigma}^z_{5} - 2\hat{\sigma}^z_{27}\hat{\sigma}^z_{6} - 2\hat{\sigma}^z_{27}\hat{\sigma}^z_{7} + 2\hat{\sigma}^z_{27}\hat{\sigma}^z_{8} + 2\hat{\sigma}^z_{27}\hat{\sigma}^z_{9} + 2\hat{\sigma}^z_{3}\hat{\sigma}^z_{4} + 2\hat{\sigma}^z_{3}\hat{\sigma}^z_{5} + 2\hat{\sigma}^z_{3}\hat{\sigma}^z_{6} + 2\hat{\sigma}^z_{3}\hat{\sigma}^z_{7} - 2\hat{\sigma}^z_{3}\hat{\sigma}^z_{8} - 2\hat{\sigma}^z_{3}\hat{\sigma}^z_{9} - 2\hat{\sigma}^z_{4}\hat{\sigma}^z_{5} - 2\hat{\sigma}^z_{4}\hat{\sigma}^z_{6} - 2\hat{\sigma}^z_{4}\hat{\sigma}^z_{7} + 2\hat{\sigma}^z_{4}\hat{\sigma}^z_{8} + 2\hat{\sigma}^z_{4}\hat{\sigma}^z_{9} - 2\hat{\sigma}^z_{5}\hat{\sigma}^z_{6} - 2\hat{\sigma}^z_{5}\hat{\sigma}^z_{7} + 2\hat{\sigma}^z_{5}\hat{\sigma}^z_{8} + 2\hat{\sigma}^z_{5}\hat{\sigma}^z_{9} - 2\hat{\sigma}^z_{6}\hat{\sigma}^z_{7} + 2\hat{\sigma}^z_{6}\hat{\sigma}^z_{8} + 2\hat{\sigma}^z_{6}\hat{\sigma}^z_{9} + 2\hat{\sigma}^z_{7}\hat{\sigma}^z_{8} + 2\hat{\sigma}^z_{7}\hat{\sigma}^z_{9} - 2\hat{\sigma}^z_{8}\hat{\sigma}^z_{9} + 756 $

\end{document}